\documentclass[twocolumn,superscriptaddress,showpacs,prd,aps,amsmath,amssymb,nofootinbib]{revtex4-1}
\usepackage{graphicx,color}
\usepackage{amsmath,amssymb}
\usepackage{verbatim}
\usepackage{float}
\usepackage{wasysym}
\usepackage{amssymb,graphicx}
\usepackage{epsfig}
\usepackage{psfrag}
\usepackage{dsfont}
\usepackage{amsfonts}
\usepackage{mathrsfs}
\usepackage{multirow}
\usepackage{bm}
\usepackage{hyperref}
\hypersetup{
  colorlinks=true,        
  linkcolor=blue,         
  citecolor=cyan,         
}
\begin{document}
\title{Jet Launching from Merging Magnetized Binary Neutron Stars with Realistic Equations of State}
\author{Milton Ruiz}
\affiliation{Department of Physics, University of Illinois at
  Urbana-Champaign, Urbana, IL 61801}
\author{Antonios Tsokaros}
\affiliation{Department of Physics, University of Illinois at
  Urbana-Champaign, Urbana, IL 61801}
\author{Stuart L. Shapiro}
\affiliation{Department of Physics, University of Illinois at
  Urbana-Champaign, Urbana, IL 61801}
\affiliation{Department of Astronomy \& NCSA, University of
  Illinois at Urbana-Champaign, Urbana, IL 61801}
%
%
\begin{abstract}
  We perform general relativistic, magnetohydrodynamic (GRMHD) simulations  of binary neutron stars
  in quasi-circular orbit that merge and undergo delayed or prompt collapse to a black hole (BH).
  The stars are irrotational and  modeled using an SLy or an H4 nuclear equation of state.
  To assess the impact of the initial magnetic field configuration on jet launching, we
  endow the stars with a purely poloidal magnetic field that is initially unimportant dynamically
  and is either confined to the stellar interior or extends from the interior into the
  exterior as in typical pulsars. Consistent with our previous results, we find that only the BH + disk
  remnants originating from binaries that form hypermassive neutron stars (HMNSs) and undergo
  delayed collapse can drive magnetically-powered jets. We find that the closer the total
  mass of the binary is to the threshold value for prompt collapse, the shorter is the time delay
  between the gravitational wave peak amplitude and jet launching. This time delay also strongly
  depends on the initial magnetic field configuration. We also find that seed magnetic fields
  confined to the stellar interior can launch a jet over~$\sim 25\,\rm ms$ later than those with
  pulsar-like magnetic fields. The lifetime of the jet [$\Delta t\lesssim 150\,\rm ms$] and its
  outgoing Poynting luminosity [$L_{\rm EM}\sim 10^{52\pm 1}\rm erg/s$] are consistent with typical
  short gamma-ray burst central engine lifetimes, as well as with the Blandford--Znajek mechanism
  for launching jets and their associated Poynting luminosities. Our numerical results also suggest
  that the dynamical ejection of matter can be enhanced by the magnetic field. Therefore, GRMHD
  studies are required to fully understand kilonova signals from GW170818-like events.
\end{abstract}

\pacs{04.25.D-, 04.25.dk, 04.30.-w, 47.75.+f}
\maketitle


\section{Introduction}
The coincident detection of  gravitational waves (GWs) with counterparts across the electromagnetic
(EM) spectrum from GW170817, whose source has been officially classified as a merging binary neutron
star~\cite{TheLIGOScientific:2017qsa,FERMI2017GCN,Savchenko:2017ffs}, triggered the beginning of 
multimessenger astronomy. This single multimesssenger event provides: {{i})} the first direct
evidence that compact binary mergers where at least one of the  companions is a neutron star can
be the progenitors of the central engine that powers short gamma-ray bursts (sGRBs). This conclusion
was anticipated in~\cite{NaPaPi,EiLiPiSc,Pac86ApJ} and confirmed by self-consistent simulations in
full general relativistic, magnetohydrodynamics (GRMHD) of merging binary neutron stars (NSNSs)
\cite{Ruiz:2016rai,Ruiz:2017inq}, and binary black hole-neutron stars (BHNSs)~\cite{prs15,Ruiz:2018wah};
ii) tight constraints on the equation of state (EOS)
at supranuclear densities~\cite{LIGOScientific:2018cki,Radice:2017lry}; iii) limits to the maximum
mass of neutron stars~\cite{Margalit:2017dij,Ruiz:2017due,Rezzolla:2017aly,Shibata:2019ctb};
iv) evidence of ejecta masses of $\approx 0.01-0.05 M_\odot$ with velocities~$\approx 0.1-0.3\rm\,c$
\cite{Cowperthwaite:2017dyu,Kasliwal:2017ngb,Smartt:2017fuw}. This ejecta is roughly consistent
with the estimated r-process production rate required to explain the Milky Way r-process
abundances; and v) an independent measure for the expansion of the Universe
\cite{LIGOScientific:2017adf,Dietrich:2020efo}. GW170817 also demonstrated that to understand
multimessenger observations and, in particular, to understand the physics of matter under extreme
conditions, it is crucial to compare them to predictions  from theoretical modeling. Due to the
complexity of the underlying physical phenomena, such modeling is largely numerical in nature.

Our GRMHD numerical simulations in~\cite{Ruiz:2016rai,Ruiz:2017inq,prs15,Ruiz:2018wah},
which model the NSs by a simple $\Gamma=2$, polytropic EOS, provided an existence proof for
jet launching.\footnote{We define an incipient jet as a tightly collimated, mildly relativistic
  outflow which is driven by a tightly wound, helical, force-free magnetic field
  (i.e.~$B^2/(8\,\pi\rho_0)\gg 1$, where $\rho_0$ is the rest-mass density, and $B^2=B_i\,B^i$,
  with $B^i$ the magnetic field~\cite{prs15}.)} However, to date there are no self-consistent, GRMHD
calculations of NSNS or BHNS mergers involving realistic EOSs, or detailed microphysical processes,
confirming these results. Such jets are believed to be crucial for launching a GRB~(see~e.g.
\cite{1993NYASA.688..321P,1994ApJ...424L.131M,1998AIPC..428.....M}).
The numerical studies of NSNS mergers undergoing delayed collapse to a BH reported in~\cite{Kawamura:2016nmk,
  Ciolfi:2017uak}, in which different EOSs, mass ratios, and orientations of a poloidal magnetic field confined
to the stellar interior, were explored  found no evidences of magnetically-driven outflows or incipient jets,
though their results confirm the formation of an ordered magnetic field above the
BH poles~(see~e.g.~Fig.~7~in~\cite{Kawamura:2016nmk}). It is likely that longer evolutions
and/or higher resolutions are required to properly capture the undergoing magnetic instabilities
for the emergence of an incipient jet. The formation of jet-like structures has been also reported
in~\cite{rgbgka11}.
On the other hand, the very high resolution
studies of NSNS mergers in~\cite{Kiuchi:2014hja} found no jets nor an ordered poloidal field
above the BH poles
after $\sim 39\,\rm ms$ following merger. By $t\sim 26\rm\,ms$ following BH formation,
there is still material in the  atmosphere  that is being accreted, so that the strong ram pressure
of the fall-back debris chokes the emergence of a magnetically-driven outflow. An incipient jet might
be launched once the baryon-loaded surrounding debris above the BH poles becomes sparse.
However,  the emergence of an incipient jet  may be possible only for EOSs for which
the fall-back timescale is shorter than the accretion disk lifetime~\cite{Paschalidis:2016agf}. 
On the other hand, microphysical processes may also  have  strong impact on the final
outcome of compact binary mergers and jets~(see~e.g.~\cite{Foucart:2021ikp,Mosta:2020hlh}).
For example, neutrino processes in BH + disk systems may extract a significant
amount of energy from inner regions of disk to power jets~\cite{Popham:1998ab,Matteo:2002ck,
  Chen:2006rra,Lei2013ApJ,Just:2015dba}. However, they cannot explain the duration of
typical sGRBs~(see~e.g.~\cite{Zou2009,Just:2015dba}). Nevertheless, it has been suggested that,  
in slow spinning BH + disk systems, jets can be initially triggered by neutrino pair
annihilation and then powered by the Blandford–Znajek (BZ) mechanism
\cite{Lei:2017zro,BZeffect77}, leading to a transition from a thermally dominated
fireball to a Poynting dominated outflow, as observed in some GRBs such as GRB 160625B
\cite{Dirirsa:2017pgm}.

As a crucial step in solidifying the role of NSNSs as multimessenger sources, we survey
in this paper configurations that undergo either delayed or prompt collapse and treat
different representative EOSs and  initial geometries of their magnetic field to probe their
impact on jet launching and the dynamical ejection of matter. 
In particular, the NSs are modeled using a piecewise polytropic representation of the nuclear 
SLy~\cite{Douchin01} and H4 EOSs~\cite{PhysRevLett.67.2414}, as in~\cite{Read:2008iy}.
We adopt these as representative of realistic candidate EOSs that are broadly consistent with
current data~\cite{Fonseca:2016tux,Antoniadis:2013pzd,NANOGrav:2019jur}.
For comparison, we also consider the binaries we considered in~\cite{Ruiz:2016rai,Ruiz:2017inq,Ruiz:2019ezy}
in which the stars are modeled using a simple polytropic EOS with $\Gamma=2$. For
the magnetic field, we endow the stars initially with a purely poloidal magnetic field that
either extends from the interior of the NSs into its exterior, as in pulsars, or that is
confined to the stellar interior. We also evolve unmagnetized configurations to assess the
impact of the magnetic field on the ejecta.  As  NSNS mergers tend to create very
baryon-loaded environments~(see e.g.~\cite{Just:2015dba,Ciolfi:2019fie}), we consider
binaries whose merger outcome is a short- ($1\lesssim\tau_{\rm HMNS}\lesssim 5\rm ms$),
medium- ($5\lesssim\tau_{\rm HMNS}\lesssim 20 \rm ms$) or long-lived ($\tau_{\rm HMNS}
\gtrsim 20\rm ms$) hypermassive neutron star (HMNS) followed by delayed collapse to a BH.
These choices will allow us  to probe the impact of light vs. heavy matter environments on
the physical properties of the incipient jet. Here $\tau_{\rm HMNS}$ is the lifetime of
the HMNS.

Consistent with our results reported in~\cite{Ruiz:2016rai,Ruiz:2017inq}, we find that incipient
jets only  emerge from binary remnants that undergo delayed collapse, regardless of the EOS. The
lifetime of the jet~[$\Delta t\sim 92-150\,\rm ms$] and its corresponding outgoing EM Poynting
luminosity~[$L_{\rm EM}\sim 10^{52\pm 1}\rm  erg/s$] are consistent with the lifetime of the  sGRB
central engine~\cite{Beniamini:2020adb,Bhat:2016odd,Lien:2016zny,Svinkin:2016fho}, as well as with
the Blandford-Znajek (BZ) mechanism for launching jets and their associated Poynting luminosities
\cite{BZeffect77}.  Our results can be summarized as follows:
i) the closer the total mass of the binary is to the threshold value for prompt collapse,
the shorter is the time delay between the GW peak amplitude (our definition of the moment of
coalescence) and the jet launching time. In particular, we find that the BH + disk
remnant of a stiff H4 NSNS configuration in which~$\tau_{\rm HMNS}~\sim 2.5\,\rm ms$ (labeled
as H4M3.0P in the discussion below) launches an incipient jet after $\sim 19\rm \,ms$ following BH
formation, while that of an H4 NSNS configuration in which~$\tau_{\rm HMNS}~\sim 9.6\,\rm ms$
(labeled H4M2.8P) launches a jet after $\sim 27\rm \,ms$ following BH formation;
ii) the jet launching time strongly depends  on the initial geometry of the seed
magnetic field. We observe that the BH + disk remnant of a soft SLy binary initially endowed
with a pulsar-like poloidal magnetic field (labeled as SLyM2.7P)
launches an incipient jet at~$t\sim 20\,\rm ms$  following BH formation, while the
same binary endowed with a poloidal magnetic field confined to the stellar interior (labeled as SLyM2.7I)
launches it at~$t~\sim 60\,\rm ms$ following BH formation. As we discuss later, during the BH + disk
phase the magnetic energy in the latter is a factor of $\sim 20$ times smaller than in the former.
Therefore, the BH + disk remnant in SLyM2.7I requires more time for magnetic pressure
gradients to overcome the ram-pressure of the falling-back debris and eventually 
launch an incipient jet.
On the other hand, in the prompt collapse case, the absence of an extended HMNS epoch prevents
the magnetic field from reaching equipartition strength above the remnant BH poles, thereby
preventing magnetic field collimation and a magnetically supported outflow;
iii) the dynamical ejection  of matter following merger strongly depends on the initial
magnetic field geometry.  Note that it has been suggested that the magnetic field lines
of a rotating compact object may accelerate fluid elements due to a
magnetocentrifugal mechanism~\cite{1982MNRAS.199..883B}. These results suggest
that magnetic field is a key ingredient in explaining kilonova signatures from
GW170817-like events. We use an analytical model recently derived in~\cite{Perego:2021dpw}
to compute the peak EM luminosity, the rise time and an effective temperature of the
potential kilonova. We find that the bolometric luminosity  is $L_{\rm knova}=10^{40.6\pm0.5}\rm erg/s$
with rise times~$\tau_{\rm peak}\sim 0.4-5.1$~days~and an effective temperature~$T_{\rm peak}\sim 10^{3.5}\rm\,K$.
This temperature can be translated in a peak wavelength $\lambda_{\rm peak}=1.35\times
{10^3}\,{\rm nm}\,({T_{\rm peak}}/10^{3.33}\,\rm{K})^{-1}$~\cite{Perego:2021dpw}, implying
$\lambda_{\rm peak}\sim 730-1830\,\rm nm$. The associated emission may be detected with
current or planned telescopes such as ALMA  or the Vera C. Rubin
Observatory~\cite{2018PASP..130a5002M,Chen:2020zoq}; and finally iv) using the GW match function
(see ~e.g.~\cite{Harry2018}), we  find that the imprints of the magnetic field on the gravitational
radiation can be observed by current based-ground detectors (aLIGO/Virgo/KAGRA) only if the GW event
occurs within at a distance $\lesssim 6.0\,\rm Mpc$. We recall that GW170817, the closest GW signal
detected to date, had a luminosity distance of $40^{+8}_{-14}\,\rm Mpc$
\cite{TheLIGOScientific:2017qsa}. If the GW event occurs within a distance $\lesssim 50\,\rm Mpc$,
these imprints can be observed only with next generation of GW observatories, such as the Einstein
Telescope or Cosmic Explorer~(see~e.g.~\cite{LIGOScientific:2016wof}),  with a sigma-to-noise ratio
(SNR) $\gtrsim 30$.

The remainder of the paper is organized as follows. A short summary of the numerical
methods and their implementation is presented in Sec.~\ref{subsec:Methods}. 
A detailed description of the adopted initial data and the grid structure used to evolve
the GRMHD equations numerically are given in Sec.~\ref{subsec:idata} and Sec.~\ref{subsec:grid},
respectively. A suite of diagnostics used to verify the reliability of our
numerical calculations is summarized in Sec.~\ref{subsec:diagnostics}. We present our results
in Sec.~\ref{sec:results}. Finally, we summarize our findings and conclusions in
Sec.~\ref{sec:conclusion}. Throughout the paper we adopt geometrized units ($G=c=1$)  except where
stated explicitly. Greek indices denote all four spacetime dimensions, while Latin indices
imply spatial parts only.

\section{Numerical Setup}
\subsection{Methods}
\label{subsec:Methods}
Much of the numerical approach employed here has been  extensively
discussed in previous works~(see~e.g.~\cite{Ruiz:2016rai,Ruiz:2020via,Tsokaros:2019lnx}).
Therefore, in the following we only summarize the basics aspects, referring the reader
to those references  for further details and code tests.

We use our well-tested {\tt Illinois GRMHD} code~\cite{Etienne:2010ui} which is embedded on the
{\tt Cactus} infrastructure~\cite{CactusConfigs} and employs the {\tt Carpet} code
\cite{Carpet,carpetweb} for its moving-box mesh capability. Our code evolves the
Baumgarte--Shapiro--Shibata--Nakamura (BSSN) equations~\cite{shibnak95,BS}, coupled
with puncture gauge conditions cast in first order form (see~Eq.~(2)-(4) in
\cite{Etienne:2007jg}), using  fourth order centered spatial differencing, except
on shift advection terms, where a fourth order upwind differencing is used.
Fifth order Kreiss-Oliger dissipation~\cite{goddard06} is also added in the BSSN
evolution equations. The matter and magnetic fields  are evolved using
the equations of ideal GRMHD, which are cast in a conservative scheme, via a high-resolution
shock capturing method (see~Eqs.~(27)-(29) in~\cite{Etienne:2010ui}). To ensure the magnetic
field remains divergenceless during the whole evolution, we integrate the magnetic induction
equation using a vector potential~$\mathcal{A}^\mu$ (see Eqs. (19)-(20) in~\cite{Etienne:2010ui}).
We adopt the generalized Lorenz gauge described in~\cite{Farris:2012ux} to avoid the
build up  of spurious magnetic fields~\cite{Etienne:2011re}. The time integration is
performed via the method of lines using a fourth-order accurate Runge-Kutta integration
scheme with a Courant-Friedrichs-Lewy (CFL) factor set to $0.5$. 
%
\subsection{Initial data}
\label{subsec:idata}

\paragraph{\bf Equation of State (EOS):}
We consider NSNS configurations  in a quasiequilibrium circular orbit that inspiral, merge and
undergo either delayed or prompt collapse to a BH. The binaries consist of two identical irrotational
NSs, modeled by a piecewise polytropic representation of the nuclear EOSs SLy (soft)~\cite{Douchin01} and
H4~(stiff)~\cite{PhysRevLett.67.2414}, as in~\cite{Read:2008iy}.  The initial binary data are computed using the
Compact Object CALculator~({\tt COCAL})~\cite{Tsokaros:2015fea,Tsokaros:2018dqs}, and their properties
are summarized in Table~\ref{table:NSNS_ID}.

We note that these representative EOSs broadly satisfy the current observational constraints on NSs. For
example, the
maximum mass configuration of an isolated star predicted by SLy is $M_{\rm sph}^{\rm max}=
2.06M_\odot$, while  that predicted by H4 is $M_{\rm sph}^{\rm max}=2.03M_\odot$.
Both are consistent with: i) $M_{\rm sph}^{\rm max}>2.072^{+0.067}_{-0.066}M_\odot$ from the NICER and XMM
analysis of PSR J0740+6620~\cite{Riley:2021pdl}; ii) $M_{\rm sph}^{\rm max}>2.01^{+0.017}_{-0.017}M_\odot$ from 
the NANOGrav analysis of PSR J1614-2230~\cite{Fonseca:2016tux}; iii) $M_{\rm sph}^{\rm max}>2.01^{+0.14}_{-0.14}
M_\odot$ from the pulsar timing analysis of PSR J0348+0432~\cite{Antoniadis:2013pzd}; and $M_{\rm sph}^{\rm max}
>2.14^{+0.20}_{-0.18}M_\odot$ from the NANOGrav and the Green Bank Telescope~\cite{NANOGrav:2019jur}.
We also note that SLy predicts that a star with a mass of $1.4M_\odot$ has a radius of $R= 11.46\,\rm km$,
consistent with the value $R = 11.94^{+0.76}_{-0.87}\rm\, km$ obtained by a combined analysis of X-ray and
GW measurements of PSR J0740+6620~\cite{Pang:2021jta}. Such a star modeled with a H4 EOS has a
radius of $R = 13.55\,\rm km$, which is just marginally outside of the above constraint. Consistent with this,
the combined analysis of the LIGO/Virgo scientific collaboration (LVSC) of the  progenitors of GW170817
with the radio-timing observations of the pulsar J0348+0432
\cite{LIGOScientific:2018cki,Antoniadis:2013pzd} constrain the radius of a NS with mass in the range
$1.16-1.6M_\odot$ to be $11.9^{+1.4}_{-1.4}\,\rm km$  at the  $90\%$ credible level. However, the NICER analysis
of PSR J0030+0451~\cite{Miller:2019cac} constrains the radius of a NS with mass of $1.44^{+0.15}_{-0.14}M_\odot$
to be $R= 13.02^{+1.24}_{-1.06}\, \rm km$. SLy and H4 EOSs predict that a $1.44M_\odot$ star has a radius of
$11.45\,\rm km$ and $13.54\,\rm km$, respectively. The former is hence slightly below the NICER constraint.
Furthermore, the LVSC analysis of GW170817 predicts that the tidal deformability of a $1.4\,M_\odot$
NS is $\Lambda_{1.4}= 190^{+390}_{-120}$ at the  $90\%$ credible level~\cite{LIGOScientific:2018cki}.
Such a star has $\Lambda_{1.4}=306.4$ and $\Lambda_{1.4}=886.6$ for SLy and H4, respectively. 
Therefore, the NICER analysis favors stiff (e.g.~H4) over soft EOSs (e.g.~SLy), while that of the LVSC favors
soft over stiff EOSs~\cite{Tsokaros:2020hli}.
%
%
\begin{center}
  \begin{table}
    \caption{Summary of the initial properties of the NSNS configurations.
      We list the EOS employed to model the NSs, the rest-mass $M_0$, the
       equatorial coordinate radius $R_x$ toward the companion of each star,
      the compactness $\mathcal{C}$  and the tidal deformability $\Lambda=
      (2/3)\,k_2\,{\cal C}^{-5}$ ( where $k_2$ is the second Love number), the
      ADM mass $M$, and the angular velocity $\Omega$, for an initial binary
      coordinate separation of $\sim 45\,\rm km$. The tag for each configuration
      is composed of the EOS followed by the binary ADM mass. 
      We also consider the
      $\Gamma=2$ NSNS configurations treated previously in~\cite{Ruiz:2016rai,Ruiz:2017inq,Ruiz:2019ezy}.
      \label{table:NSNS_ID}}
    \begin{tabular}{cccccccccccc}
      \hline\hline
          Model            & EOS         &$M_0\,[\rm M_\odot]$  &$R_x\,[\rm km]$ &  $\mathcal{C}$ & $\Lambda$ &$M\,[\rm M_\odot]$  & $M\,\Omega$ \\  
          {SLyM$2.6$}      & SLy         &  $1.45$              & $9.14$         &     $0.169$    &   $463$   &$2.6$               & $0.024$     \\
          {SLyM$2.7$}      & SLy         &  $1.51$              & $9.04$         &     $0.175$    &   $367$   &$2.7$               & $0.023$     \\          
          {SLyM$3.0$}      & SLy         &  $1.71$              & $8.72$         &     $0.197$    &   $173$   &$3.0$               & $0.031$     \\
          \hline
          {H4M$2.8$}       & H4          & $1.55$               & $11.12$        &     $0.155$    &   $818$    &$2.8$              & $0.025$     \\
          {H4M$3.0$}       & H4          & $1.67$               & $10.81$        &     $0.166$    &   $524$    &$3.0$              & $0.031$     \\
          \hline 
          {$\Gamma2$M2.8}  & $\Gamma=2$  & $1.51$               & $12.74$        &     $0.140$     &  $979$   &$2.8$              & $0.027$      \\
          {$\Gamma2$M3.0}  & $\Gamma=2$  & $1.67$               & $11.52$        &     $0.160$     &  $359$   &$3.0$              & $0.030$      \\
          \hline\hline
    \end{tabular}
  \end{table}
\end{center}

For comparison, we also consider the $\Gamma=2$ NSNS configurations we treated previously
in~\cite{Ruiz:2016rai,Ruiz:2017inq,Ruiz:2019ezy} that allow us to cover a large set of stellar
compactions. Note that $\Gamma=2$ is the stiffest EOS considered in this survey.
When using an $\Gamma$-law EOS,  we have the freedom to scale the masses and distance
to any value. Therefore,  we rescale the rest-mass of the $\Gamma=2$ star companions to match those
of SLyM2.7 (i.e.~$M_{0} = 1.51M_\odot\,(\kappa/\kappa_{L})^{1/2}$, where $\kappa_L=
232.93\,\rm km^2$ is the polytropic constant employed to compute the initial data).  We note that the
maximum mass predicted by the $\Gamma=2$ EOS is $M^{\rm max}_{\rm sph}=1.69(k/k_{\rm L})^{1/2} M_\odot$
and, that a $\Gamma=2$ star with a mass of $1.4\,M_\odot$ has a radius of $R=22.72(k/k_{\rm L})^{1/2}\,
\rm km$.

The NSNS merger outcome may be a remnant that can either  settle into a transient HMNS or promptly
form a highly spinning BH  upon merger. Given an EOS, this outcome depends strongly on the total
mass of the binary and it is independent of the mass-ratio~\cite{Shibata:2006nm}. The threshold mass
$M^{\rm thres}_{}$ for prompt collapse for a $\Gamma=2$ EOS is $\simeq 2.88(k/k_{\rm L})^{1/2} M_\odot$,
while for  SLy and H4   EOSs are  $M^{\rm thres}_{}\simeq 2.82M_\odot$ and $M^{\rm thres}_{}\simeq3.12M_\odot$,
respectively~\cite{Bauswein:2020aag}. Thus, it is expected that only the NSNS configurations SLyM3.0 and
$\Gamma2$M3.0 (see Table~\ref{table:NSNS_ID}) undergo prompt collapse to a BH. The merger outcome
in all other cases is a highly differentially rotating HMNS~\cite{BaShSh}.

A cold EOS is  adequate to model the NS prior to merger. However, during merger
considerable shock heating  increases the internal energy. To account for this, we adopt an EOS
that has both a thermal and cold contribution. The total pressure can be expressed as
\begin{equation}
  P=P_{\rm th} + P_{\rm cold}\,,
\end{equation}
where $P_{\rm cold}=\kappa_{i}\,\rho_0^{\Gamma_i}$, with $\kappa_i$ and $\Gamma_i$ the corresponding
polytropic constant and the polytropic exponent in the rest-mass density range $\rho_{0,i-1}\leq \rho_0
\leq \rho_{0,i}$, respectively (see~e.g.~\cite{Read:2008iy}), and  the thermal pressure is given by
\begin{equation}
  P_{\rm th}=(\Gamma_{\rm th}-1)\,\rho_0\,(\epsilon-\epsilon_{\rm cold})\,,
\end{equation}
where
\begin{equation}
\epsilon_{\rm cold} =-\int P_{\rm cold}\,d(1/\rho_0)\,,
\end{equation}
and $\Gamma_{\rm th}$ a constant that we set to $1.66\simeq 5/3$ in all our simulations.
This value is  appropriate for  ideal nonrelativistic baryons~\cite{Bauswein:2010dn,Paschalidis:2011ez}.

\begin{figure*}
  \centering
  \includegraphics[width=0.495\textwidth]{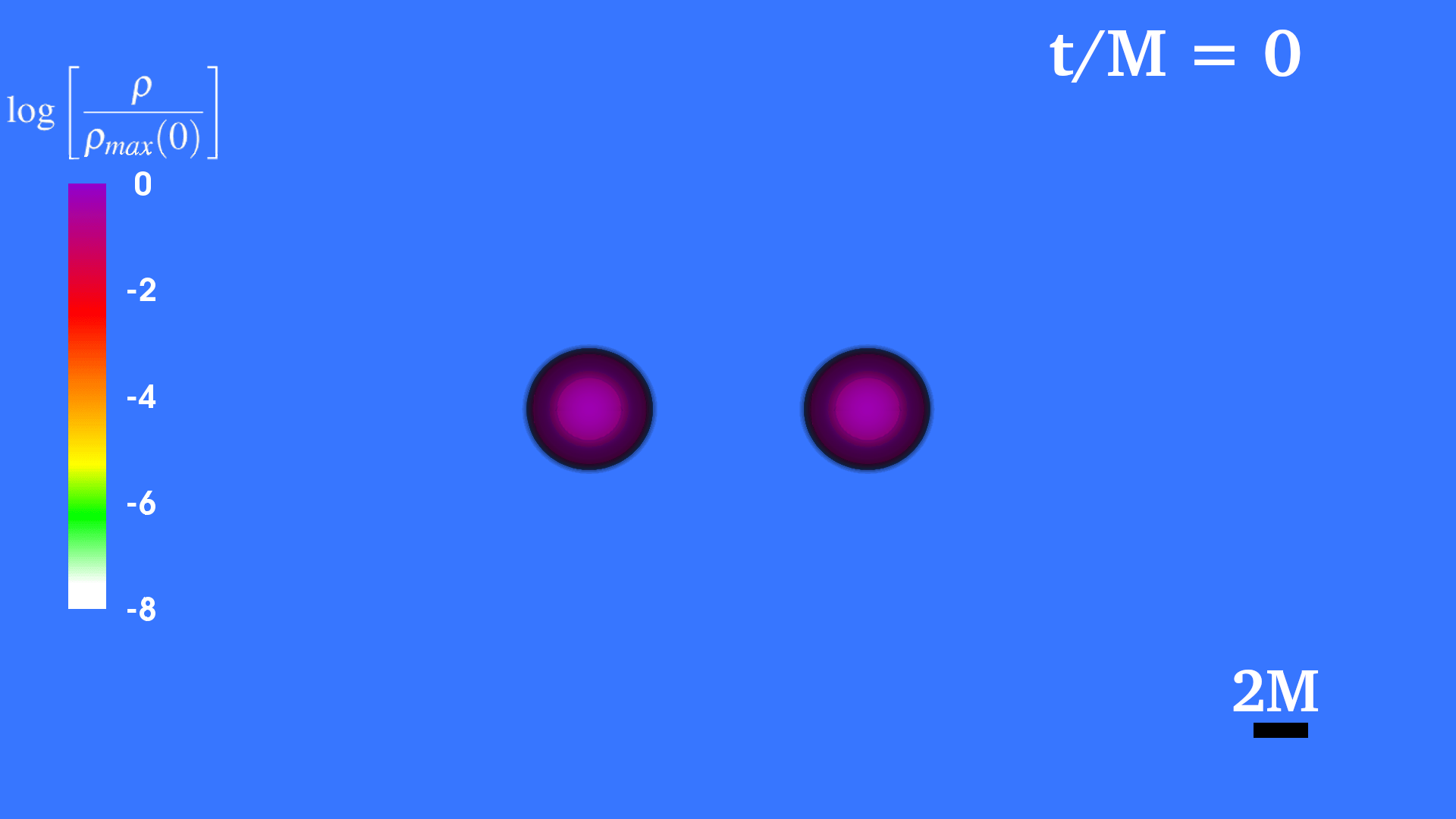}
  \includegraphics[width=0.495\textwidth]{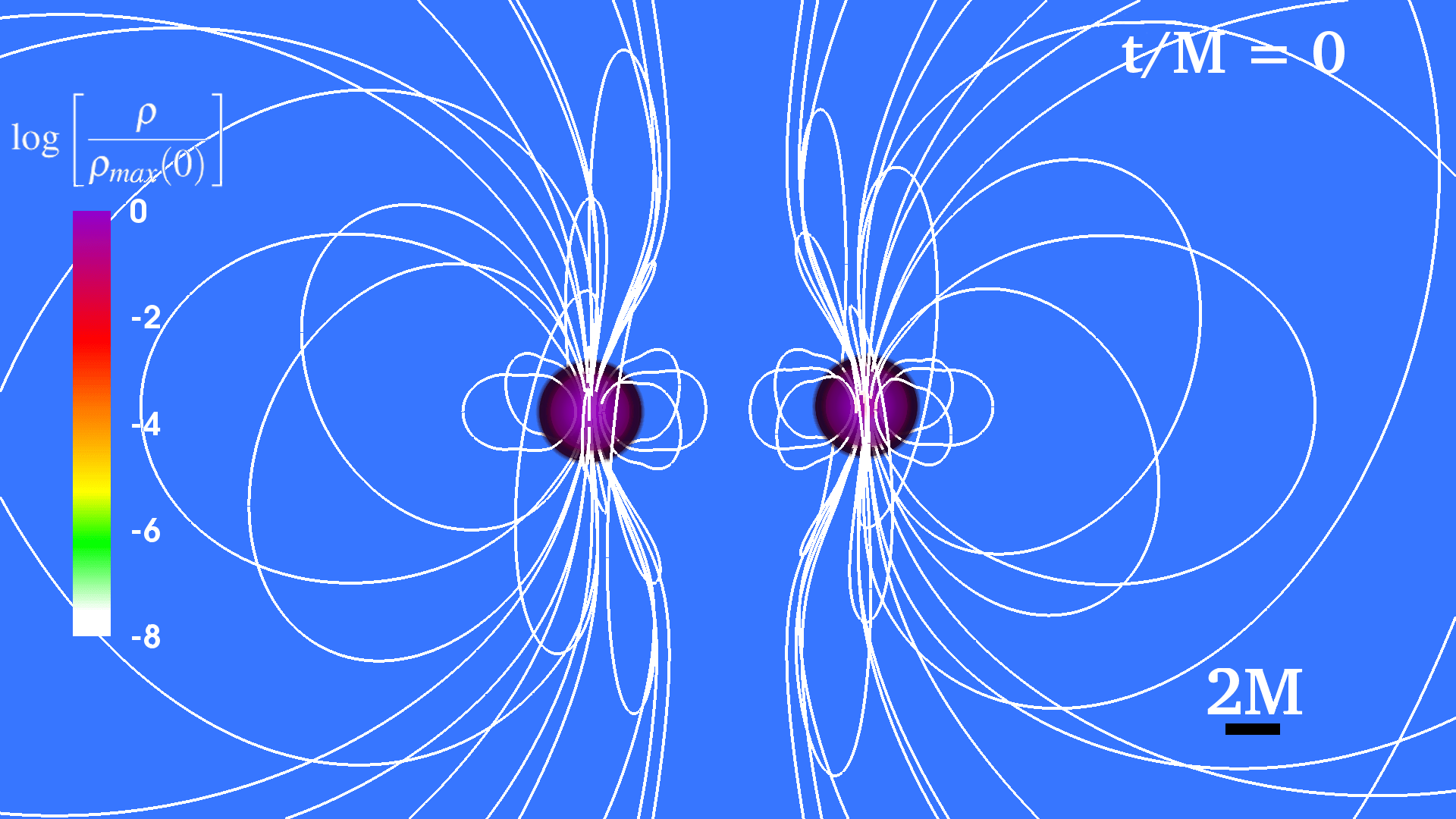}
  \includegraphics[width=0.495\textwidth]{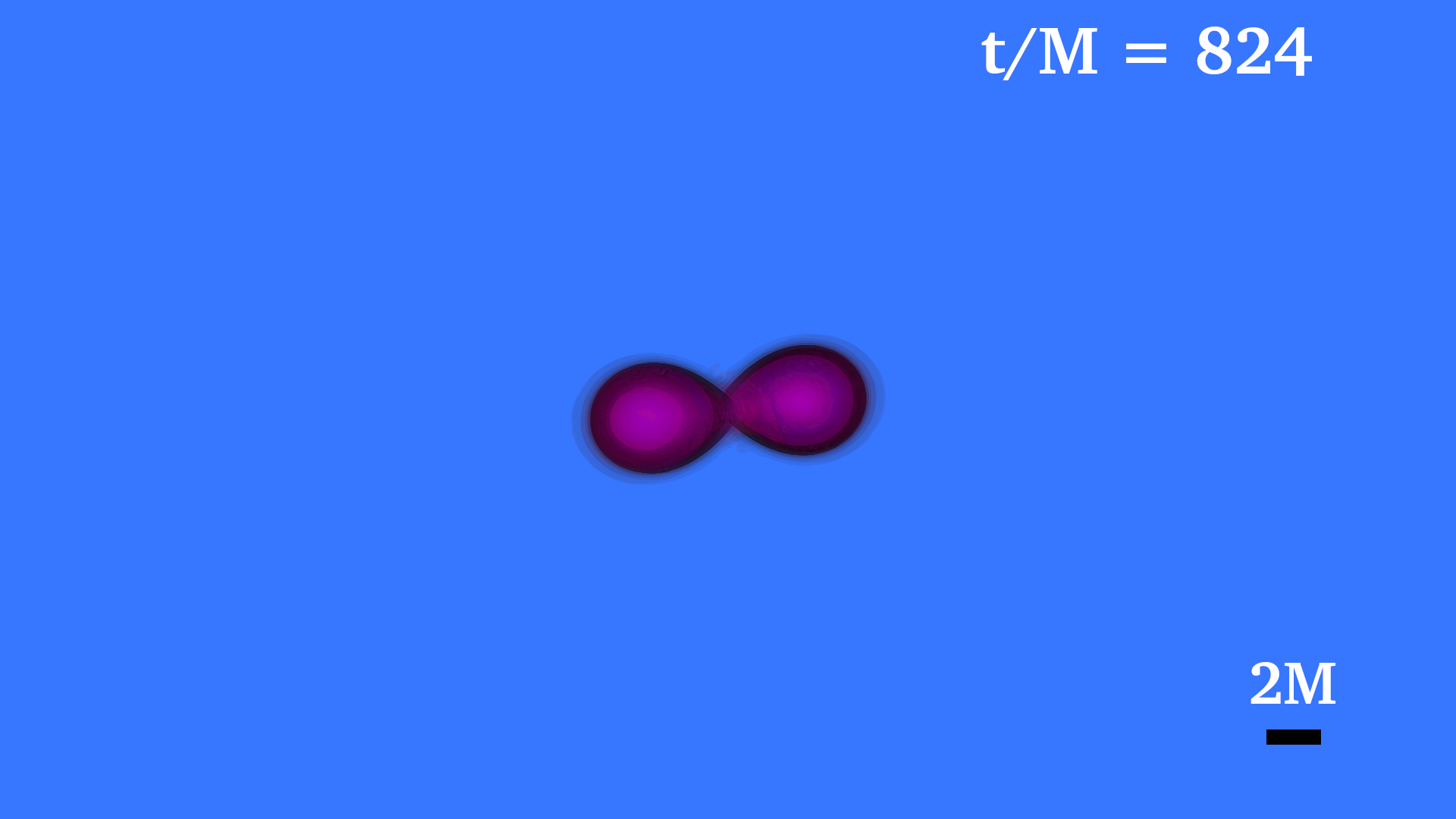}
  \includegraphics[width=0.495\textwidth]{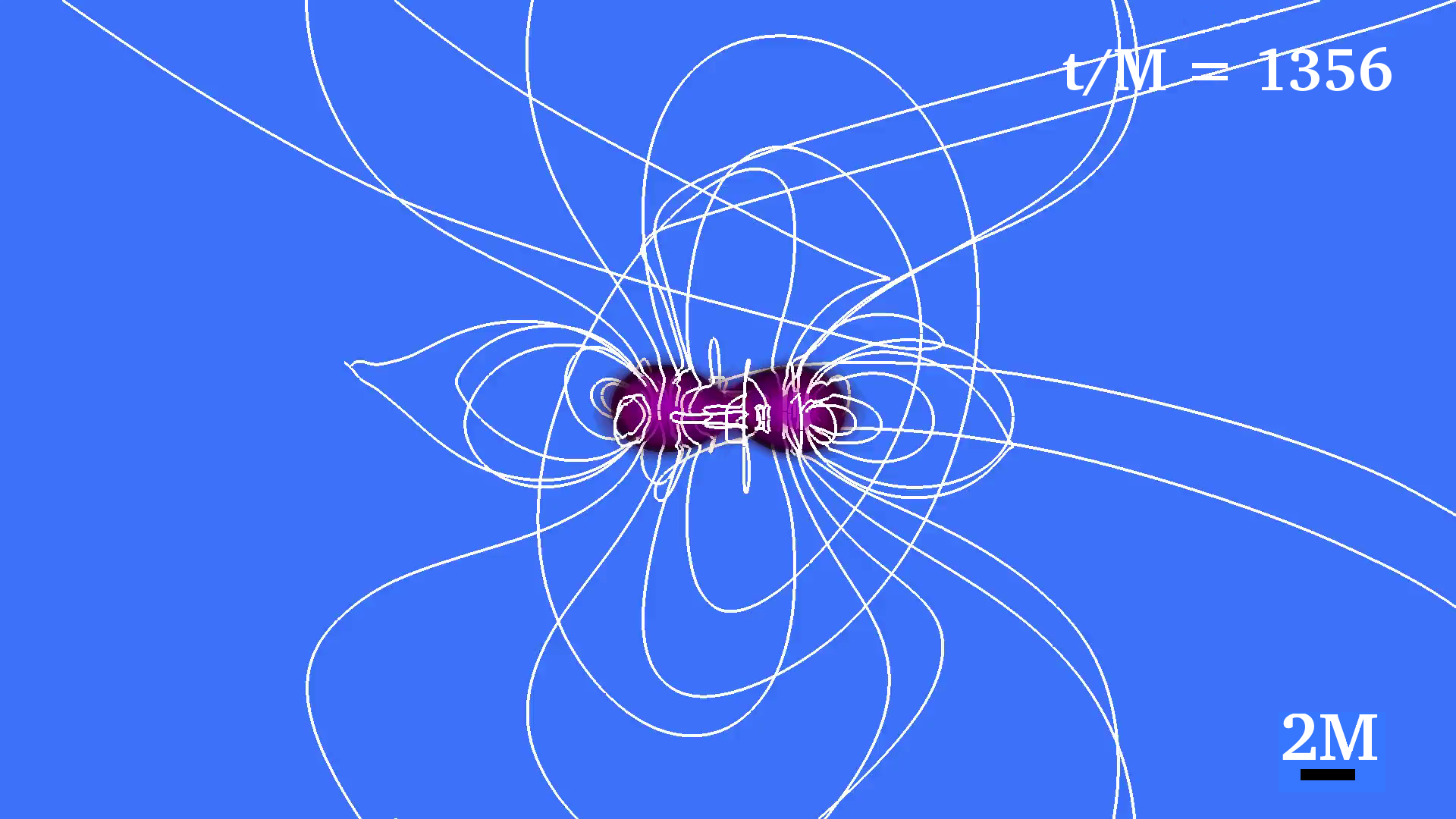}
  \includegraphics[width=0.495\textwidth]{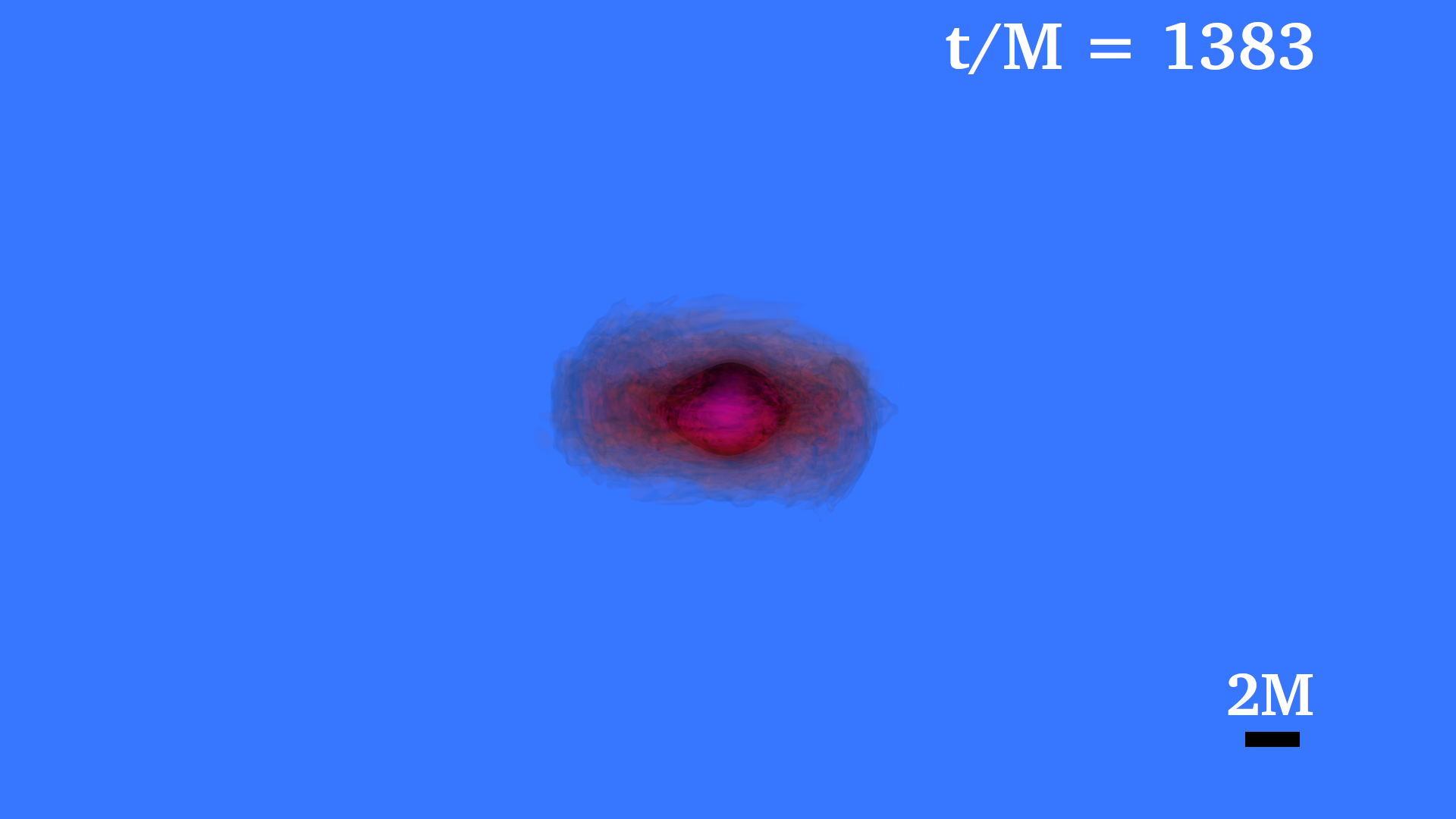}
  \includegraphics[width=0.495\textwidth]{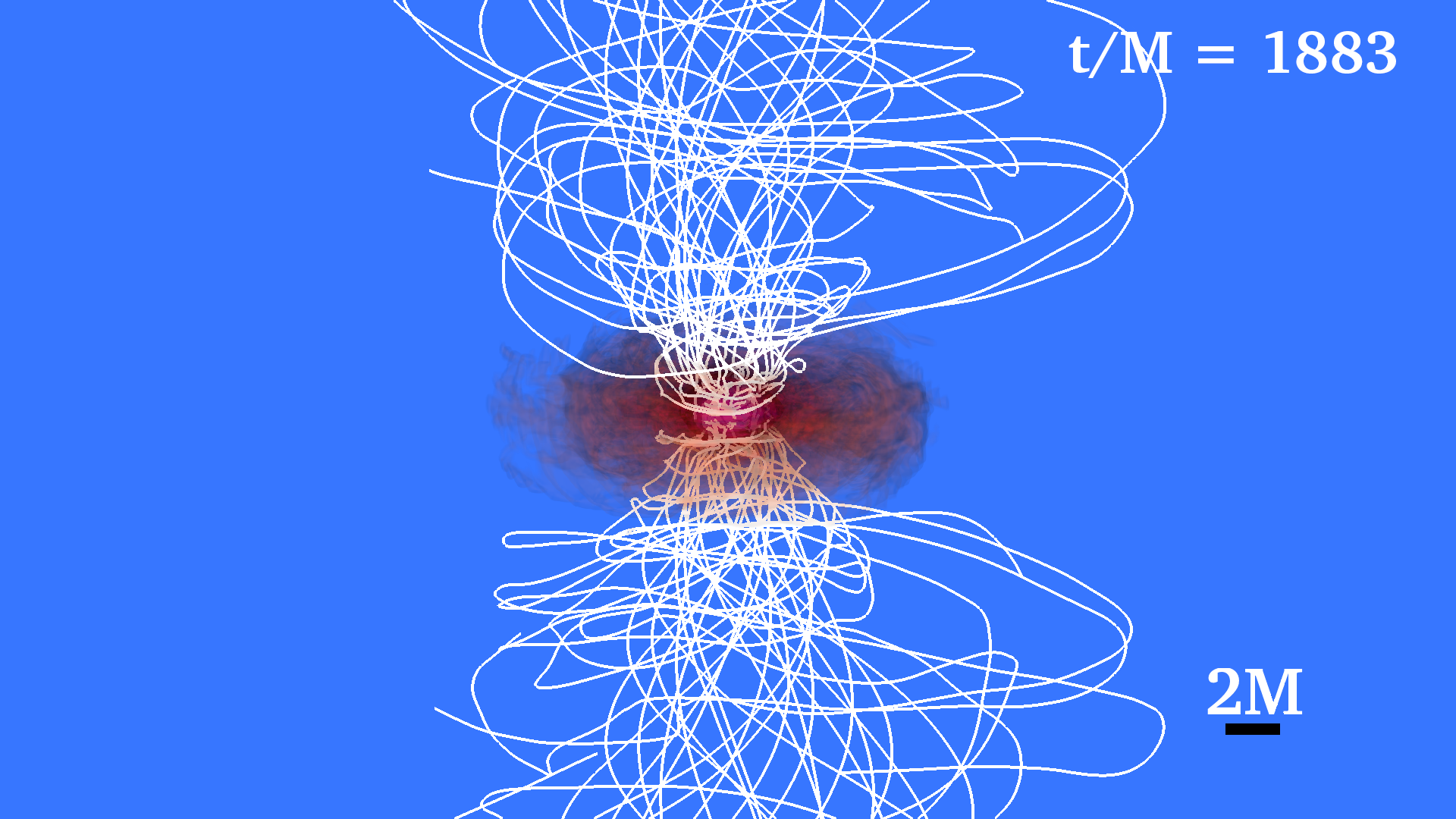}
  \includegraphics[width=0.495\textwidth]{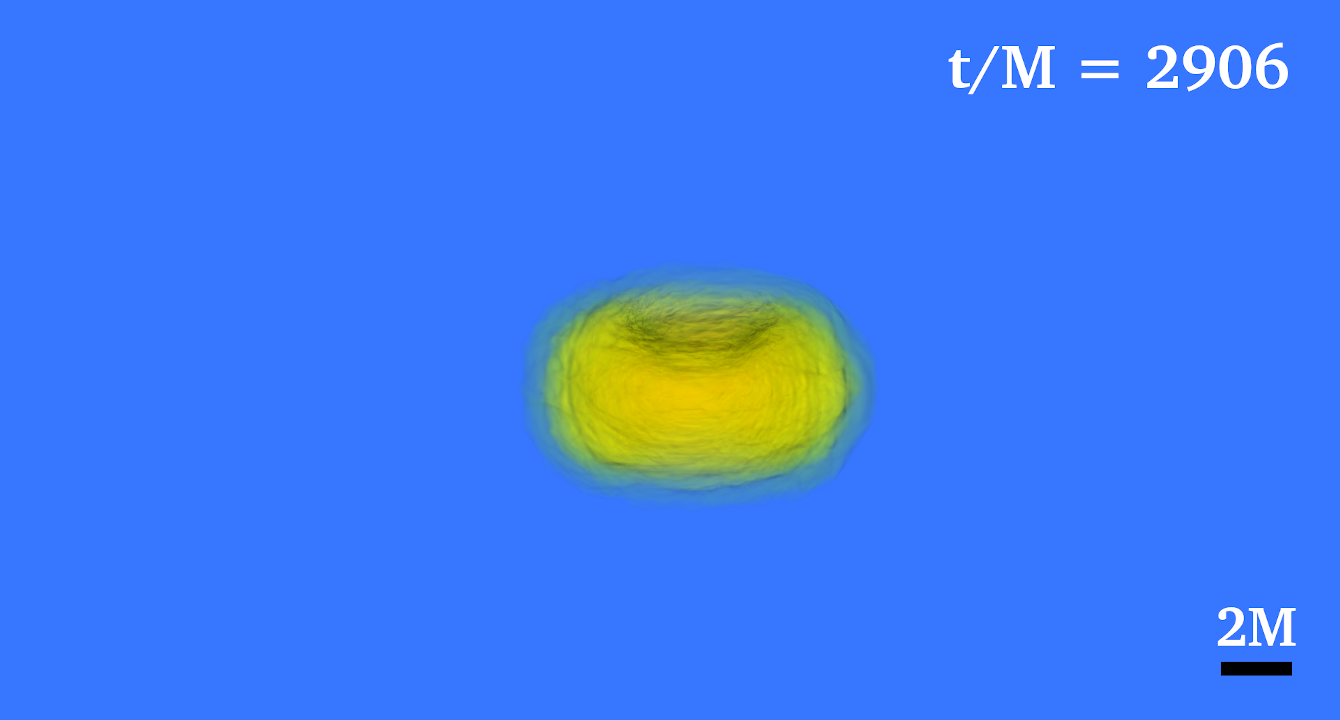}
  \includegraphics[width=0.495\textwidth]{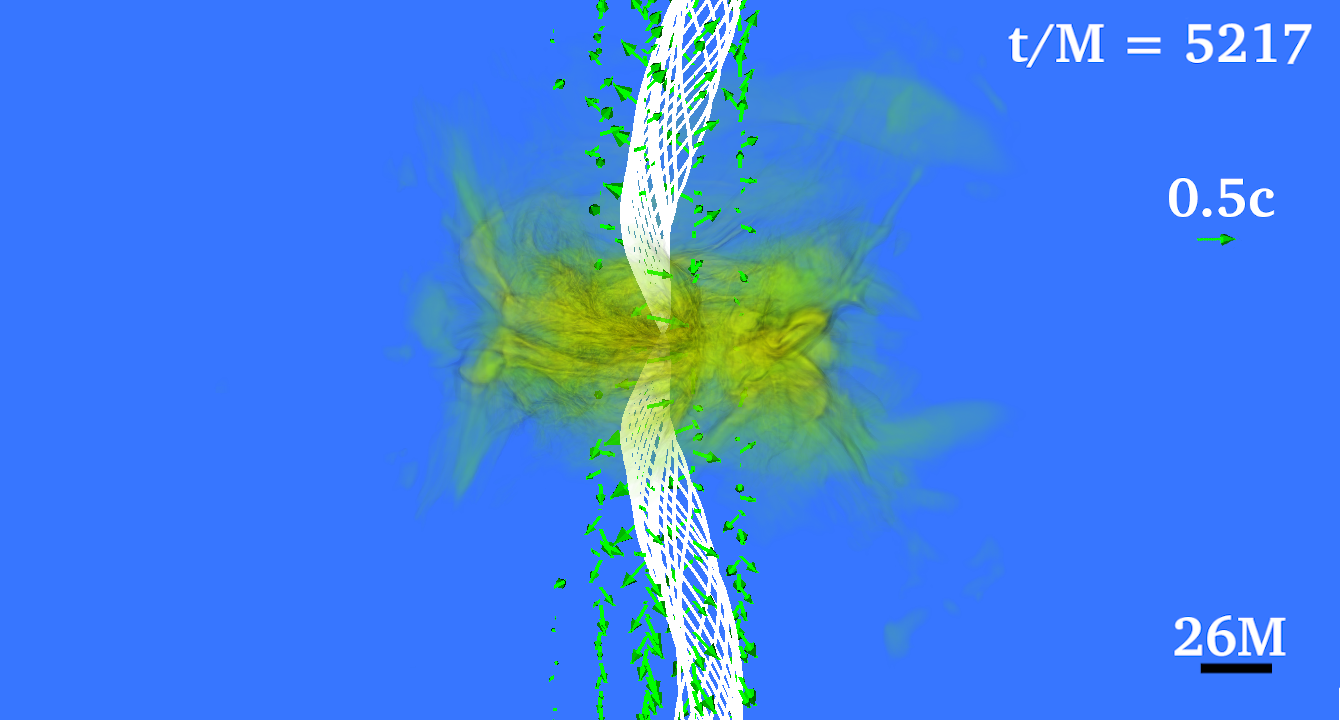}
  \caption{Volume rendering of the rest-mass density $\rho_0$ normalized to its initial NS maximum value
    (log scale) at selected times for H4M2.8H (left column) and SLyH2.6P (right column) cases~(see
    Table~\ref{table:key_results_NSNS}). White lines shows the magnetic field lines and the arrows
    indicate plasma velocities. The bottom right panel highlights the system after an incipient jet is launched.
    Here $M=2.8M_\odot (\rm left\,column)$ or $M=2.6M_\odot (\rm right\, column)$, hence $M\sim 1.3\times
    10^{-2}\,\rm ms \sim 4\,\rm km$.
    \label{fig:sly_ID}}
\end{figure*}

We further simplify our notation hereafter by dropping the factor $(k/k_{\rm L})$
in all physical quantities quoted for $\Gamma=2$ models.  The scale-free property of these results
can be easily recovered by restoring this factor.
%
\paragraph{\bf Magnetic field configuration:}
Following~\cite{Ruiz:2016rai}, we initially seed the star with a  dynamically unimportant,
purely poloidal magnetic field  that extends from the interior of the NSs into the exterior
(see top right panel in Fig.~\ref{fig:sly_ID}), and that approximately corresponds to that
generated by  an interior current loop with radius $r_0$ and current $I_0$ (see Eq.~6
in~\cite{Ruiz:2017inq}). In all our configurations, we choose $I_0$ and $r_0$ such that the
maximum value of the magnetic-to-gas-pressure ratio in the NS interior is $P_{\rm mag}/P_{\rm gas}=
0.003125$. The resulting magnetic field strength at the NS pole measured 
by a normal observer  is ${B}_{\rm pole}\simeq 10^{15.3}\,\rm G$. This choice allows us to mimic
the exponential growth of the magnetic field observed  in high-resolution simulations arising from 
magnetic instabilities (mainly the Kelvin-Helmholtz (KH) instability) triggered during the binary
merger. The NSNS simulations reported in~\cite{Aguilera-Miret:2020dhz} found that, with a local
numerical resolution of $\Delta x~\sim 37\,\rm m$, a pure poloidal magnetic field confined in the NSs
with a strength of $\sim 10^{11}\,\rm G$ can be  amplified to rms values of~$\sim 10^{16}\,\rm G$
within the first $\sim 5\,\rm ms$
following merger. Similar results have been reported in very-high-resolution NSNS simulations
in~\cite{Kiuchi:2015sga} employing a local resolution of $\Delta x~\sim 17.5\,\rm m$.
These values are beyond  the resolutions of this broad survey~(see Table~\ref{table:grid}). 
%
%
\begin{table}
  \caption{Grid hierarchy for models listed in Table~\ref{table:NSNS_ID}. The computational
    mesh consists of two sets of nine nested refinement boxes for the binaries modeled
    with a nuclear EOS, and seven nested refinement boxes for those modeled with a $\Gamma=2$
    EOS. The finest boxes are centered on each star and  have a half-length of $\sim 1.2\,R_{\rm NS}$,
    where $R_{\rm NS}$ is the initial equatorial  stellar radius. The number of grid points covering
    the equatorial radius of NS is denoted by $N_{\rm NS}$. In terms of grid points per NS radius, the
    resolution used here is a factor of $\sim 1.4$ finer than that in~\cite{Ruiz:2016rai}. In all cases,
    we impose reflection (equatorial) symmetry about the orbital plane.}
\begin{tabular}{ccccc}
  \hline \hline
  Model &  Grid Hierarchy$^{\dag}$    & Max. Resolution  & $N_{\rm NS}$ \\
  \hline \hline
  {SLyM$2.6$}          &  $2835.26\,\rm km/2^{n-1}$     & $110.75\,\rm m$  &  83\\
  {SLyM$2.7$}          &  $2835.26\,\rm km/2^{n-1}$     & $110.75\,\rm m$  &  82\\
  {SLyM$3.0$}          &  $2835.26\,\rm km/2^{n-1}$     & $110.75\,\rm m$  &  79\\
  \hline
   {H4M$2.8$}          &  $3477.92\,\rm km/2^{n-1}$     & $92\,\rm m$      &  121\\
   {H4M$3.0$}          &  $3477.92\,\rm km/2^{n-1}$     & $92\,\rm m$      &  118\\
  \hline
   {$\Gamma2$M2.8$^{\ddag}$}     &  $1015.02\,\rm km/2^{n-1}$    & $206.27\,\rm m$  &  62\\
   {$\Gamma2$M3.0$^{\ddag}$}     &  $954.06\,\rm km/2^{n-1}$     & $133.10\,\rm m$  &  87\\

  \hline \hline
\end{tabular}
\begin{flushleft}
  $^{\dag}$ Box half-length.\\
  $^{\ddag}$ Configurations treated previously~in~\cite{Ruiz:2016rai,Ruiz:2017inq,Ruiz:2019ezy}.
\end{flushleft}
\label{table:grid}
\end{table}
%
\begin{figure}
  \centering
  \includegraphics[width=0.47\textwidth]{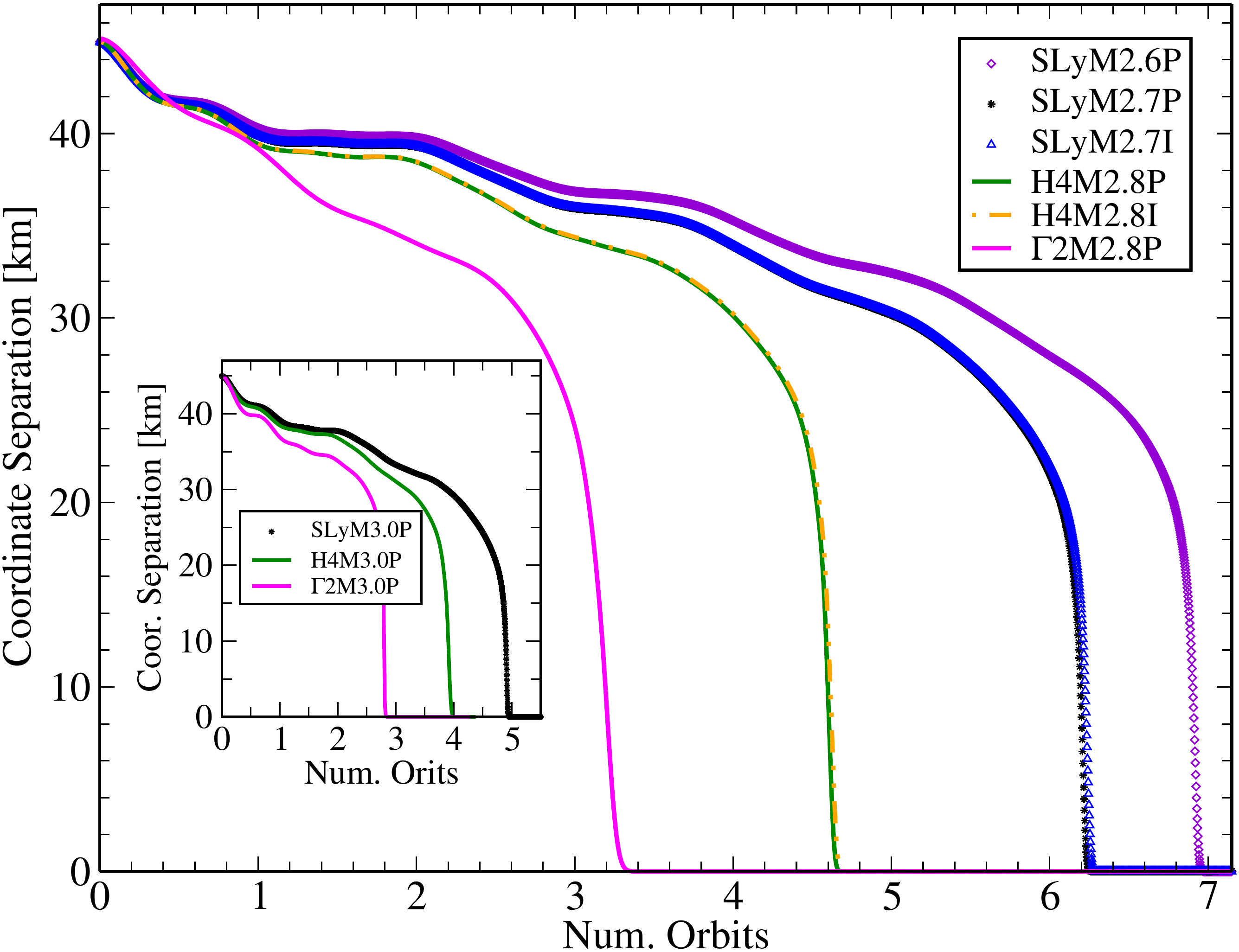}
  \caption{Binary coordinate separation between NS centroids, defined as the position of the maximum value of
    the rest-mass density, for the magnetized cases displayed in Table~\ref{table:key_results_NSNS}.
    The inset shows the binary separation for cases where the BH forms within the first $\sim 3
    \,\rm ms$ following merger (i.e. prompt collapse and short-lived HMNS undergoing collapse).
    \label{fig:coor_Sep}} 
\end{figure}
%
\begin{figure*}
  \centering
  \includegraphics[width=0.49\textwidth]{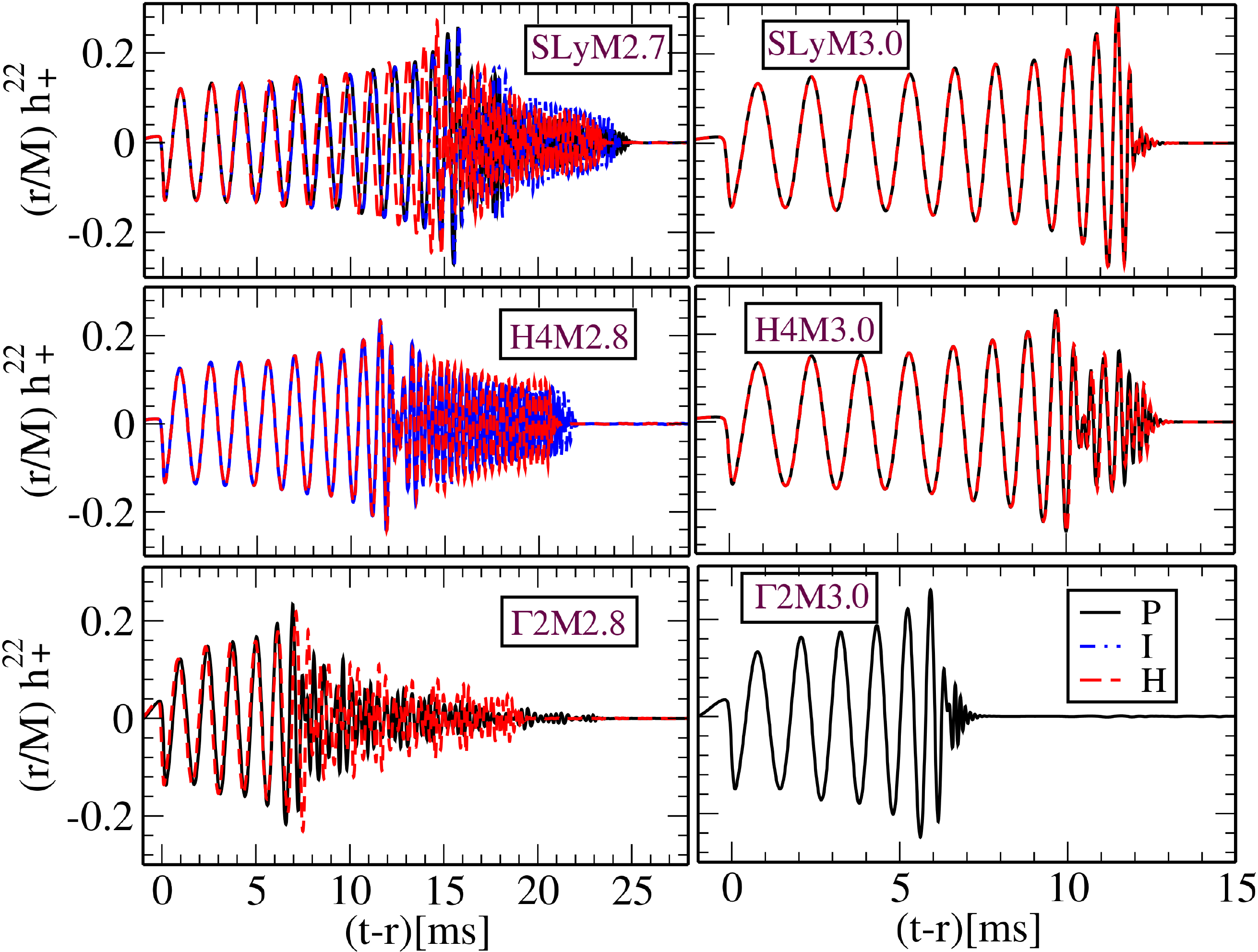}
  \includegraphics[width=0.48\textwidth]{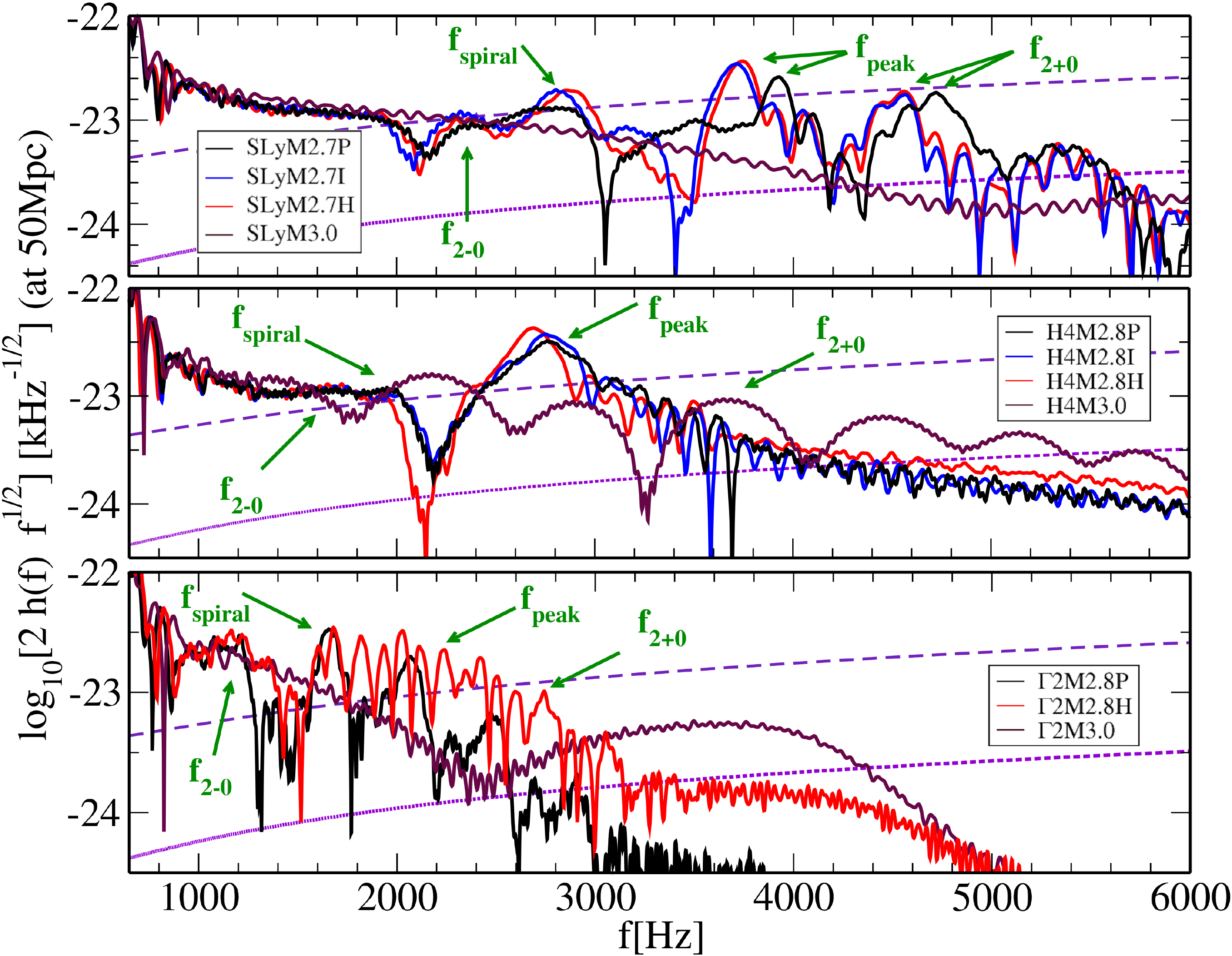}
    \caption{GW strain $h_+^{22}$ (dominant mode) as functions of retarded time (left) at  extracted
      coordinate radius $r_{\rm ext}\approx 100M$, and GW power spectrum of the dominant mode (right) at a
      source distance of 50Mpc along with the aLIGO  (dashed line) and Einstein Telescope (dotted line) noise
      curves of configurations in~\cite{LIGOScientific:2016wof} for cases listed in Table~\ref{table:key_results_NSNS}.
      Main spectral frequencies are denoted as $f_{2\pm 0}$, $f_{\rm spiral}$, and $f_{\rm peak}$ (see Table~\ref{table:freq_GW}).
    \label{fig:hydro_magGW}} 
\end{figure*}

To reliably evolve  the exterior magnetic field  and simultaneously model magnetic-pressure
dominance that characterizes the force-free pulsar-like magnetosphere, we initially enforced a
variable and low atmosphere in regions where magnetic field stresses dominate over the fluid
pressure gradient such that the magnetic-to-gas-pressure ratio in the NS exterior is $P_{\rm mag}/
P_{\rm gas}=100$ (see~Eq.~4~in~\cite{Ruiz:2018wah} for implementation details). This choice
increases the total rest-mass of the system in less than  $0.5\%$.

Finally, to assess the impact of the initial magnetic field configuration, and to compare
with previous studies, we also consider cases where seed the NSs with a poloidal magnetic field confined to the
stellar interior. This field  is generated via the vector potential~$A_\phi$~\cite{Etienne:2011ea}
\begin{eqnarray}
\mathcal{A}_i &=& \left( -\frac{y-y_c}{\varpi_{\rm c}^2}\delta^x{}_i
+ \frac{x-x_{\rm c}}{\varpi^2_{\rm c}}\delta^y{}_i\right)\,\mathcal{A}_\varphi\,, \\
\mathcal{A}_\varphi &=& \mathcal{A}_b\,\varpi^2_{\rm c}\,\max(P_{\rm gas}- P_{\rm cut},0)^{n_b}\,,
\label{ini:Aphi_int}
\end{eqnarray}
where $(x_{\rm c},y_{\rm c},0)$ is the coordinate position of the center of mass of the NS,
$\varpi^2_{\rm c}=(x-x_{\rm c})^2+(y-y_{\rm c})^2$, and $A_b$, $n_p$ and $P_{\rm cut}$ are
free parameters. The cutoff pressure parameter $P_{\rm cut}$ confines the magnetic field
inside the NS within the region where $P_{\rm gas}>P_{\rm cut}$. The parameter $n_b$ determines the degree
of central condensation of the magnetic field. In our evolutions, we choose $P_{\rm cut}$
to be $1\%$ of the maximum pressure and $n_b=1$, while the value of $A_b$ is chosen such as the
magnetic-to-gas-pressure ratio at the NS center matches that in our pulsar-like case
(i.e.~$P_{\rm mag}/P_{\rm gas}=0.003125$).
%
\subsection{Grid structure}
\label{subsec:grid}
The numerical grid hierarchy is summarized in Table~\ref{table:grid}. It consists of two
sets of nested refinement boxes centered on each star. Each of them contains nine boxes
that differ in size and in resolution by factors of two. When two boxes overlap they
are replaced by a common box centered on the center of mass of the NSNS. The finest box
around the NS has a~side half-length of $\sim 1.2\,R$, where~$R$ is the
initial NS equatorial radius. This choice allows us to initially resolve the equatorial
NS radius by $\sim 120$ grid points~(see Table~\ref{table:NSNS_ID}). We impose reflection
(equatorial) symmetry across the orbital plane. In terms of grid points per
NS radius, the resolution employed in these studies is a factor of $\sim 1.4$ finer than
that in~\cite{Ruiz:2016rai,Ruiz:2017inq} where NSNSs are modeled by a $\Gamma=2$ EOS~(see
Table~\ref{table:NSNS_ID}).
%
\subsection{Diagnostics}
\label{subsec:diagnostics}
To validate our evolutions, we monitor the $L_2$ norm of the normalized Hamiltonian and momentum
constraints computed through Eqs.~(40)-(43)~in~\cite{Etienne:2007jg}. In all simulations,  the
Hamiltonian constraint violations remain smaller than $0.06\%$ during the inspiral, peak at
$\sim 0.3\%$ during BH formation and then relax to $\sim 0.02\%$  once the BH + disk remnant
settles into a steady state. The normalized momentum constraint violations
remain smaller than $0.8\%$ during the inspiral, peak at $4.2\%$ during BH formation and subsequently
relax to~$\sim 0.18\%$ during steady state. We note that these values are similar to those
previously reported in our long-term, pure hydrodynamic simulations of spinning NSNSs modeled by
SLy and ALF2 EOSs~\cite{Tsokaros:2019anx}. 

After the catastrophic collapse of the merger-outcome-remnant, we use the~{\tt AHFinderDirect}
thorn~\cite{ahfinderdirect} to track the BH apparent horizon. In addition,
we estimate the BH mass  $M_{\rm BH}$ and its dimensionless spin~$a/M_{\rm BH}$ using the isolated
horizon formalism as in~\cite{dkss03}. We use the {\tt Psikadelia} thorn to compute the Weyl
scalar $\Psi_4$ which is decomposed into $s=-2$~spin-weighted spherical harmonic modes.
We use Eqs.~(2.8), (2.11) and (2.13) in~\cite{Ruiz:2007yx} at ten different extraction radii
between~$r_{\rm min}\approx 50M$ and $r_{\rm max}\approx 320M$ to compute
the total flux of energy and angular momentum transported away by GWs. We  find that between
$\sim 0.15\%$ and $\sim 3.6\%$ of the total energy of our NSNS models is radiated away during the
evolution in form of gravitational radiation, while between $\sim 13.3\%$ and $\sim 31.4\%$ of
the angular momentum is radiated (see~Table~\ref{table:key_results_NSNS}).
We measure the dynamical ejection of matter via
%
$M_{\rm esc} = \int_{r > r_0}\rho_*\,d^3x$,
where $\rho_*=-\sqrt{\gamma}\,\rho_0\,n_\mu u^\mu$,
on the conditions that: i) the  specific energy $E = -u_t -1$~of the outgoing material is
always positive (unbound material), and  ii) the radial velocity of the outgoing material
$v^r > 0$. Here $\gamma$ is the determinant
of the three-metric, $n_\mu$ is the timelike future pointing unit (i.e. normal) vector, and $u^\mu$ is the
four-velocity of the fluid. We vary the coordinate radius $r_0$ between  $r_{\rm min}=30M$
and $r_{\rm max}= 100M$  to verify that  the measure of the ejecta is $r_0$-independent.
Depending on the stiffness of the EOS and the magnetic field we find that the rest-mass
ejected following merger  ranges between $\lesssim 10^{-4}M_\odot$  and $\sim 10^{-2}
M_\odot$ consistent with values previously reported in~e.g.~\cite{Shibata:2019wef,Radice:2018pdn}.
We note that if ejected material with masses~$\gtrsim 10^{-2}M_\odot$ is converted to
$r$-process elements, GW170817-like events could account for
the amount of heavy elements observed in the Milky Way~\cite{LIGOScientific:2017pwl,Cote:2017evr}.

We monitor the conservation of the ADM mass and ADM angular momentum computed throughout 
Eqs. (19)-(22)~in~\cite{Etienne:2011ea}. Consistent with~\cite{Tsokaros:2019anx}, in all our evolved
configurations we find that the ADM mass is conserved to within~$\lesssim 1\%$ and angular momentum to within
$\lesssim 6\%$. In addition, we monitor the conservation of the rest mass $M_0=\int \rho_* d^3x$, which
is conserved to within~$\lesssim 0.05\%$, as well as the magnetic energy growth outside the BH
apparent horizon~$\mathcal{M} =\int u^\mu u^\nu T^{(EM)}_{\mu\nu}\,dV$ as measured by a
comoving observer~\cite{Ruiz:2017inq}. 

To probe MHD turbulence  in our systems, we compute the effective
Shakura--Sunyaev $\alpha_{\rm SS}$ parameter~\cite{Shakura73} associated  with the effective
viscosity due to magnetic stresses~$\alpha_{\rm SS}\sim T^{\rm EM}_{\hat{r}\hat{\phi}}/P$
(see~Eq. 26~in~\cite{FASTEST_GROWING_MRI_WAVELENGTH}). To check if the magnetorotational instability (MRI) can
be captured in our evolution following the NSNS merger, we compute the $\lambda_{\rm MRI}$-quality factor
$Q_{\rm MRI}\equiv\lambda_{\rm MRI}/dx$, which  measures the number of grid points
per fastest growing MRI mode. Here $\lambda_{\rm MRI}$ (see Eq.~1 in~\cite{UIUC_PAPER2}) is the
fastest-growing MRI wavelength and $dx$ is the local grid spacing.  The MRI can be properly captured
if: i) the quality factor $Q_{\rm MRI}\gtrsim 10$; and ii) $\lambda_{\rm MRI}$ fits inside 
the remnant~\cite{Sano:2003bf,Shiokawa:2011ih}.

We  compute the outgoing EM Poynting luminosity $  L=-\int T^{r(EM)}_t\,\sqrt{-g}\,d\mathcal{S}$
across spherical surfaces of coordinate radii between $r_{\rm ext}=50M$ and $350M$.
A summary of the above results  is displayed in Table~\ref{table:key_results_NSNS}. 
Note that we add an ``P'', ``I'' or ``H'' at the end of the tag name  of a given configuration to
denote the initial magnetic field configuration (i.e.~P $=$ magnetic
field that   extends from the stellar interior into the pulsar-like exterior, I $=$ magnetic field confined
in the stellar interior, or H $=$ purely hydrodynamic (i.e.~unmagnetized)).
%
%
\begin{turnpage}
\begin{table*}[]
  \begin{center}
    \caption{Summary of key results. Here $t_{\rm GW}$, $\Delta t_{\rm BH}$ and $t_{\rm sim}$ are the  merger
      time at GW peak amplitude, the BH formation time measured as $t_{BH}-t_{\rm GW}$, and the full simulation time,
      respectively, all in $\rm ms$. The mass  and the dimensionless spin parameter
      of the BH remnant are given by $M_{\rm BH}\,[M_\odot]$
      and $\tilde{a}=a_{\rm BH}/M_{\rm BH}$, both computed using the isolated horizon formalism. $M_{\rm disk}\,[M_\odot]$ denotes
      the rest-mass of the accretion disk, $\dot{M}\,[M_\odot/s]$ is the rest-mass accretion rate computed via Eq.
      (A11)~in~\cite{Farris:2009mt}. These two quantities are measured when 
      $\dot{M}$ begins to settle into a steady state. The disk lifetime is $\tau_{\rm disk}[\rm ms]\equiv M_{\rm disk}/\dot{M}$,
      $M_{\rm esc}$ denotes the rest-mass fraction of escaping matter (ejecta) following
      GW peak amplitude. The fraction of energy and angular momentum carried off by gravitational
      radiation are given by $\Delta \bar{E}_{\rm GW}\equiv\Delta E_{\rm GW}/M_{ADM}$
      and $\Delta \bar{J}_{\rm GW}\equiv\Delta J_{\rm GW}/J_{ADM}$, respectively. $\alpha_{\rm SS}$ is the
      Shakura--Sunyaev viscosity parameter,  $B_{\rm rms}\,[G]$ is the rms value of the magnetic field
      at the HMNS pole just before collapse. The Poynting luminosity driven by the jet, and time-averaged over the last $\sim 5\,
      \rm ms$ before the termination of our simulations, is denoted as $L_{\rm EM}\,[\rm erg/s]$.
      $\Gamma_L$ denotes the
      maximum fluid Lorentz factor at $t_{\rm sim}$. $L_{\rm knova}\,[\rm erg/s]$, $\tau_{\rm peak}\,[\rm days]$,
      and $T_{\rm peak}[\rm K]$ are the peak EM luminosity, the rise time, and the temperature of the potential kilonova, respectively,
      and finally, the fate of the binary merger.  
      An ``P'', ``I'' or ``H'' at the end of the
      tag name for each configuration denotes the initial magnetic field configuration, i.e. P= pulsar-like (interior + exterior) magnetic field,
      I= interior magnetic field, or H=hydrodynamic (unmagnetized). A dash symbol denotes ``not applicable''.
     \label{table:key_results_NSNS}}
    \scalebox{0.98}{
    \begin{tabular}{cccccccccccccccccccc}
      \hline\hline
          {Model} & $t_{\rm GW}$  & $\Delta t_{\rm BH}$ &$t_{\rm sim}$ &$M_{\rm BH}$ &  $\tilde{a}$      & $M_{\rm disk}/{M_0}^{(\dag)}$     & $\dot{M}$  &  $\tau_{\rm disk}$ &
          $M_{\rm esc}/{M_0}$     & $\Delta \bar{E}_{\rm GW}$ & $\Delta\bar{J}_{\rm GW}$  & $\alpha_{\rm SS}$          &$B_{\rm rms}$  & $L_{\rm EM}$  & $\Gamma_L$ & $L_{\rm knova}$ &
          $\tau_{\rm peak}$ & $T_{\rm peak}$ & Fate\\

          &   &  &  &  &    & $\times 10^{-2}$     &   &  & $\times 10^{-2}$     & $\times 10^{-2}$ & $\times 10^{-1}$  &  &  &   &  &  &  &  & \\
          
          \hline
SLyM2.6P & 17.5 &41.0& 68.6  &2.26 & 0.47   & 8.71 & 1.93   & 130.88      & $0.870$  &  $2.61$ & $3.14$  & 0.01-0.04     &$10^{16.2}$  &  $10^{52.3}$ & 1.24 &$10^{40.7}$&1.61
         & $10^{3.4}$ & HMNS $\rightarrow$ delayed col. \\
SLyM2.6H & 17.6 & -  & 60.0  &-    & -      &       -      & -      & -  & $0.870$  &  $3.64$ & $3.07$  & -             &  -          &      -       &  -   &$10^{40.8}$&1.67
         & $10^{3.4}$ & HMNS $\rightarrow$ delayed col.\\
          \hline 
SLyM2.7P & 15.7 &8.9 & 48.2  &2.45 & 0.62   & $6.16$ & 2.01   & 92.55    & $0.251$  &  $2.64$ & $2.91$  & 0.02-0.07   &$10^{16.0}$  &  $10^{52.8}$ & 1.26 &$10^{41.2}$ & 5.13
          &$10^{3.2}$ & HMNS $\rightarrow$ delayed col.\\
SLyM2.7I & 15.8 &8.2 & 100.2  &2.50 & 0.69   & $3.08$ & 0.96  & 96.89   & $0.501$  &  $3.18$ & $3.32$  & 0.01-0.03     &$10^{15.8}$  &  $10^{51.3}$ & 1.25 &$10^{40.9}$ & 2.31
& $10^{3.3}$ & HMNS $\rightarrow$ delayed col.\\
SLyM2.7H & 14.6 &9.0 & 32.0  &2.51 & 0.70   & $2.24$ &  0.94  & 71.97    & $0.316$  &  $3.20$ & $3.13$  & -             &  -          &   -          & -    &$10^{40.8}$ & 1.91
& $10^{3.4}$ & HMNS $\rightarrow$ delayed col.\\
          \hline 
SLyM3.0P & 11.5 &0.6& 24.4  &2.96 & 0.81   & $0.25$  & $10^{-2.4}$ & -  & $0.050$ &  $1.31$ & $2.07$ & -               &$10^{16.1}$  &  $10^{45.7}$ & -   &$10^{40.5}$ & 0.93
& $10^{3.5}$ & prompt col.\\
SLyM3.0H & 11.5 &0.6& 23.5  & 2.95 & 0.80  & $0.40$  & $10^{-2.3}$ & -  & $0.013$ &  $1.32$ & $2.08$ & -         &    -        &  -       & - &$10^{40.3}$ &0.47 &
$10^{3.6}$  & prompt col.\\
          \hline 
H4M2.8P  &11.5 &  9.6& 56.3 &2.64 & 0.71     & $3.50$ & 1.11    & 97.75 & $1.000$     & $2.21$& $2.68$    & 0.01-0.09 &  $10^{15.9}$  & $10^{52.5}$   & 1.30 &$10^{41.1}$ &3.20
&$10^{3.3}$ & HMNS $\rightarrow$ delayed col.\\
H4M2.8I  &11.5 & 13.3& 46.5 &2.69 & 0.73     & $2.21$ & 1.20    & 57.10 & $0.063$     & $2.44$& $2.89$    & 0.02-0.04 &  $10^{15.7}$  &    -          &  -  &$10^{40.6}$  &0.85
& $10^{3.5}$ & HMNS $\rightarrow$ delayed col.\\
H4M2.8H  &11.5 &  9.1& 40.2 &2.70 & 0.73     & $1.51$ & 0.37    & 120.65& $0.016$     & $2.63$& $3.03$    & -         &       -       &     -   &  -  &$10^{40.4}$ &0.43
& $10^{3.6}$ & HMNS $\rightarrow$ delayed col.\\
          \hline
H4M3.0P  &9.70 & 2.5& 40.0 &2.93 & 0.80   & $1.04$ & $0.23$  & 152.23& $0.016$     & $1.43$&  $1.92$   & $10^{-3.5}$-0.02 &  $10^{16.0}$  &  $10^{52.2}$ & 1.19 &$10^{40.4}$&0.43
          & $10^{3.6}$ & HMNS $\rightarrow$ delayed col.\\
H4M3.0H  &9.70 & 2.6& 34.8 &2.95 & 0.81   & $0.54$ & $0.19$  & 94.50& $0.016$      & $1.34$&  $1.87$   & -         &      -        &     -            &  -  &$10^{40.4}  $&0.42
& $10^{3.5}$ & HMNS $\rightarrow$ delayed col.\\
          \hline
          \hline
$\Gamma$2M2.8P$^{(\ddag)}$ & 6.9 & 16.9 & 70.0 &2.65  & 0.74 & $2.82$  & 0.48  & 92.43 & - & $0.81$ & $1.34$ & 0.04-0.08 & $10^{15.9}$&$10^{51.3}$ & 1.25 & - & -& - & HMNS $\rightarrow$ delayed col. \\
$\Gamma$2M2.8H$^{(\ddag)}$ & 7.2 & 14.5 & 48.8 &2.65  & 0.73  & $0.98$   & 0.48    & 74.86 & $0.79$&  $1.29$ & $1.31$ & -         &          -  &    -  &  - &$10^{40.4}$ &0.38
          & $10^{3.6}$ & HMNS $\rightarrow$ delayed col.\\
          \hline
$\Gamma$2M3.0P$^{(\ddag)}$ & 4.2 & 0.51 & 15.0&2.80    & 0.83  & $0.47$   & 0.34    &  -    &         -   & $0.78$ & $1.33$ &   -   &  $10^{15.9}$  &    -  &  - & - & -& -
          & prompt col. \\
          \hline\hline
    \end{tabular}
    }
  \end{center}
  \begin{flushleft}
    $^{(\dag)}$ $M_0$ denotes the initial total rest-mass of the system.\\
    $^{(\ddag)}$ Configurations  treated previously in~\cite{Ruiz:2016rai,Ruiz:2017inq,Ruiz:2019ezy}.
  \end{flushleft}
\end{table*}
\end{turnpage}
%
\begin{figure*}
  \centering
  \includegraphics[width=0.47\textwidth]{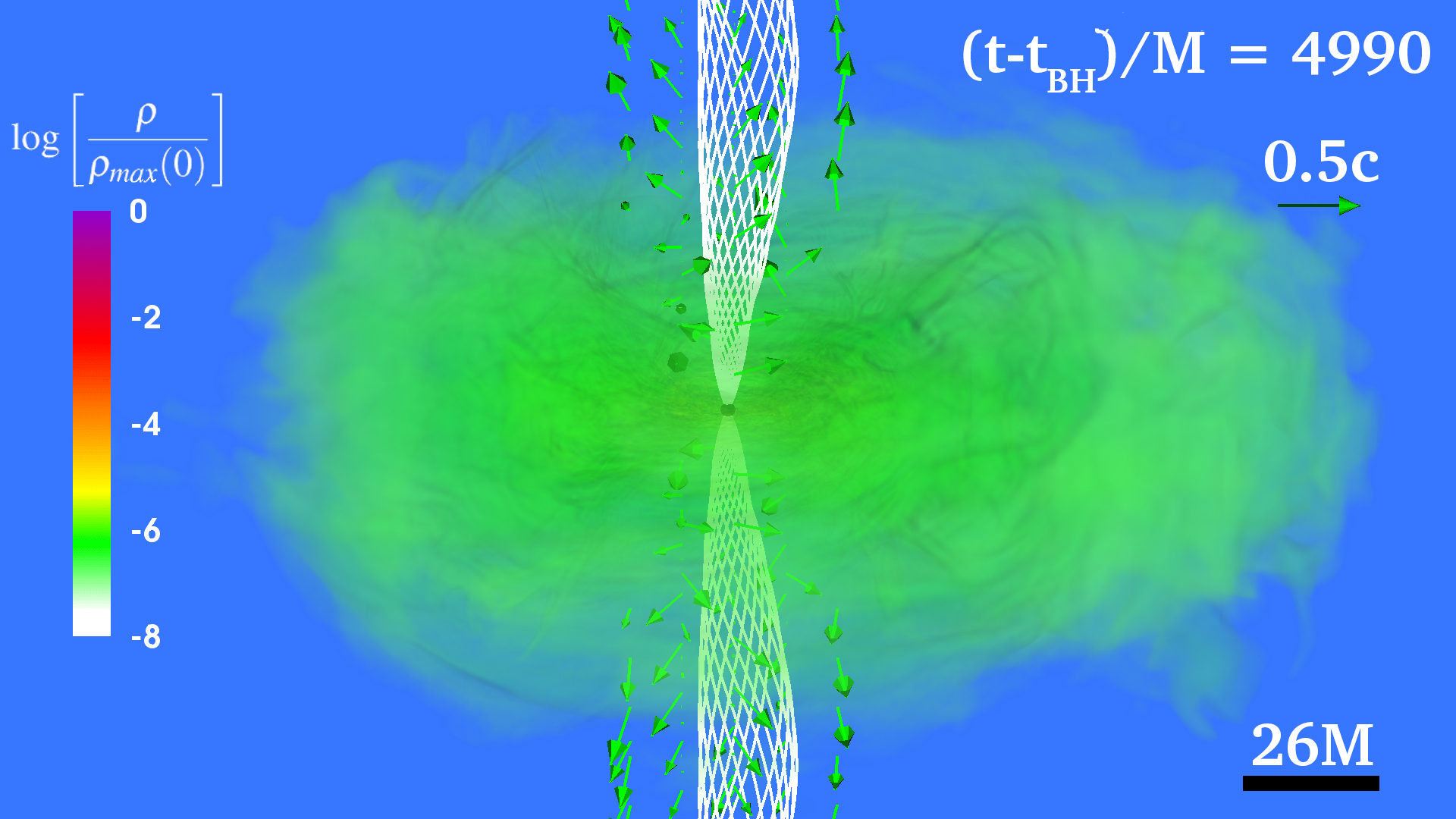}
  \includegraphics[width=0.47\textwidth]{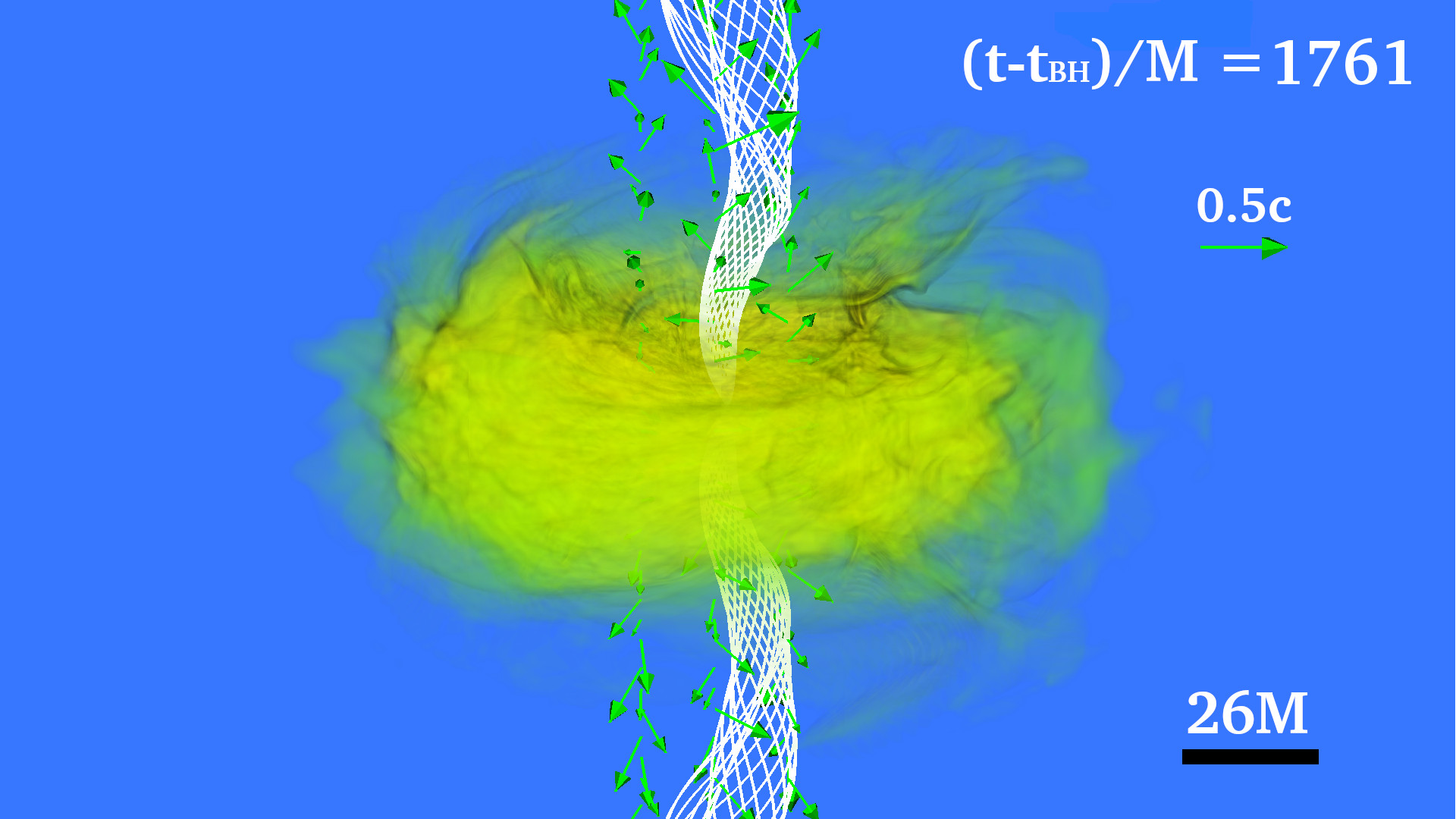}
  \includegraphics[width=0.47\textwidth]{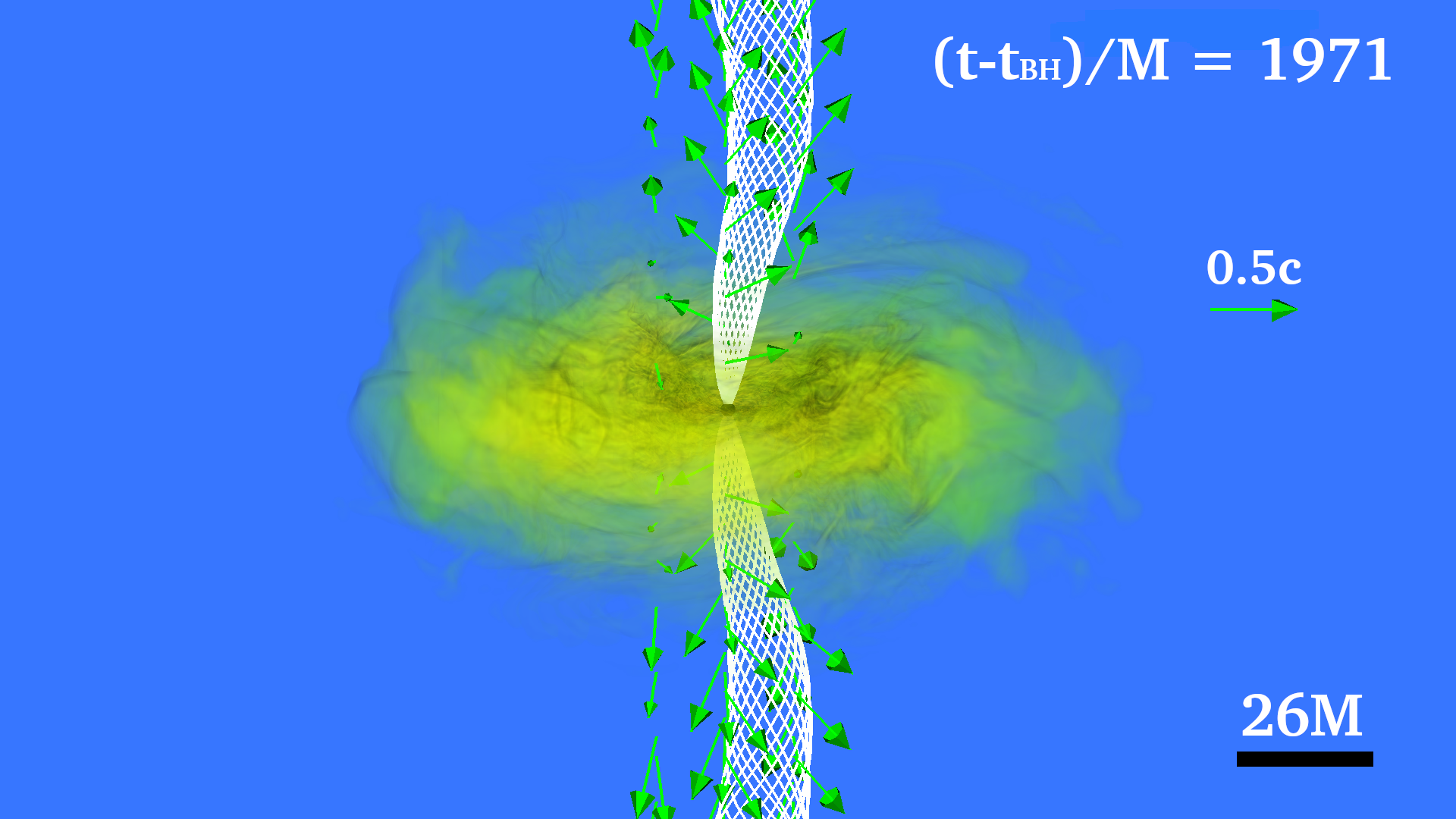}
  \includegraphics[width=0.47\textwidth]{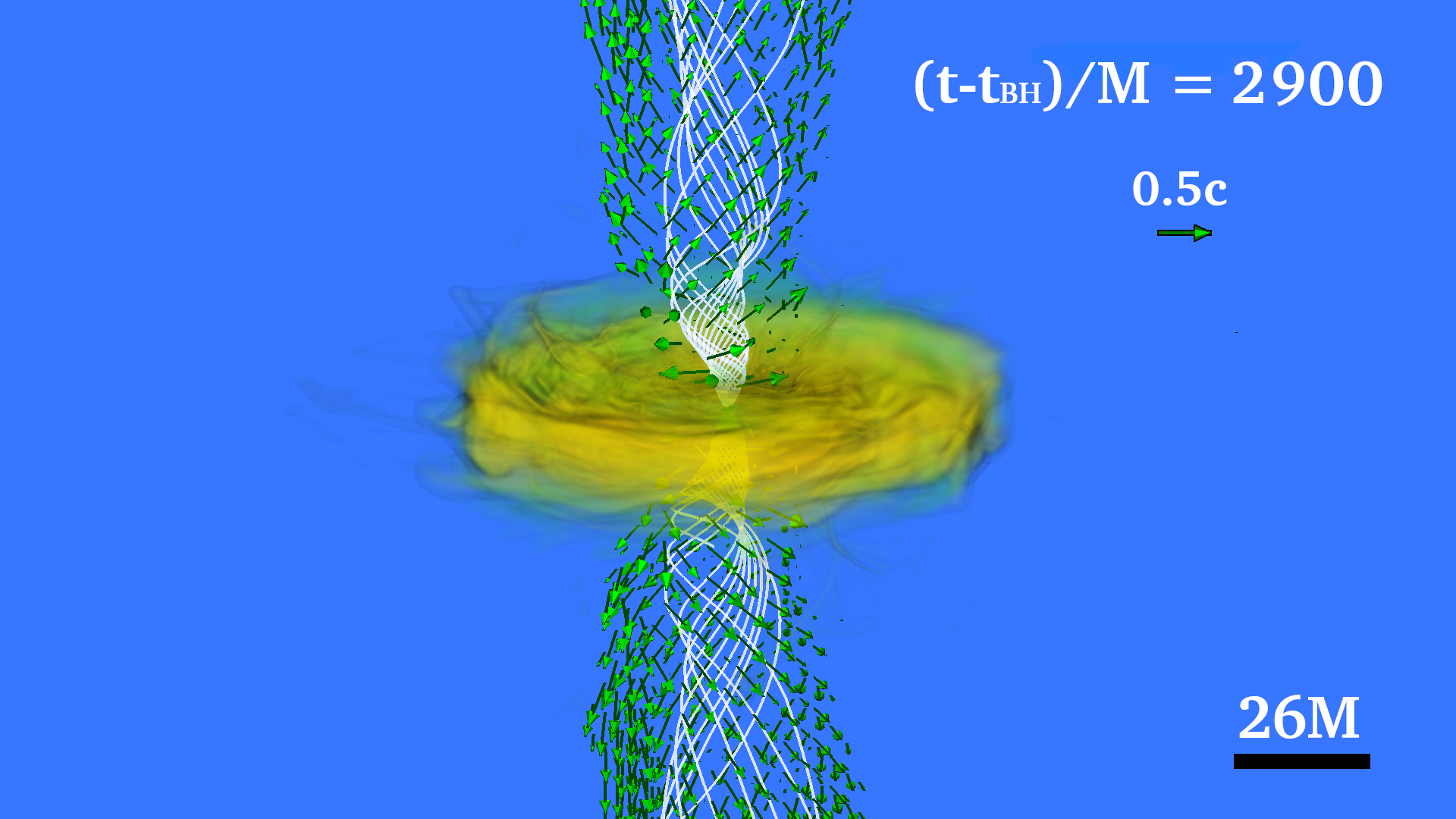}
  \caption{Volume rendering of the remnant  BH + disk configuration 
    for cases SLyM2.7I (top left), SLyM2.7P (top right), H4M2.8P (bottom left)
    and $\Gamma$2M2.8 (bottom right). The rest-mass density is normalized to the 
    NS initial maximum value. White lines depict the magnetic field lines while the
    arrows display fluid velocities. The BH apparent horizon is displayed as a black surface.
    Here $M\sim 1.3\times 10^{-2}\,\rm ms\sim 4\,\rm km$.
    \label{fig:NSNS_Sly_H4_G2}}
\end{figure*}
%
%
\section{Results}
\label{sec:results}
The basic dynamics and final outcome of the NSNS models listed in Table~\ref{table:key_results_NSNS} can be
summarized in Fig.~\ref{fig:sly_ID}. Left and right columns display  the evolution of
the unmagnetized and magnetized representative cases at selected times, respectively. Gravitational
radiation losses cause the orbital separation to shrink  driving the binary merger (see second
rows). Fig.~\ref{fig:coor_Sep}  displays the coordinate separation between NS centroids,
defined as the coordinate position of the maximum value of the rest-mass density, for the magnetized
cases.  We note  that the softer the EOS (see Sec.~\ref{subsec:idata}), the longer  the inspiral phase. In
particular, SLyM2.7P and $\Gamma$2M2.8P, binaries with the same initial rest-mass~(see~Table~\ref{table:NSNS_ID})
merge after $\sim 6.2$ and~$\sim 3.2$~orbits, respectively, while H4M2.8P, the binary with
the same ADM mass as the latter, merges after~$\sim 4.6$ orbits. 
As shown in the left panel of~Fig.~\ref{fig:hydro_magGW}, similar behavior is observed in the unmagnetized cases.
This is anticipated because the seed magnetic field is initially unimportant dynamically  
(we recall that the initial magnetic pressure is only $\sim 0.3\%$ of $P_{\rm gas}$)
and hence cannot have a strong impact during the inspiral.

\begin{figure*}
  \centering
  \includegraphics[width=0.32\textwidth]{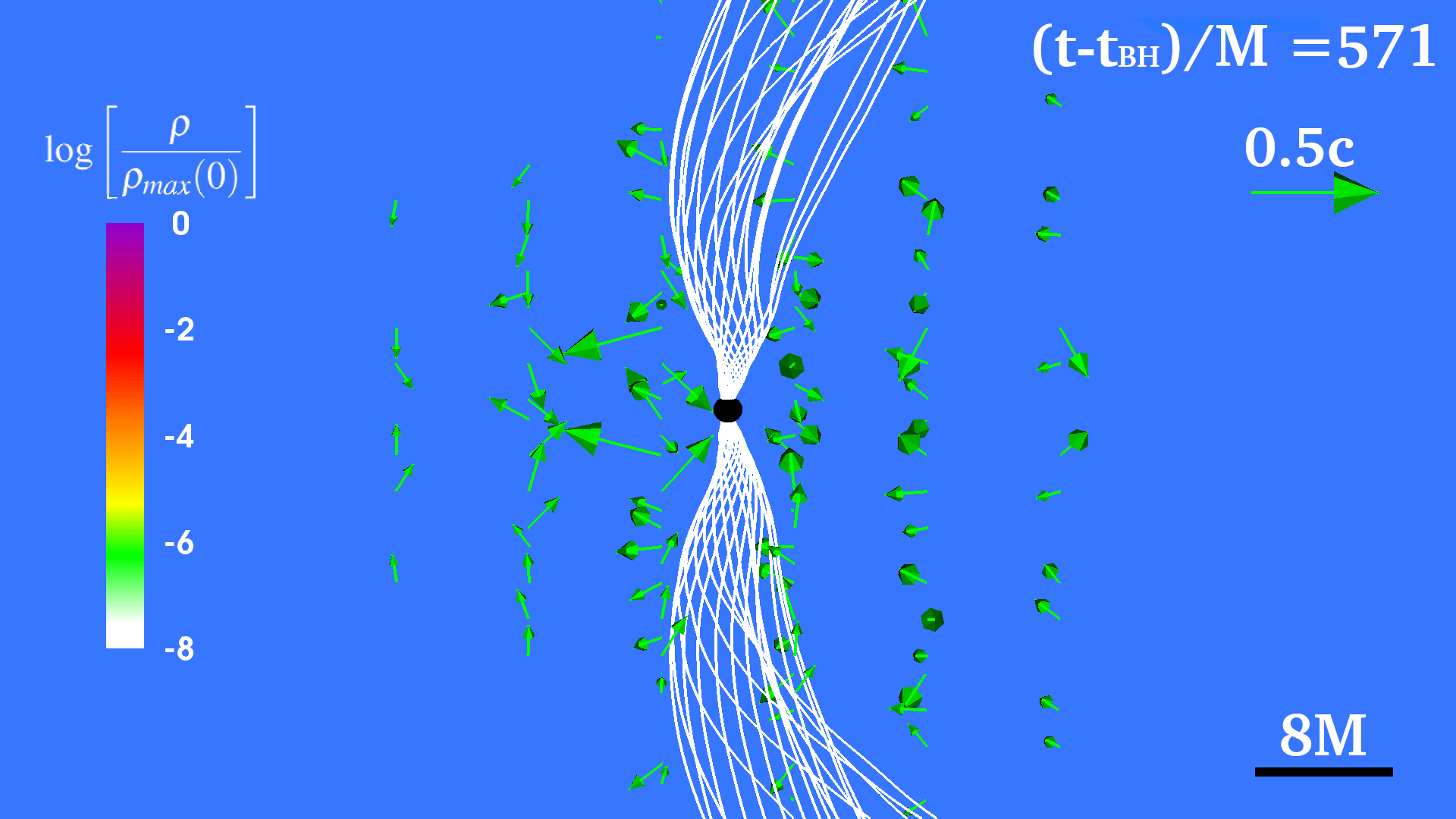}
  \includegraphics[width=0.32\textwidth]{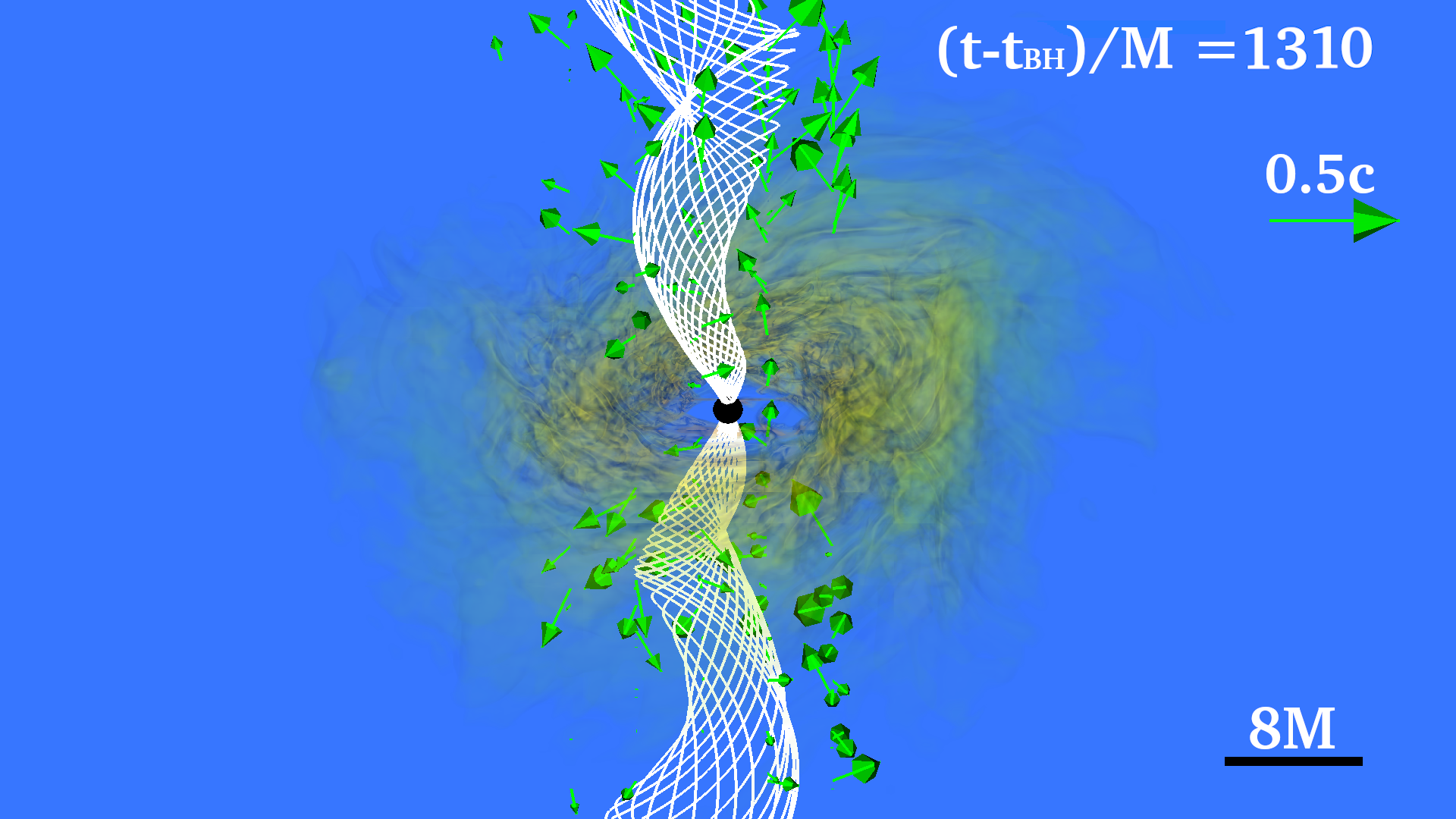}
  \includegraphics[width=0.32\textwidth]{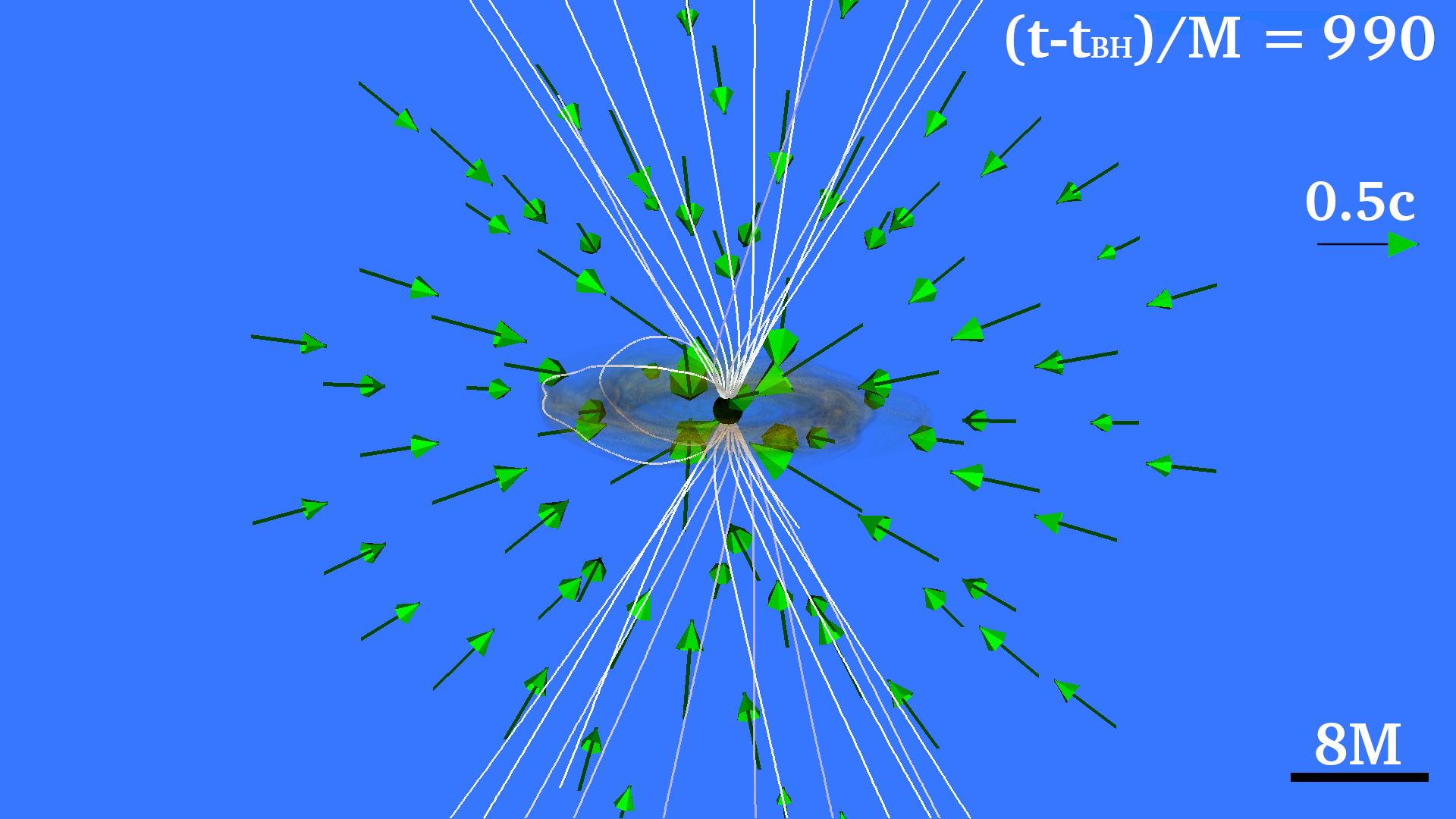}
  \caption{Volume rendering of the remnant  BH + disk configuration 
    for cases SLyM3.0I (left), H4M3.0P (middle), and $\Gamma$2M3.1 (right).
    The rest-mass density is normalized to the  NS initial maximum value.
      White lines depict the magnetic field lines while the arrows display fluid velocities.
      The BH apparent horizon is displayed as a black surface. 
      Here $M\sim 1.5\times 10^{-2}\,\rm ms\sim 4.5\,\rm km$.
    \label{fig:NSNS_Sly_int}}
\end{figure*}

Following the merger, we note that:
\begin{itemize}
\item If the ADM mass of the binary is $ \lesssim M^{\rm thres}$, then
  the merger outcome can be a short- medium- or long-lived HMNS that undergoes delayed collapse to a BH
  immersed in an accretion disk~(see second and third rows in Fig.~\ref{fig:sly_ID}).
  Although the masses of the NSNSs are slightly different~(see Table~\ref{table:NSNS_ID}),
  there is an impact of the EOS on the lifetime of the transient: the
  softer the EOS, the shorter the HMNS lifetime~$\tau_{\rm HMNS}$. In particular, the HMNS remnant in
  cases SLyM2.7P, H4M2.8P, and $\Gamma$2M2.8P lasts $\sim 8.9\,\rm ms$, $\sim 9.3\,\rm ms$, and $\sim 17\,
  \rm ms$, respectively (see~Table~\ref{table:key_results_NSNS} for other cases). In addition,
  the softer the EOS, the larger the amount of energy and angular momentum
  carried away by GWs. Note that the sensitivity of the HMNS lifetime to the magnetic field is
  physical and has been previously reported in~\cite{BrunoMagNSNS}. However, as pointed out
  in~\cite{Ruiz:2016rai}, $\tau_{\rm HMNS}$ depends on numerical resolution, even in unmagnetized
  simulations. High resolution studies may be required to accurately determine the HMNS lifetime. 
  
  Following the collapse of the HMNS,  material with high angular momentum wraps around the BH, forming
  an accretion disk~(see bottom panels in Fig.~\ref{fig:sly_ID}). If the accretion disk is magnetized
  then an incipient jet --i.e. a mildly relativistic outflow confined in a tightly wound, helical
  magnetic field~\cite{Ruiz:2016rai}- is launched once the ratio $B^2/(8\,\pi\rho_0)\gtrsim 1$
  above  the BH poles. The incipient jet emerges  regardless of the EOS or the initial  magnetic field
  configuration (see~Figs.~\ref{fig:sly_ID} and~\ref{fig:NSNS_Sly_H4_G2} and middle panel in
  Fig.~\ref{fig:NSNS_Sly_int}). These preliminary studies suggest  that an incipient jet is the
  typical outcome of magnetized NSNS mergers.

  We  note that the jet launching time strongly depends on how close the total mass of the binary is
  to the threshold value for prompt collapse, which in turn depends on the maximum mass
  configuration of a given EOS. For realistic EOSs, such as SLy or H4, the threshold mass for
  prompt collapse is
  $\sim1.3-1.5\,M_{\rm sph}\simeq 2.82M_\odot$ while for polytropic EOSs like $\Gamma=2$ it is~$\sim1.7
  \,M_{\rm sph}\simeq 2.88M_\odot$~\cite{Shibata:2002jb,STU1,ST,Bauswein:2020aag}). The same values also  apply
  to configurations with dynamically weak initial magnetic fields.
  In particular, we observe that in SLyM2.7P a magnetically-driven jet is launched
  after $t-t_{\rm BH}\sim 20\,\rm ms$ while in SLyM2.7I it is launched after $\gtrsim 60\,\rm ms$
  (see below).
  
\item  If the ADM mass of the binary  is $> M^{\rm thres}_{}$, then it 
  undergoes prompt collapse to a BH surrounded by a small a disk with rest-mass of
  $\lesssim 0.5\%$ of the total rest-mass of the binary~(see Fig.~\ref{fig:NSNS_Sly_int}).
  Consistent with the results reported in~\cite{Ruiz:2017inq}, we do not observe a persistent
  outflow or tight magnetic field collimation in these cases (see left and right panels in Fig.~\ref{fig:NSNS_Sly_int}).
\end{itemize}

We summarize below  the outcome of our binary merger simulations during the HMNS and BH + disks
phases of the evolution. We recall that our configurations differ in the EOS and seed magnetic field (magnitude
and initial geometry). Key results from these simulations are displayed in Table~\ref{table:key_results_NSNS}. 
%
%
%
%
\subsection{Delayed collapse}
\label{sec:delayed}
\begin{figure}
  \centering
  \includegraphics[width=0.49\textwidth]{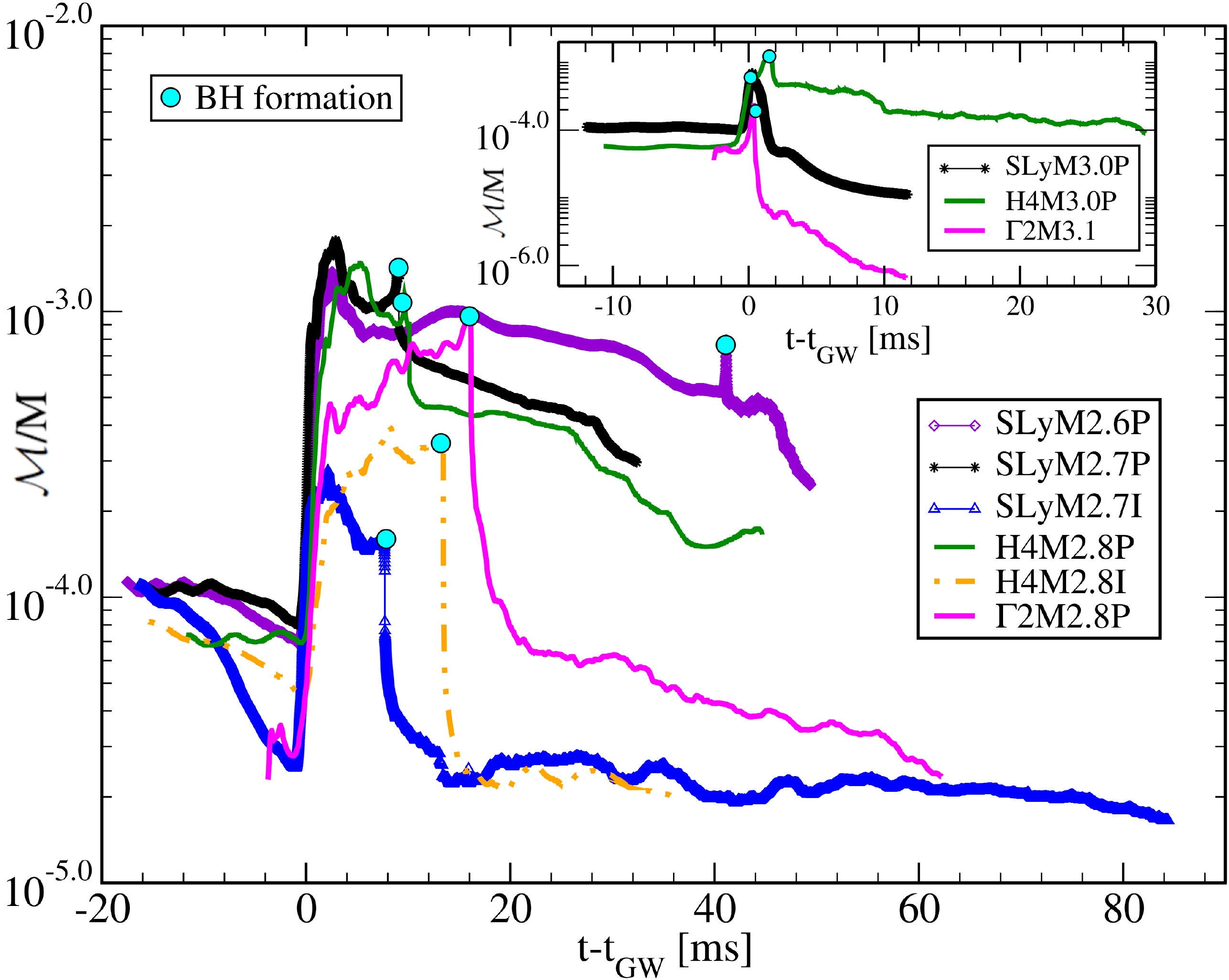}
  \caption{Total magnetic energy $\mathcal{M}$ normalized by the corresponding ADM mass
    ($M\sim 5.\times 10^{54}\,\rm erg$) versus coordinate time for cases listed in Table
    \ref{table:key_results_NSNS}. The inset displays the short-lived HMNS and prompt collapse cases.  
    Dots mark  the BH formation time $t_{\rm BH}$;  $t_{\rm GW}$ is the merger time.
    \label{fig:total_EM}}
\end{figure}

Fig.~\ref{fig:total_EM} shows the evolution of the magnetic energy for all cases in Table
\ref{table:key_results_NSNS}. During the inspiral the magnetic energy~$\mathcal{M}$ either
decreases or at most is slightly amplified until merger. This behavior depends on the NS
compactness  and the initial seed magnetic field geometry. In our extreme cases~(see
Table~\ref{table:NSNS_ID}), we observe that at merger $\mathcal{M}$
has decreased by a factor of $\sim 1.5$ in SLyM2.6P, while in $\Gamma2$M2.8P it has been amplified
by roughly the same factor. Our simulations show that until merger, the frozen-in magnetic field
lines, which are anchored to the fluid elements, are simply advected
(see second row, right panel in Fig.~\ref{fig:sly_ID}). Neither spurious magnetic fields or appreciable
changes in the internal structure of the stars are observed. The relative changes in the central NS rest-mass
density during  inspiral is $\lesssim 0.2\%$.  
It has been suggested that pure poloidal magnetic field configurations
are unstable on an Alfv\'en time scale
(see~e.g.~\cite{Stable_NS_cannot_have_poloidal_fields2,Stable_NS_cannot_have_poloidal_fields1}).
In our models, the Alfv\'en timescale can be estimated as (see Eq.~2~in~\cite{Sun:2018gcl})
\begin{eqnarray}
  \tau_{\rm Alfven}\sim\frac{R}{v_{A}}=\frac{\sqrt{4\pi\rho_0}\,R}{|B|}\sim 
  10\,|B|^{-1}_{15}\,R_{10}\,\rho^2_{0,14}\,{\rm ms}\,,
  \label{eq:alvfen_w}
\end{eqnarray}
where $|B|_{15}={|B|}/{(10^{15}\,\rm G)}$, $R_{10}={R}/(10\,\rm km)$ and
$\rho_{0,14}={\rho_0}/(10^{14}\,\rm g/cm^{3})$ are the characteristic strength of the magnetic
field, the radius, and rest-mass of the NS, respectively. In all our cases, we find that the
central Alfv\'en time is $\tau_{\rm Alfven}\lesssim 7.5\,\rm ms$, and hence their inspiral phase
last at least one Alfv\'en time (see Table~\ref{table:key_results_NSNS}). 
In addition to magnetic instabilities, tidal effects, which can drive fluid motion
in the stars, may also change the magnetic energy~\cite{Ciolfi:2017uak}.
We are currently investigating this effect.

During the next $\sim 3\,\rm ms$ following merger, the magnetic energy is exponentially
amplified  due mainly to the KH instability. Such an effect has been found in high
resolution studies reported in~\cite{Aguilera-Miret:2020dhz,Kiuchi:2015sga}.
We note that in binaries endowed  with a pulsar-like interior + exterior magnetic field, $\mathcal{M}$ is
amplified by a factor of $\sim 20$~(see~Fig.~\ref{fig:total_EM}). In contrast,
$\mathcal{M}$ is only amplified by a factor $\sim 10$ in those endowed with a magnetic
field confined to the NS. Further amplification during the HMNS phase is only observed in
$\Gamma2$M2.8P. In all other cases, $\mathcal{M}$ slightly decreases.

\begin{figure*}
  \centering
  \includegraphics[width=0.495\textwidth]{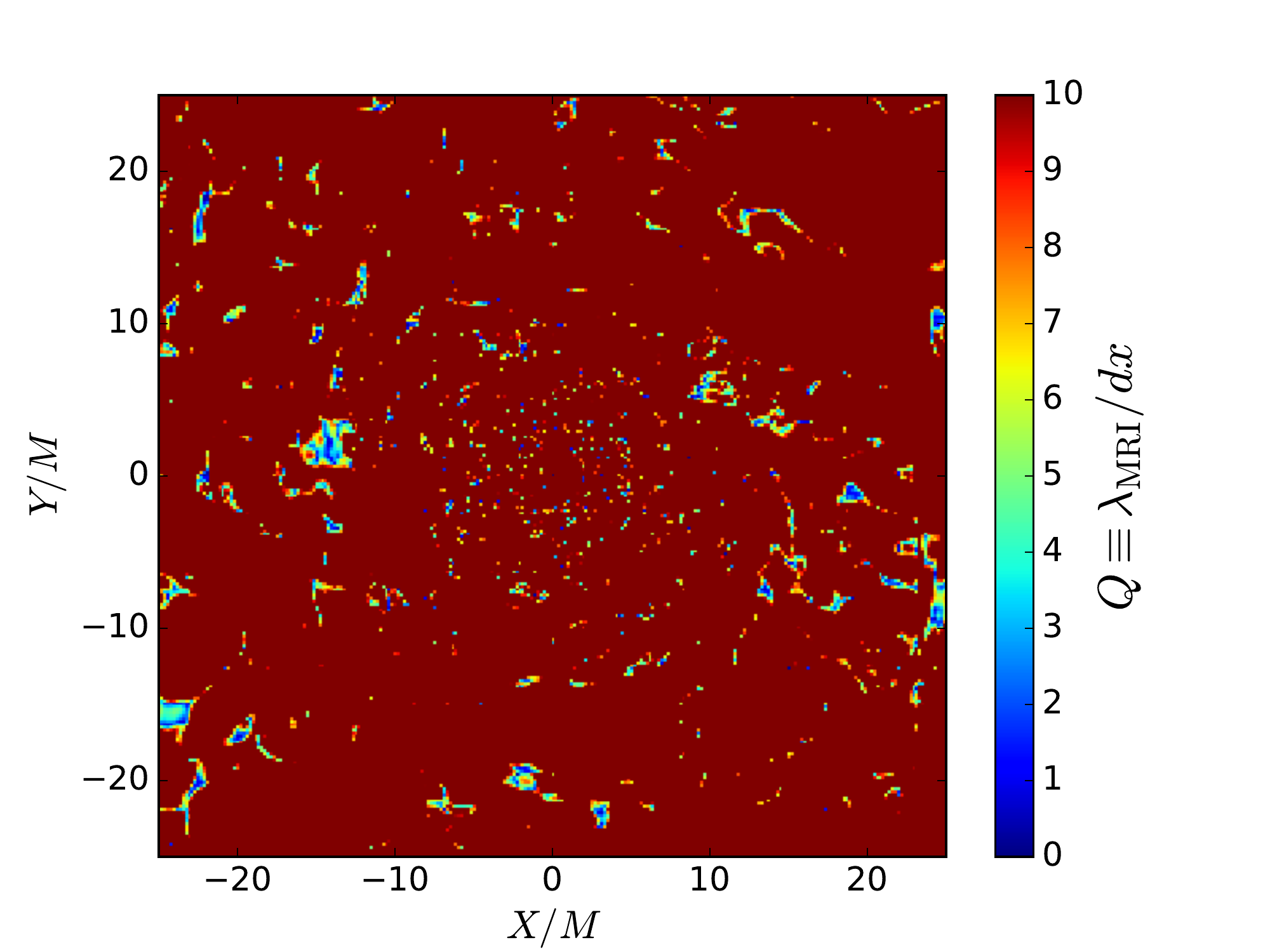}
  \includegraphics[width=0.495\textwidth]{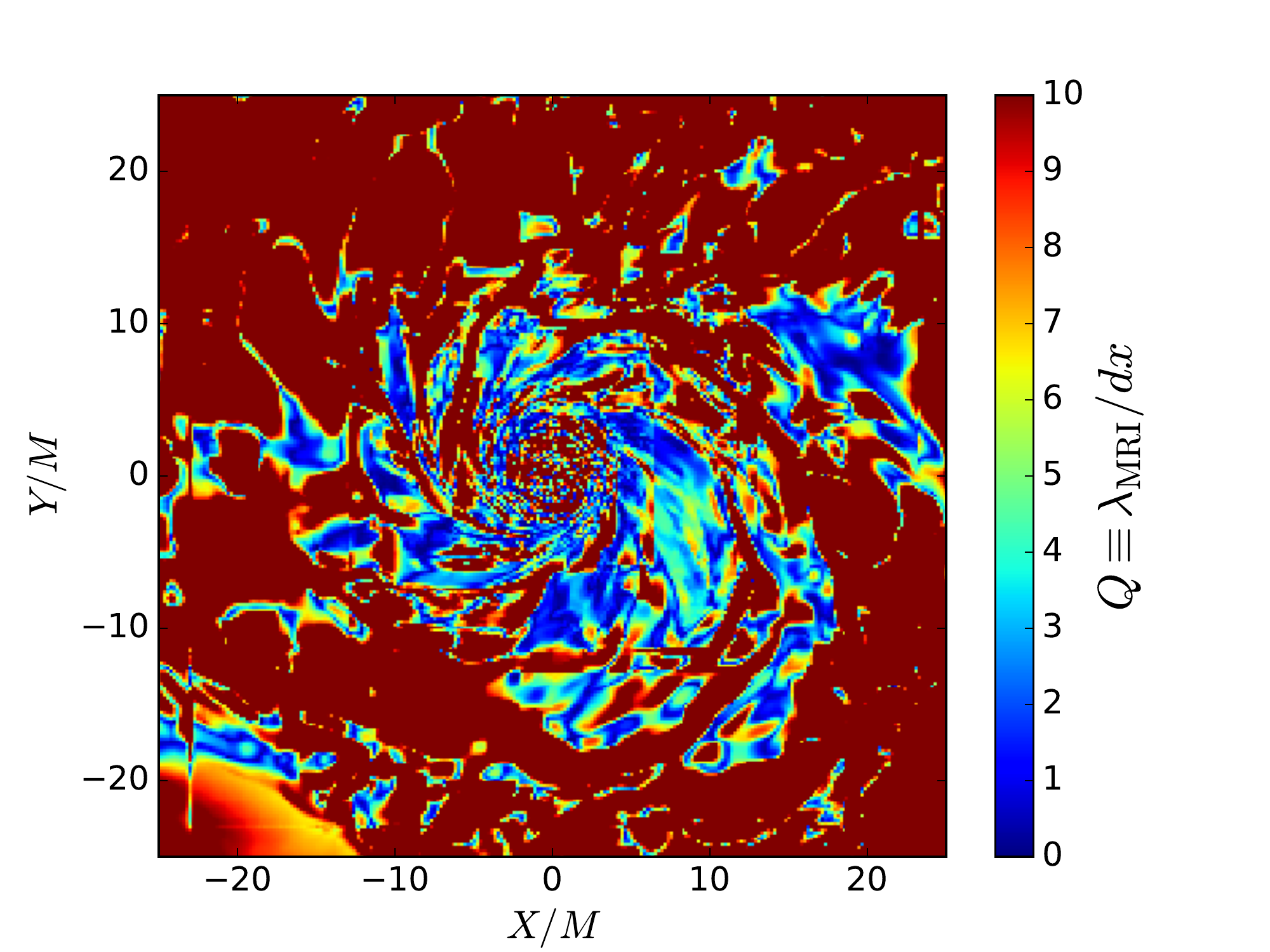}
  \includegraphics[width=0.495\textwidth]{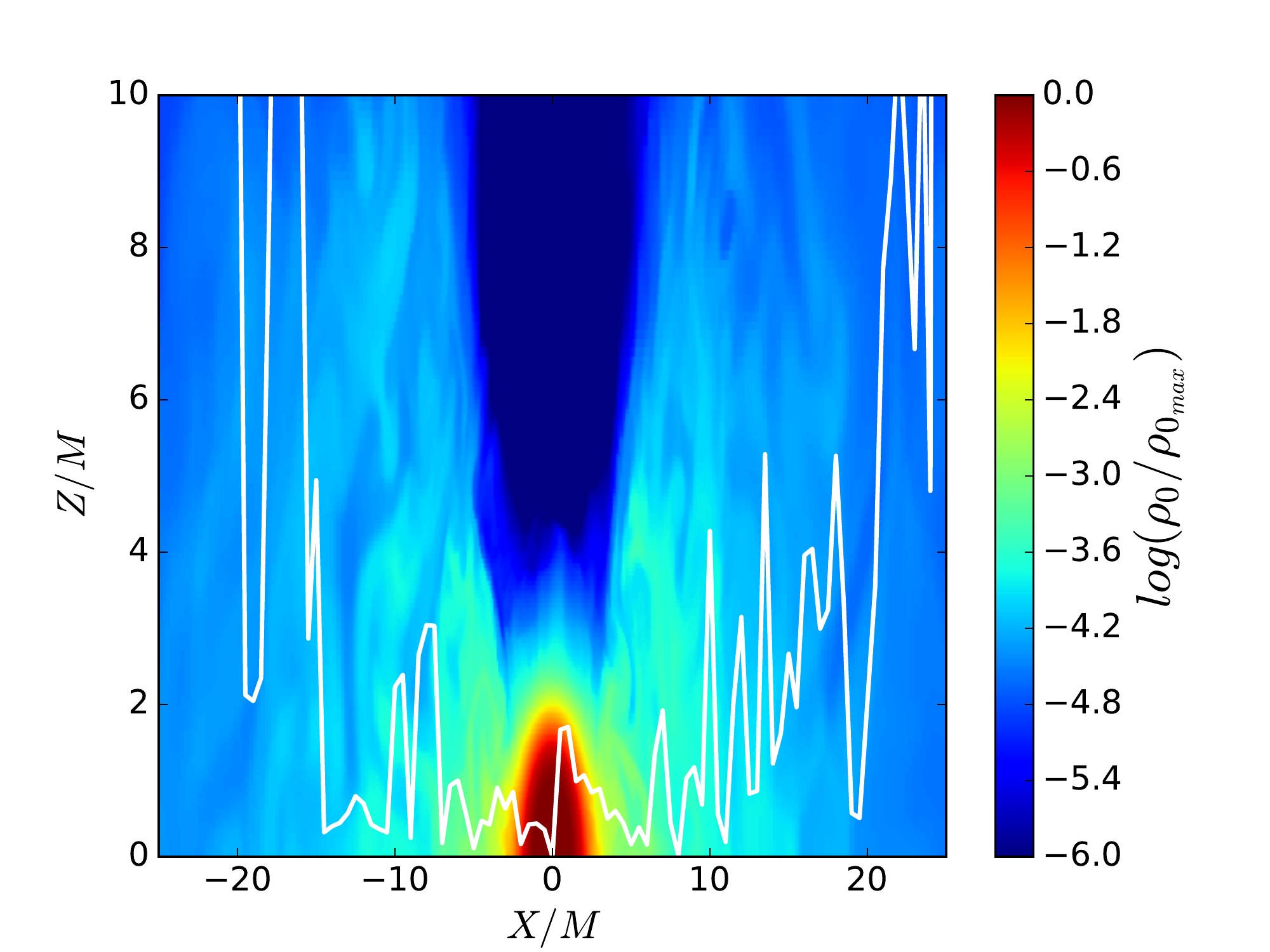}
  \includegraphics[width=0.495\textwidth]{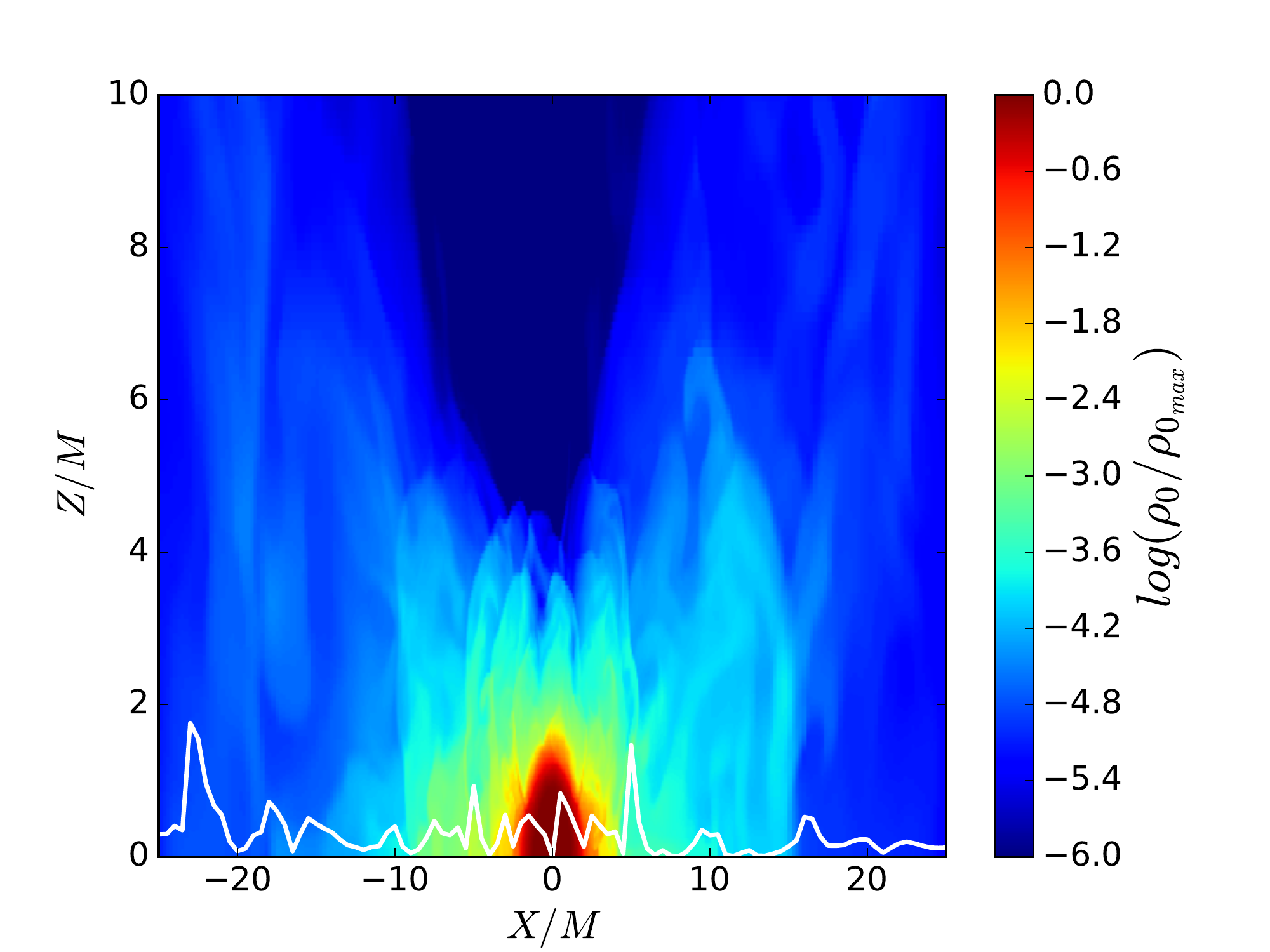}
  \caption{
    Contours of the quality factor $Q=\lambda_{\rm MRI}/dx$ on the equatorial plane (top),
    and the rest-mass density of the transient HMNS normalized to its initial maximum value
    (log scale) along with $\lambda_{\rm MRI}$ (white line) on the meridional plane (bottom),
    at $t-t_{\rm GW}\sim 5.2\,\rm ms$ for SLyM2.7P (left) and SLy2.7I (right). Similar
    behavior is observed in all other cases in~Table~\ref{table:key_results_NSNS}.
    \label{mri_alig}}
\end{figure*}

Top panels in Fig.~\ref{mri_alig} display the contours of the $\lambda_{\rm MRI}$-quality factor
$Q_{\rm MRI}$ on the equatorial plane for SLyM2.7P (left) and SLyM2.7I (right) after the transient
HMNS has settled down ($t-t_{\rm GW}\sim 5\,\rm ms$). This parameter must be $\gtrsim 10$
in order to resolve MRI.
We note that $Q_{\rm MRI}\gtrsim 10$ over a
rather lager portion of the HMNS remnant (central core + cloud of matter that has wrapped
around it)  of SLyM2.7P. By contrast, $Q_{\rm MRI}$ is larger than 10 
only in the bulk of the central core of the HMNS of SLyM2.7I, and by at most 4 
in the external cloud of matter (low-density regions). Bottom panels
in Fig.~\ref{mri_alig} show the rest-mass density of the HMNS on the meridional plane
overlaid by $\lambda_{\rm MRI}$ along the $x$ coordinate. For the MRI to be
unstable at a given position, the HMNS must extend to a local height above $\lambda_{\rm MRI}$.
In the above cases, $\lambda_{\rm MRI}$ fits within a region where the rest-mass
density is $\gtrsim 10^{11.3}\rm g/cm^3$, well above of the floor density. Based on the above
results, we conclude that MRI can operate all over the transient remnant in SLyM2.7P, but 
only in the central core of the HMNS of SLyM2.7I. This is expected because
right after merger, a double-core structure  is formed by the two central cores of the merging NS.
These two cores collide and bounce repeatedly until they eventually merge, forming a single central
core. During this process, the external layers of the merging stars gain angular momentum due
to orbital angular momentum advection and to
torques arising from the nonaxisymmetric structure of the double-core, as well as to  magnetic
instabilities (MRI and magnetic winding).  This causes the external layers to expand and
simultaneously the central core to shrink, forming a massive central core immersed in a low density
cloud.
In the HMNS remnant of SLyM2.7I, the cloud of matter is formed by low-magnetized material.
We recall that in this case,
the  magnetic field is initially confined  within interior regions with pressure larger than $1\%$
of the initial maximum pressure of the NS~(see Sec.~\ref{subsec:idata}), and hence its
outermost layers are initially unmagnetized. Consistent with this, once the central
core of the HMNS collapses, the magnetic energy in SLyM2.7I decreases and settles into a steady
state faster than in SLyM2.7P (see Fig.~\ref{fig:total_EM}).
Similar behavior is observed in all cases in Table~\ref{table:key_results_NSNS}; MRI
is operating all over the HMNS formed after the merger of an NSNS initially
seeded with a pulsar-like interior + exterior magnetic field, and is only operating in the central core
of the HMNS of those binaries seeded with a magnetic field confined inside the star.
Such behavior has been reported before~(see e.g~ Fig.~17~in~\cite{Ciolfi:2019fie}). Notice that
very high-resolutions are required to properly capture MRI in the low density regions of the
HMNS formed after the collapse of stars initially endowed with magnetic field confined to
their interior.

Calculating the effective Shakura–Sunyaev $\alpha_{\rm SS}$ parameter in the HMNS,
we find that $\alpha_{\rm SS}$ ranges between~$\sim 10^{-3.5}$ to $\sim 10^{-2}$ (see Table
\ref{table:key_results_NSNS}). Similar values  were reported in previous, high-resolution NSNS merger
studies~\cite{Kiuchi:2017zzg}. Therefore, we expect that  magnetic turbulence, which  grows on an
effective viscous time scale 
$\tau_{\rm vis} \sim{R^{3/2}_{\rm HMNS}}\,{M_{\rm HMNS}^{-1/2}\,\alpha_{\rm SS}^{-1}}\sim 1-10\,\rm ms$
(see Eq.~3~in~\cite{Sun:2018gcl}) where $R^{3/2}_{\rm HMNS}$ and $M_{\rm HMNS}$ are the characteristic
radius and mass of the HMNS, is also operating in the remnant. We confirm this on meridional slices
of the transient star, where we see evidence for turbulent magnetic fields. Magnetic turbulence
strongly depends on resolution. The results in~\cite{Kiuchi:2017zzg} show that numerical (diffusion)
artifacts can suppress a sustained magnetic turbulence, hence values of $\alpha_{\rm SS}$ quoted in
Table~\ref{table:key_results_NSNS} may be underestimated. Higher resolutions studies may shorten the
viscous time scale~$\tau_{\rm vis}$.

\begin{figure*}
  \centering
  \includegraphics[width=0.495\textwidth]{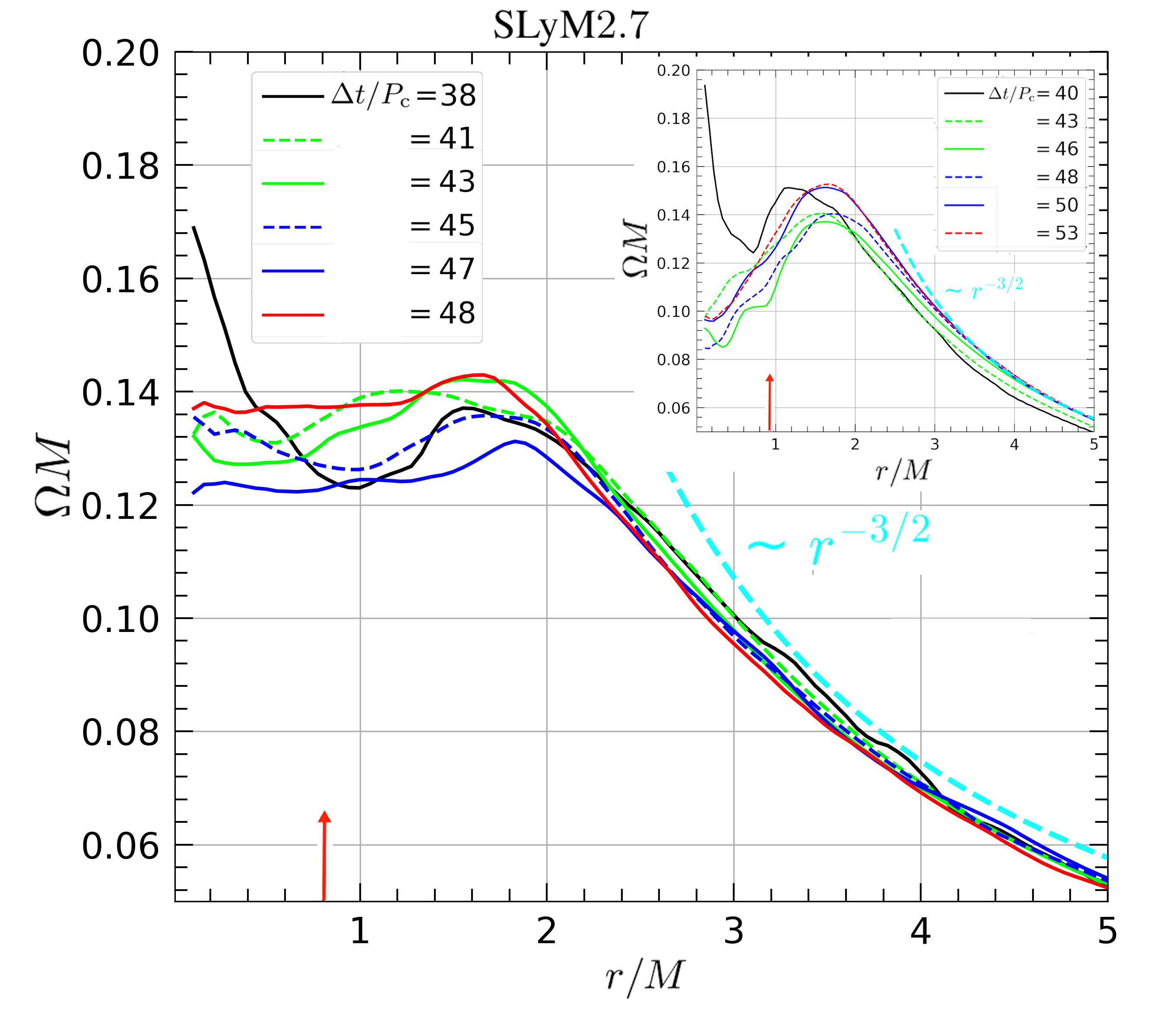}  
  \includegraphics[width=0.495\textwidth]{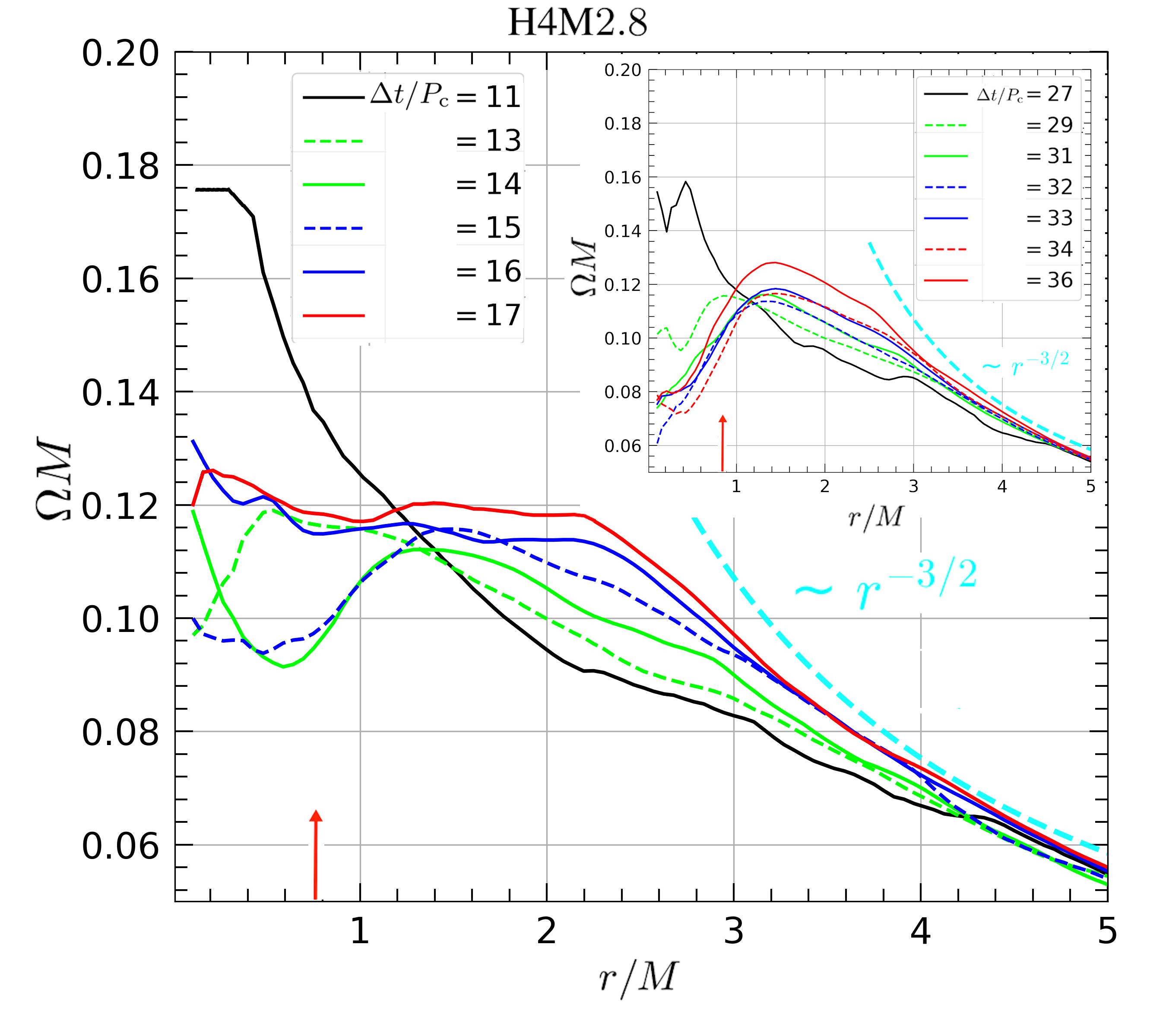}
  \caption{Average rotation profile of the HMNS~(see Eq.~2 in~\cite{Ruiz:2019ezy}) for
    magnetized cases SLyM2.7P (left) and H4M2.8P (right) in the equatorial plane at $\Delta t =
    t-t_{\rm HMNS}$, together with a Keplerian angular velocity  profile. The inset displays
    the corresponding unmagnetized cases (see~Table~\ref{table:key_results_NSNS}).
    The arrow marks the coordinate radius containing~$\sim 50\%$ of the total rest-mass
    of the transient remnant.  Here $t_{\rm HMNS}$ is  the HMNS formation time, with  $P_{\rm c}$
    the central  HMNS period at $t=t_{\rm HMNS}$, which is  $P_{\rm c}\sim 0.25\,\rm ms$ and
    $P_{\rm c}\sim 0.51\,\rm ms$ for SLy and H4 cases, respectively.
    \label{fig:rotatio_prof}}
\end{figure*}

Following merger, nonaxisymmetric oscillation modes of the HMNS, which persist until stellar collapse
to a BH, trigger the emission of  quasiperiodic GWs~(see~Fig.~\ref{fig:hydro_magGW}).
The dissipation of angular momentum due to
GW radiation is more efficient in the unmagnetized cases. In particular, we find that in SLy2.6H around
$17\%$ of the total angular momentum is carried away by gravitational radiation, while in SLy2.6P it is
around $14\%$~(see Table~\ref{table:key_results_NSNS} for other cases). However, centrifugal support
due to differential rotation allows the HMNS remnant of SLy2.6H to survive for more than $\sim 43\,\rm ms$
after merger, time at which we terminate its evolution. By contrast, magnetic turbulence
in the  HMNS remnant damps the centrifugal support, driving the collapse to a BH roughly at
$\sim 41\,\rm ms$ following merger. This effect is shown in Fig.~\ref{fig:rotatio_prof} in which the
averaged angular rotation profiles of the HMNSs for SLyM2.7 and H4M2.8 are plotted  at regular times
(in multiples of the initial central period
of the HMNS). In contrast to the unmagnetized cases (see~insets~in Fig.~\ref{fig:rotatio_prof}) where the angular velocities 
roughly maintain their initial profiles, in the magnetized cases the central core of
the HMNS becomes almost uniformly rotating over an Alfv\'en timescale (see Eq.~\ref{eq:alvfen_w}),
while its external, low density envelope maintains a Keplerian rotation profile. These results demonstrate
that angular momentum redistribution from the inner to the outer regions due to nonaxisymmetric torques
and GW losses are inefficient mechanisms compared to magnetic fields in removing the added centrifugal
support provided by differential rotation~\cite{Sun:2018gcl}.

\begin{figure}
  \centering
  \includegraphics[width=0.49\textwidth]{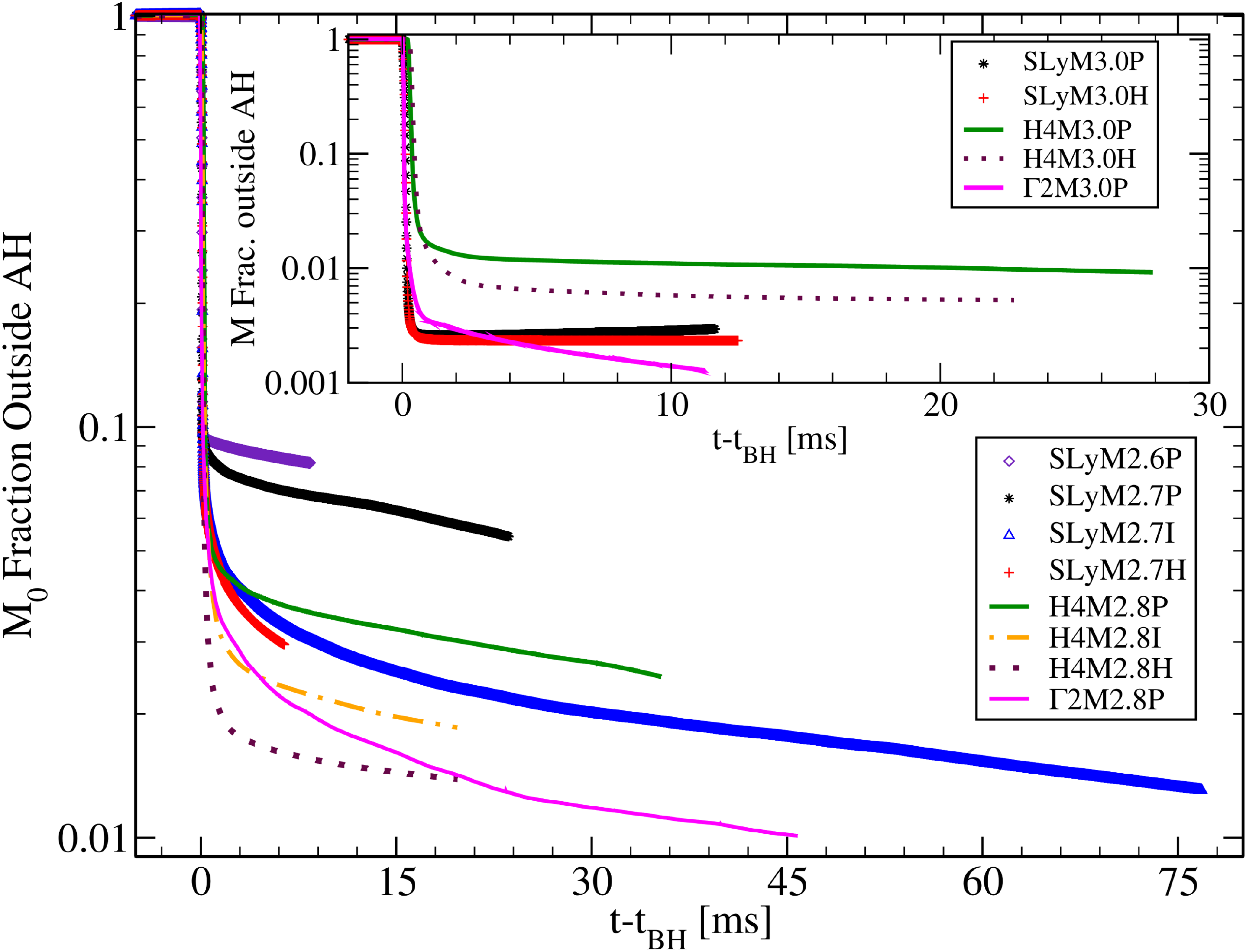}
  \caption{Rest-mass fraction outside the BH apparent horizon versus coordinate time for all cases listed in
    Table~\ref{table:key_results_NSNS}.
    \label{fig:M0_outside}}
\end{figure}

Angular momentum redistribution eventually triggers HMNS collapse to a highly-spinning BH
immersed in an accretion disk~(see bottom panels in~Fig.~\ref{fig:sly_ID}).
The stiffer the EOS, the heavier the BH remnant and the higher its spin parameter. In particular, the
mass and spin of the BH remnant in SLyM2.7P are $\sim 2.45M_\odot$ and $a/M_{\rm BH}\sim 0.62$, respectively,
while in $\Gamma2$M2.8P they are
$\sim 2.65M_\odot$ and~$a/M_{\rm BH}\sim 0.74$ (see Table~\ref{table:key_results_NSNS} for other cases).
As mentioned before, after merger angular momentum  is transferred from the inner to the outer layers of
the stars, causing the latter to expand. In the soft EOS (SLy) a significant fraction of the matter from the
outer layers of the star gains enough angular momentum to expand and remain outside the ISCO once the bulk
of the star collapses. Eventually this material wraps around the BH forming the accretion disk. In the stiff EOS
($\Gamma=2$), the external layers remain closer to the bulk of the star and, hence, when the BH forms,
most of them get caught inside the ISCO and eventually are swallowed by the BH,  leaving less material
to form the accretion disk.
%
%
Consistent with this, Fig.~\ref{fig:M0_outside} shows the
rest-mass of the accretion disk~for all cases in Table~\ref{table:key_results_NSNS}, which
ranges between $3\%$ in $\Gamma2$M2.8P (stiff EOS) to $7\%$ in SLyM2.7P  (soft EOS)
of the total mass of the system.  

In our long-lived HMNS (SLy2.6P) case, in which the transient remnant has a lifetime
of~$\sim 41\,\rm ms$, angular momentum transfer
operates for many rotation periods, and hence more material from the external layers
can be released in the surroundings of the central core, forming  a puffy disk that wraps around
it. Consistent with this, Fig.~\ref{fig:M0_outside} demonstrates that the accretion disk of the
SLy2.6P remnant  is the most massive one of our cases, and hence it has the  lightest BH
($\sim 2.26M_\odot$) with the lowest dimensionless spin ($a/M_{\rm BH}\sim 0.47$).

Right before the collapse of the HMNS,  the rms magnetic field strength ranges between
$\sim 10^{15.8}$ and $10^{16.2}\,\rm G$~(see~Table~\ref{table:key_results_NSNS}), consistent with the
values reported in~\cite{Kiuchi:2015sga,Aguilera-Miret:2020dhz}.
Shortly after BH formation, the inner region of the star, which contains most of the magnetic
energy, is promptly swallowed by the BH, causing the magnetic energy to drop quickly in $\Delta
t\lesssim 3\,\rm ms$ and then slightly decrease thereafter as the accretion proceeds (see~Fig.~\ref{fig:total_EM}).
During the BH + disk evolution phase we do not find evidence of magnetic field enhancement.
In all cases, the rms value of the  magnetic field in the bulk of the disk remnant is $\lesssim
10^{15}\,\rm G$. 
%
%

\begin{figure}
  \centering
  \includegraphics[width=0.49\textwidth]{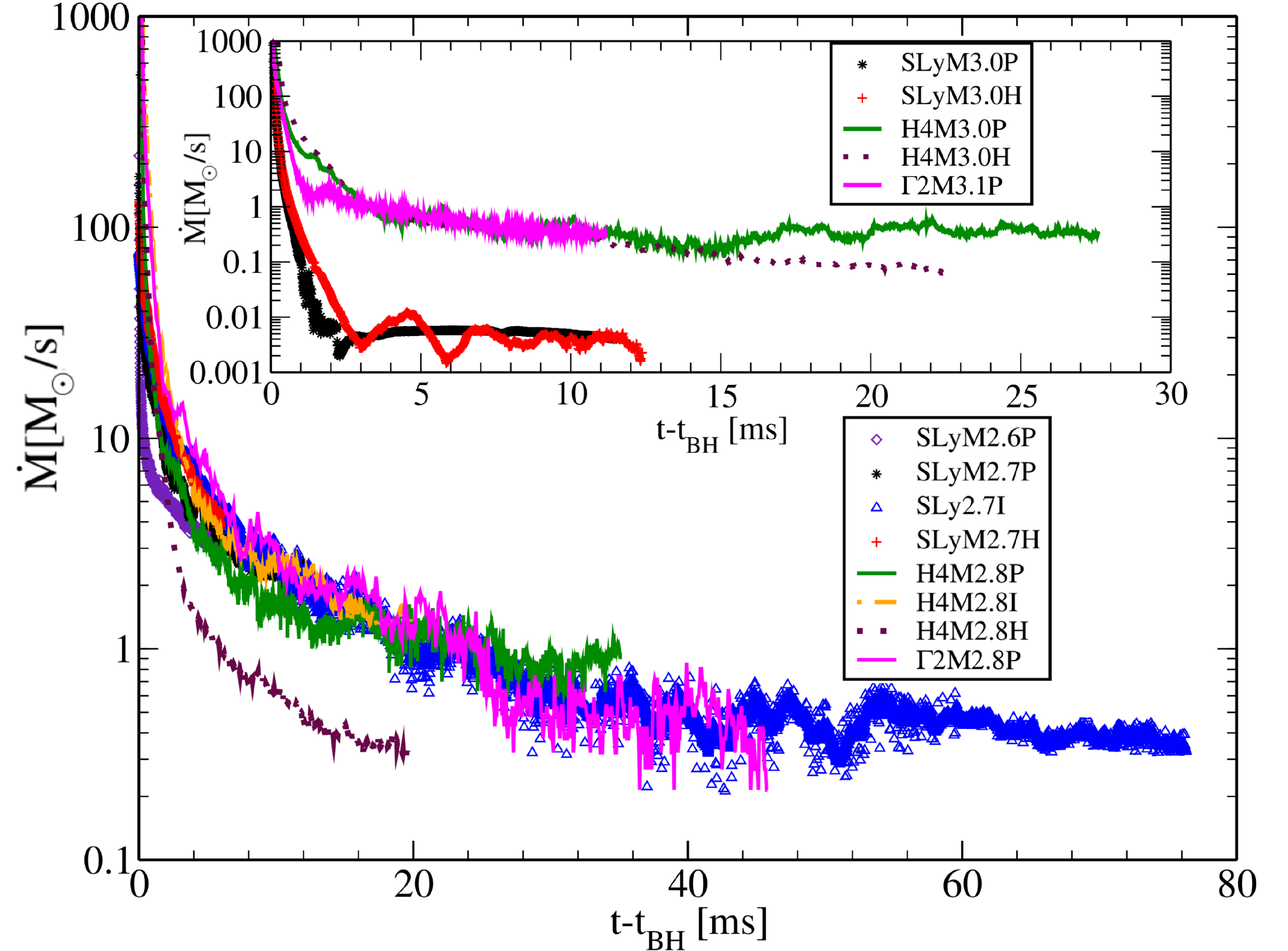}
  \caption{Rest-mass accretion rate for all cases listed in Table~\ref{table:key_results_NSNS}
    computed via Eq.~(A11) in~\cite{Farris:2009mt}. 
    \label{fig:M0_dot}}
\end{figure}

Shortly after BH formation, material ejected during merger and HMNS formation
begins to fall back, increasing the ram-pressure. This pressure is so strong that it prevents
the launching of a wind~\cite{kks12}. However, as shown in the third row and right panel in
Fig.~\ref{fig:sly_ID}, magnetic winding 
has begun even before BH formation. As the accretion proceeds~(see Fig.~\ref{fig:M0_dot}), the
baryon-loaded environment in the polar region above the BH gradually becomes thinner until the
ratio $B^2/(8\,\pi\,\rho_0)$ exceeds unity. At this point, the inflow halts and eventually a
magnetically-supported outflow is triggered--the incipient jet.
 
\paragraph{\bf Magnetic field initially confined inside the star:}
During the HMNS phase, the enhancement of the magnetic energy in case H4M2.8I is a factor of $\sim 2$ larger
than that in SLyM2.7I~(see~Fig.~\ref{fig:total_EM}). However,  following  BH formation,  we note that in
both cases $\mathcal{M}$ plummets and settles into a steady state. By $t-t_{\rm BH}\sim 3\rm\,ms$  the magnetic
energy in both cases is $\mathcal{M}/M\sim 10^{-4.6}$ and roughly remains constant until the termination of our
simulations. We evolve these two cases during the next $t-t_{\rm GW}\sim 20\,\rm ms$ and, consistent with the
results reported in ~\cite{Kiuchi:2014hja},  we observe a persistent inflow toward the BH and an intermittent 
and weak helical magnetic field structure above the BH poles ($B_p\sim 10^{15.8}\rm G$), though in both cases we
note that the ratio~$B^2/(8\,\pi\,\rho_0)$ is slightly rising as the accretion takes place. As  the behavior of
the BH + disk remnant in these two cases is basically the same, given our finite computational resources
we chose to continue only the evolution of SLyM2.7I. This case has the  longest  accretion disk
lifetime~(see~Table~\ref{table:key_results_NSNS}), consistent with the sGRB engine lifetime
\cite{Beniamini:2020adb,Bhat:2016odd,Lien:2016zny,Svinkin:2016fho}.

By~$t-t_{BH}\sim 3000M\sim 40\,\rm ms$, the magnetic pressure in the regions immediately above the BH remnant 
is high enough to balance the ram pressure of the fall-back material and hence the inflow halts. Fluid
velocities then begin to turn around and magnetically dominated regions (i.e.~$B^2/(8\,\pi\rho_0)\gtrsim1$)
gradually expand. As these regions expand, the field lines tighten around them forming a helical structure,
inside of which fluid elements escape.
By~$t-t_{BH}\sim 4200M\sim 56\,\rm ms$ the outflow, which  has
been accelerated to Lorentz factor $\Gamma_L\lesssim 1.25$, reaches a height of $\sim 100M\sim 400\,\rm km$,
and so an incipient jet has formed (see top left panel in Fig.~\ref{fig:NSNS_Sly_H4_G2}). Following
\cite{Ruiz:2016rai}, we define the half-funnel opening angle $\theta_{\rm f}$ as the boundary of the region
above the BH in which $B^2/(8\,\pi\rho_0)\gtrsim 10^{-2}$. In this case, we find that the half-opening angle is
$\theta_{\rm f}\sim 25^\circ$.

We also assess if the BZ mechanism is operating in the BH + disk remnant, as we found in our previous studies.
For this,  we compare the outgoing Poynting luminosity $L_{\rm EM}$ generated in our simulations~(see
Sec.~\ref{subsec:diagnostics}) with the EM power generated by the BZ mechanism given in~\cite{Thorne86} 
\begin{eqnarray}
  L_{\rm BZ}\sim 10^{52}\,\left(\frac{\tilde{a}}{0.75}\right)^2\,
  \left(\frac{M_{\rm BH}}{2.8 M_\odot}\right)^2\,|B_{\rm p}|_{16}^2\,\rm erg/s\,,
  \label{eq:LBZ}
\end{eqnarray}
where $|B_{\rm p}|_{16}\equiv|B_{p}|/10^{16}\rm G$ is the strength of the magnetic field at the BH
poles. The Poynting luminosity, time-averaged over the last $\sim 5\,\rm ms$ before the termination of our 
simulation for SLyM2.7I, is $10^{51.3}\rm erg/s$~(see~Fig.~\ref{fig:Poynting_plot}). On the other hand,
$|B_{\rm p}|\sim~10^{15.8}\,\rm G$, and $L_{\rm BZ}\sim 10^{51.4}\rm\, erg/s$, in close agreement.

Another key feature of the BZ mechanism is that the field lines rotate at frequency $\Omega_F\approx 0.5\,
\Omega_H$ for $\tilde{a}\ll 1$ if the field has a monopole geometry and its surroundings are strongly
force-free ($B^2/(8\,\pi\rho_0)\gg 1$)~\cite{BZeffect77}. Here $\Omega_H$ is
the angular frequency of the BH. The numerical results in~\cite{Komissarov2001} showed that for a monopole magnetic
field around a spinning BH, the field lines at the BH poles rotate at a frequency of $\sim 0.5\,\Omega_H$
for a BH spin $\tilde{a}=0.1$, and between $\sim 0.52\,\Omega_H$, at the BH pole, and $\sim 0.49\,\Omega_H$,
near the equator, for a  BH spin $\tilde{a}=0.9$.  Following~\cite{prs15}, we  measure~$\Omega_F$ on
a meridional plane passing through the BH centroid and along coordinate semicircles of radii $r_{\rm BH}$
and $2\,r_{\rm BH}$~within the magnetically-dominated (or mildly force-free) region. We find that the field
lines differentially rotate with a frequency in the range $\Omega_F/\Omega_H\sim 0.2-0.6$. As pointed out
in~\cite{2004ApJ...611..977M,prs15}
the differences from the expected $\Omega_F/\Omega_H\sim 0.5$ BZ-factor may be due to  artifacts such
as the deviation from strictly force-free conditions, deviations from monopole geometry and/or
lack of resolution.  Our cumulative results nevertheless suggest that the BZ mechanism is operating in our
system, as concluded in~\cite{Ruiz:2016rai,Ruiz:2019ezy}.
%
\paragraph{\bf Pulsar-like Magnetic field:}

The magnetic energy exponentially increases during merger and slightly changes during the 
HMNS phase. As shown in Fig~\ref{fig:total_EM}, by $t-t_{\rm GW}\sim 1.5$ the magnetic energy in all these cases
is amplified roughly by the same factor, which is $\gtrsim 3$ times larger than in those seeded with
a magnetic field confined inside the star.
We note that in HMNS cases with intermediate lifetime, the decrease of the magnetic energy following BH formation
depends on the EOS. The stiffer the EOS, the faster~$\mathcal{M}$ decreases. In particular, the magnetic
energy in SLyM2.7P decreases by a factor of $\sim 2$ in $t-t_{\rm BH}\sim 3\,\rm ms$,  while in $\Gamma$2M2.8P
it decreases by a factor of $\sim 10$ during the same period of time. As shown in~Fig~\ref{fig:total_EM}, at the
termination time of our simulations, the BH + disk remnants  of  the SLy and H4 cases have a larger magnetic energy
than that in $\Gamma2$M2.8P. This is likely due to the amount of magnetized material that wraps around the BH forming
the accretion disk. As mentioned before, Fig.~\ref{fig:M0_outside} shows that  the stiffer the EOS, the smaller the
accretion disk, so less magnetized material outside of the BH.

Consistent with our previous results~\cite{Ruiz:2016rai}, 
as the magnetic field is not amplified during the BH + disk phase,  a magnetically driven jet is launched only
after the density in the polar region above the BH poles decreases to values~$\rho_0\lesssim B^2/8\pi$.
Figs.~\ref{fig:sly_ID} (right bottom panel),~\ref{fig:NSNS_Sly_H4_G2}
and~\ref{fig:NSNS_Sly_int} (middle panel) show the  helical structure of the magnetic field once the
incipient jet has reached steady state.
There is a trend in the jet launching time: the shorter the HMNS lifetime, the faster the emergence
of the incipient jet.
In particular, the HMNS lifetime in H4M3.0P is $\tau_{\rm HMNS}\sim 2.5\,\rm ms$ and the system
launches an incipient jet at time $t-t_{\rm BH}\sim 19\,\rm ms$ following BH formation. By contrast, the
HMNS lifetime in H4M2.8P is~$\tau_{\rm HMNS}\sim 9.6\,\rm ms$ and its BH + disk remnant launches a jet by
$t-t_{\rm BH}\sim 27\,\rm ms$~following BH formation. Similarly, $\tau_{\rm HMNS}\sim 17\,\rm ms$ in $\Gamma2$M2.8P
and the jet is launched by $t-t_{\rm BH}\sim 40\,\rm ms$. This time difference is likely due to the
ram-pressure of the falling-back debris toward the BH. The shorter the HMNS lifetime, the less angular momentum
deposited in the outer layers of the transient remnant, and so the less material released in the atmosphere. This
effect is translated in lighter baryon-loaded environments, which allows the ratio $B^2/(8\,\pi\rho_0)$
to grow  to  values $\gtrsim 1$ more rapidly. Consistent with this, 
the BH + disk remnant of SLyM2.7P launches an incipient jet by $t-t_{\rm BH}\sim 24\,\rm ms$ , while that in
SLyM2.6P (the case with the longest $\tau_{\rm HMNS}$) launches it by $t-t_{\rm BH}\sim 26\,\rm ms$.
In the latter case angular momentum transfer processes operate for longer times than
in the former.
This behavior is confirmed in Fig.~\ref{fig:b2rho}, which  displays the force-free parameter~$B^2/(8\,\pi\,\rho_0)$
once the incipient jet is well-developed for SLyM2.7P (left), H4M2.8P (middle), and $\Gamma$2M2.8P (right).
The shorter the $\tau_{\rm HMNS}$, which results in lighter baryon-loaded environments, the higher
the above ratio.
%

In all cases, magnetically dominated regions
($B^2/(8\,\pi\rho_0)\gtrsim 1$) extend to heights $\gtrsim 20M\approx 20\,r_{\rm BH}$ above the BH, where $r_{\rm BH}$
is the BH apparent horizon radius. As before, we use the $B^2/(8\,\pi\,\rho_0) \sim 10^{-2}$ contour as the definition
of the funnel boundary, which has an opening half-angle of $25^\circ-35^\circ$. We note that in the funnel
the maximum value of the Lorentz factor is $\Gamma_L\sim 1.2-1.25$~(see~Table~\ref{table:key_results_NSNS}) and
hence the outflow is only mildly relativistic. However, as pointed out in~\cite{Vlahakis2003}, the maximum attainable
Lorentz factor of a magnetically--powered, axisymmetric jet is $\Gamma_L \approx B^2/(8\,\pi\rho_0)$. Therefore,
we expect  that the fluid elements inside the funnel can be accelerated to values $\Gamma_L\gtrsim 100$
as required by most sGRB models~\cite{Zou2009}.

Fig.~\ref{fig:Poynting_plot} displays the outgoing Poynting luminosity $L_{\rm EM}$. We find that 
$L_{\rm EM}\sim 10^{51.3}-10^{52.8}$~(see Table~\ref{table:key_results_NSNS}). These values reside
in the narrow band of luminosities shown theoretically in~\cite{Shapiro:2017cny} to constitute a ``universal range'' for
most BH + disk + jet systems arising from compact binary mergers containing NSs, as well as from the
magnetorotational collapse of massive stars. This band also agrees with the narrow range characterizing
the observed luminosity distributions of over $400$ short and long GRBs with distances inferred from
spectroscopic redshifts or host galaxies\cite{Li:2016pes} (see also~\cite{Beniamini:2020adb}). A similar
universal range applies  to
the BH accretion rates upon jet launching, also shown in~\cite{Shapiro:2017cny}, and the range shown in
Fig.~\ref{fig:M0_dot}, is also consistent  with this expectation.  We also note that near
the end of the simulations, the magnetic field magnitude above the BH pole is $\sim 10^{15.8}-10^{16.2}\rm\,G$
and hence using Eq.~\ref{eq:LBZ} we have~$L_{\rm BZ}\sim 10^{51.4}-10^{52.1}$,  which is roughly consistent with
our numerical results as well.
As before, we compute the angular frequency of the magnetic field lines on a meridional plane passing through
the BH centroid and along coordinate semicircles of radii $r_{\rm BH}$ and $2\,r_{\rm BH}$ within the
force-free region. In all these cases the field lines differentially rotate  with a frequency in the range
$\Omega_F\sim 0.1-0.54$, and  according to our previous discussion,  it is likely that the BZ
mechanism is operating in our systems.
%
%
\subsection{Prompt collapse}
\label{sec:prompt}
The basic dynamics of NSNSs undergoing prompt collapse has been reported in~\cite{Ruiz:2017inq}, where
it was found that, shortly after the stars touch for the first time the remnant undergoes collapse to a BH.
As there is no angular momentum transport due magnetic instabilities, only the external and low density outer
layers of the stars, which are able to gain enough angular momentum due to tidal torques, are pushed away
from the bulk of the merging stars. Eventually, this material wraps around a highly spinning BH, forming
an accretion disk with a rest-mass $\lesssim 0.5\%$ of the total rest-mass of the system~(see Table
\ref{table:key_results_NSNS}).
The rest-mass of the BH remnant settles to  $M_{\rm BH}\gtrsim 2.8M_\odot$, with a dimensionless spin
parameter $a_{\rm BH}/M_{\rm BH}\sim 0.8$. The BH remnant in prompt collapse cases
is the heaviest and with the highest spin of all our cases. These results are anticipated because
basically all the material is swallowed by the nascent BH.

Consistent with the previous results, the inset in Fig.~\ref{fig:total_EM} shows that during the merger
the magnetic energy is continuously amplified until BH formation, when it then plummets. By $t-t_{\rm BH}\sim
2\,\rm ms$ it falls to values $\mathcal{M}/M\lesssim 10^{10^{-5}}$ and continuously decreases thereafter. 
We track the evolution of the BH +  disk remnant  for $t-t_{\rm BH}\sim 15\,\rm ms$. We note that on
$x$-$z$ meridional slices passing through the BH centroid, regions with  $B^2/(8\,\pi\rho_0)\sim 10^{-0.6}$
begin to expand as material in the polar region is accreted. By $t-t_{\rm BH}\sim 9\,\rm ms$, these
regions reach a height of~$\sim 10\,r_{\rm BH}$, where the magnetic pressure is high enough to
balance the ram pressure of the fall-back material.  Subsequently, we observe that these regions expand and
contract  until the termination of our simulations. As shown in the left and right panels
of Fig.~\ref{fig:NSNS_Sly_int}, we do not find evidence of an outflow or magnetic field collimation,
hence no jet.
%
%
\subsection{Dynamical ejection of matter}
\begin{figure}
  \centering
  \includegraphics[width=0.49\textwidth]{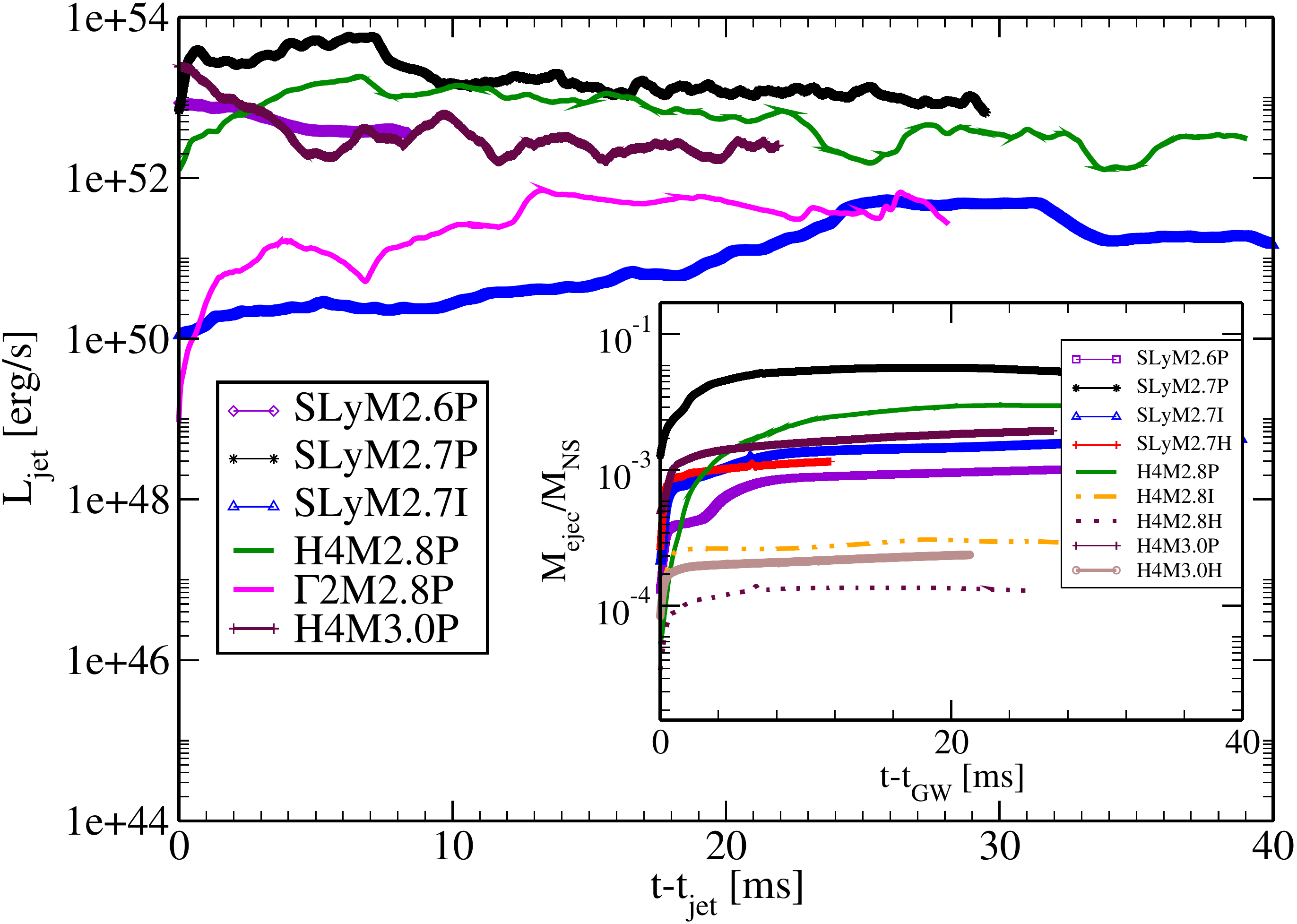}
  \caption{Outgoing EM (Poynting) luminosity following jet launching at a  coordinate sphere
    of  radius $r=160M$ for cases listed in Table~\ref{table:key_results_NSNS}. The inset shows
    the rest-mass fraction of escaping matter following the GW peak amplitude (merger).
    \label{fig:Poynting_plot}}
\end{figure}

\begin{figure*}
  \centering
  \includegraphics[width=0.32\textwidth]{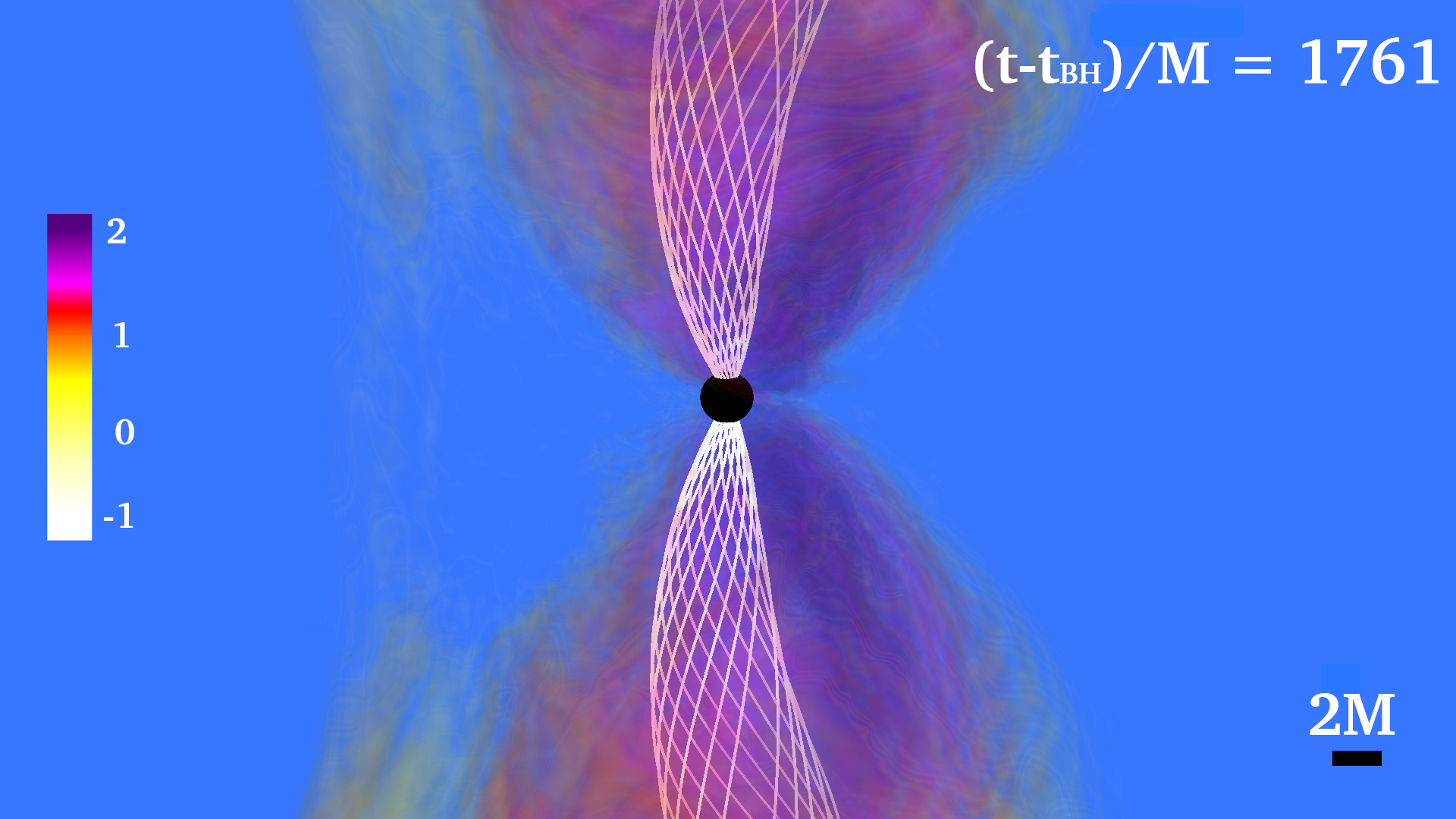}
  \includegraphics[width=0.32\textwidth]{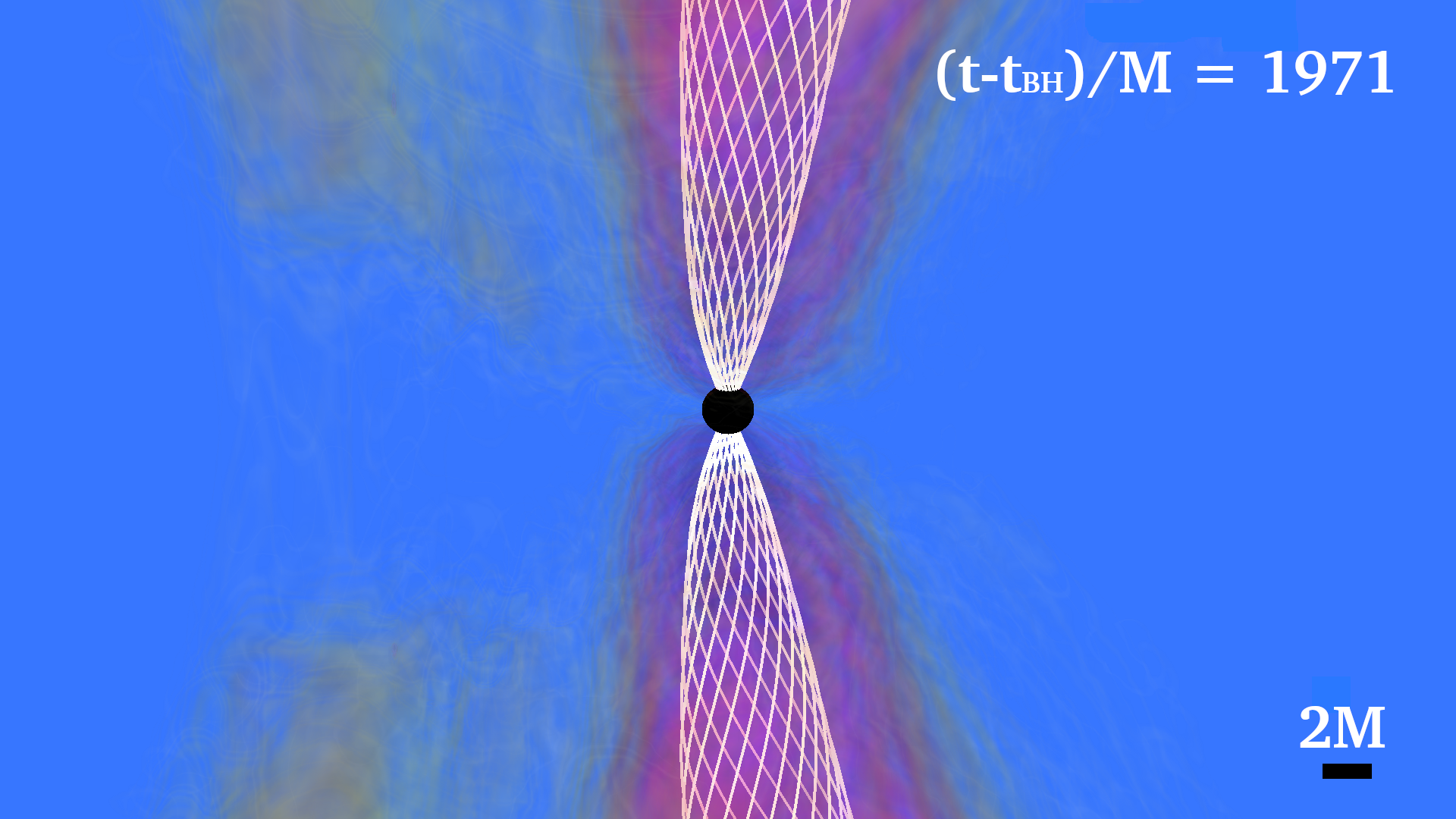}
  \includegraphics[width=0.32\textwidth]{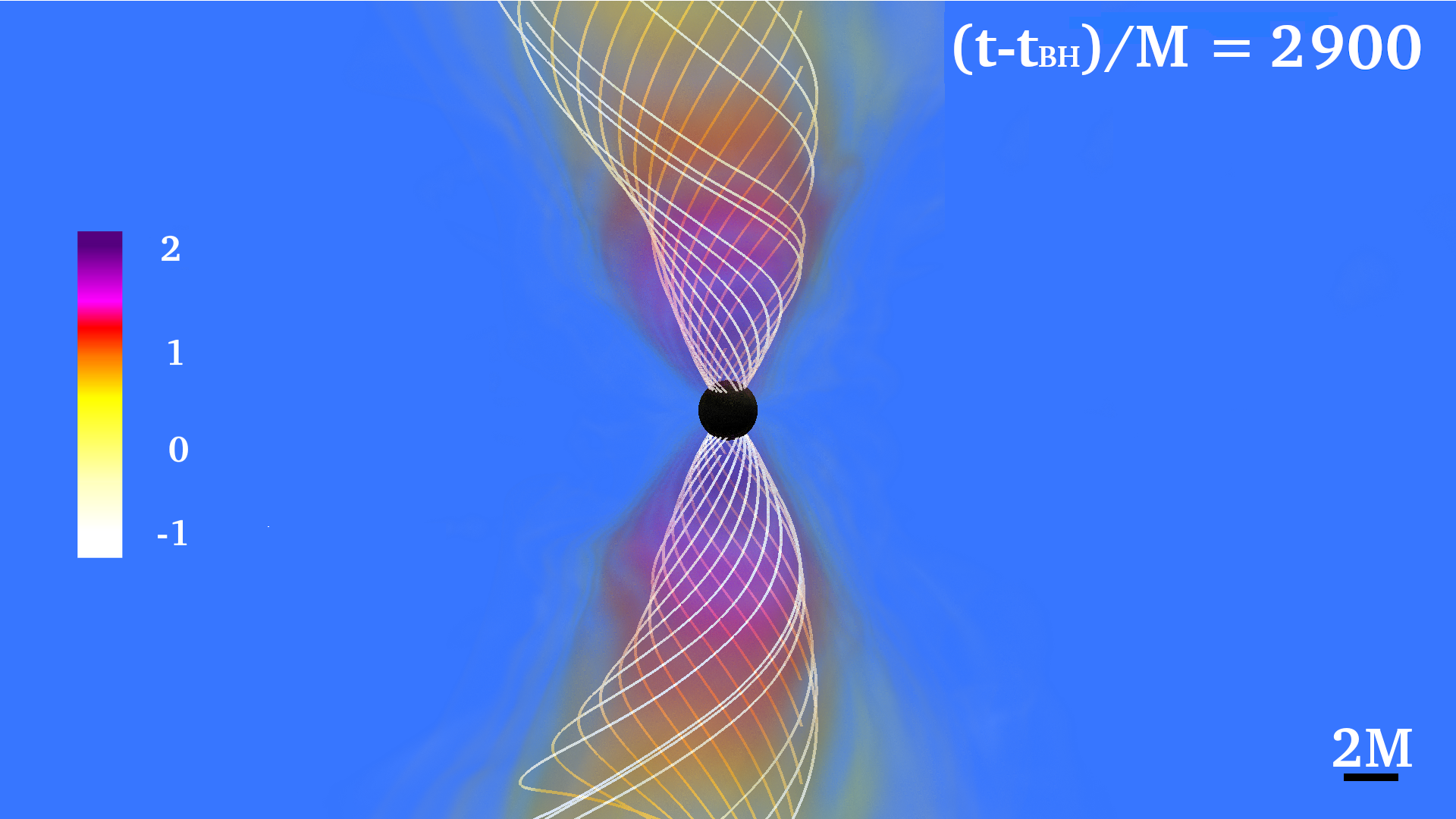}
  \caption{Volume rendering of the force-free parameter~$B^2/(8\,\pi\rho_0)$ (log scale) after the incipient jet is
    well-developed for case SLyM2.7P (left), H4M2.8P (middle), and $\Gamma2$M2.8P (right).
    Magnetic field lines (white lines) are plotted inside regions where~ this ratio is $\gtrsim 10^{-2}$
    (funnel boundary). Magnetically-dominated regions ($B^2/(8\,\pi\rho_0) \gtrsim 1$) extend to heights
    $\gtrsim 20M\sim 20\,r_{\rm BH}$ above the BH (black sphere). Here $M\sim 1.3\times 10^{-2}\,\rm ms\sim 4\,\rm km$.
    \label{fig:b2rho}}
\end{figure*}

The inset of Fig.~\ref{fig:Poynting_plot} shows the fraction of the dynamical ejection of
matter (ejecta) following the GW  peak amplitude for all cases in Table
\ref{table:key_results_NSNS}. These values are roughly consistent with those
reported previously (see~e.g.~\cite{Shibata:2019wef,Radice:2018pdn}).

We note that the EOS and the seed magnetic field (geometry and strength) have a strong impact
on the ejecta. The softer the EOS, the larger the amount of matter ejected following the NSNS merger.
In particular, the ejecta in SLyM2.7P is a factor of~$\sim 3$ larger than that in H42.8P. This
result is anticipated because, as mentioned above, a compact object modeled by a soft EOS
cannot hold high angular momentum material. During merger, orbital angular momentum transfer
by tidal torques induces the ejection of the outer layers of the stars.  A significant fraction
of this material gains enough energy to escape.
The ejecta in SLy2.7I (configuration with a poloidal magnetic field confined in the NS) and
SLy2.7H (unmagnetized configuration) is a factor of $\sim 5$ and $\sim 8$ smaller than in
SLy2.7P, respectively~(see Table~\ref{table:key_results_NSNS} for other cases).   It
has been suggested that the magnetic field lines  of a rotating compact object may accelerate
fluid elements due to a magnetocentrifugal mechanism~\cite{1982MNRAS.199..883B}.
These results collectively suggest that GRMHD studies are required to fully understand kilonova signals
from GW170817-like events.

Ejecta masses $\gtrsim 10^{-3}M_\odot$ are expected to lead to detectable, transient kilonova
signatures~(see e.g.~\cite{Metzger:2016pju}) powered by radioactive decay of unstable elements
formed by the neutron-rich material ejected during NSNS mergers~\cite{Li:1998bw,
  Metzger:2016pju}. An analytical model that computes the peak rise times,
bolometric luminosities and the effective temperatures for kilonovae was derived 
in~\cite{Perego:2021dpw}. This model assumes an ejecta of mass $M_{\rm esc}$ that is
spherically distributed and expanding homologously with an average speed
$\left<v_{\rm esc}\right>$ and characterized by a gray opacity~$\kappa_{\gamma}$. The peak time of
the kilonova emission $\tau_{\rm peak}$ can be then estimated as~\cite{Perego:2021dpw}
\begin{eqnarray}
  \tau_{\rm peak}&\sim&\sqrt{\frac{M_{\rm esc}\,\kappa_{\gamma}}{4\,\pi\,\left<v_{\rm esc}\right>\,c}}\\
  &\approx& 4.6\,{\rm days}\,\left(\frac{M_{\rm eje}}{10^{-2} M_{\odot}}\right)^{1/2}\,
  \left(\frac{\left<v_{\rm eje}\right>}{0.1\,c}\right)^{-1/2}\,,\nonumber
\label{t_peak_knove}
\end{eqnarray}
assuming a gray opacity of $\kappa_{\gamma}=10\,\rm cm^2/g$. This opacity corresponds to
ejecta containing a significant fraction of lanthanides and actinides (i.e.
ejecta with an initial electron fraction $Y_e\lesssim 0.25$)
\cite{2013ApJ...775..113T,Barnes:2013wka}. The  peak luminosity of the ejecta can be
approximated by
\begin{equation}
  L_{\rm knova}\sim 2.4\times 10^{40}\left(\frac{M_{\rm eje}}
  {0.01M_{\odot}}\right)^{0.35}\,\left(\frac{\left<v_{\rm eje}\right>}{0.1\,c}\right)^{0.65}\,
  \rm erg/s\,.
 \label{L_peak_knove}
\end{equation} 
Finally, assuming black body emission, and using the Stefan-Boltzmann law, the effective temperature 
at the peak can be be estimated as~\cite{Perego:2021dpw}
\begin{equation}
  T_{\rm peak}\sim 2.15\times 10^{3}\left(\frac{M_{\rm eje}}
  {0.01M_{\odot}}\right)^{-0.16}\,\left(\frac{\left<v_{\rm eje}\right>}{0.1\,c}\right)^{-0.09}\,
  \rm K\,.
 \label{T_peak_knove}
\end{equation} 
%

Using the above formulae, we estimate that the bolometric luminosity of potential kilonovae signals is $L_{\rm knova}=10^{40.6\pm0.5}\rm erg/s$
with rise times of $\tau_{\rm peak}\sim 0.4-5.1$~days~and an effective temperature $T_{\rm peak}\sim 10^{3.5}\rm\,K$~(see~Table
\ref{table:key_results_NSNS}). This temperature can be translated into a peak wavelength $\lambda_{\rm peak}=1.35\times
    {10^3}\,{\rm nm}\,({T_{\rm peak}}/10^{3.33}\,\rm{K})^{-1}$~\cite{Perego:2021dpw}. We found that
    $\lambda_{\rm peak}\sim 730-1830\,\rm nm$, and the emission can be detected with current or
    planned telescopes, such as ALMA  or the Vera C. Rubin
observatory~\cite{2018PASP..130a5002M,Chen:2020zoq}.
%
%
\subsection{Gravitational Wave Signals}
The left panel of Fig.~\ref{fig:hydro_magGW} shows the GW strain of the  dominant $(2,2)$ mode 
as a function of retarded time for cases in Table~\ref{table:key_results_NSNS}. We note that
that the amplitude differences between the unmagnetized and magnetized cases are 
less than $3\%$ and their GW peaks occur roughly at the same time. The largest
difference between the GW peak (merger) times of the unmagnetized and magnetized SLyM2.7 binaries
is~$\sim 83M\sim1.1\rm\, ms$. These results confirm that the seed magnetic field used here
does not significantly impact the global dynamics during inspiral.

Following merger, the GW strain either comes to an abrupt end following the
quasinormal ringdown modes of the BH, in prompt collapse cases, or  decays
as the (initially nonaxisymmetric) HMNS remnant settles prior to its collapse.  During this period,
the waveform amplitudes in SLyM2.7 and $\Gamma2$M2.8 are roughly the same, and diminish faster
than those in H4M2.8. 

To assess the impact of the magnetic field during the postmerger phase on the GW forms,
we first extend the GW spectra in the low frequency domain by appending a {\tt TaylorT1} post-Newtonian
waveform~\cite{nijidataformat} to that of our GRMHD simulations to cover the earlier inspiral phase.
As in~\cite{Ruiz:2020elr}, the hybrid waveform is obtained by minimizing
\begin{equation}
  \int_{t_{\rm i}}^{t_{\rm f}} dt\,\left[
    (h_+^{\rm NR}      - h_+^{\rm PN})^2 +
    (h_\times^{\rm NR} - h_\times^{\rm PN})^2
    \right]^{1/2}\,,
\label{eq:waveformh}
\end{equation}
using as free parameters the initial PN phase, amplitude, and orbital angular frequency.
In all cases,  the integration range was chosen to be between~$t_{\rm i}\approx 100M$
and~$t_{\rm f}\approx 600M$. The right panel of Fig.~\ref{fig:hydro_magGW} shows the spectra
of the GWs of the dominant mode $(l,m)=(2,2)$ at a source distance of $50\rm Mpc$, 
along with the aLIGO (dashed line) and the Einstein Telescope (dotted line) noise curves of
configurations in~\cite{LIGOScientific:2016wof} for the cases listed in Table~\ref{table:key_results_NSNS}.
%
%

%
\begin{center}
  \begin{table}
    \caption{Main spectral frequencies (in Hz) of the dominant mode $h^{22}_+$ for NSNS undergoing delayed
      collapse~(see Table~\ref{table:key_results_NSNS}). Binaries whose merger outcome last
      $\lesssim 2.5\,\rm ms$ do no exhibit peak frequencies (see right panel in Fig.~\ref{fig:hydro_magGW}).
      \label{table:freq_GW}}
    \begin{tabular}{cccccc}
      \hline\hline
      Model             & $\mathcal{C}$   &  $f_{2-0}$  & $f_{\rm spiral}$ &$f_{\rm peak}$       &   $f_{2+0}$\\
      {$\Gamma2$M2.8H}  & $0.140$         &  1178       &  1680            &   2027              & 2425       \\
      {$\Gamma2$M2.8P}  & $0.140$         &  1192       &  1656            &   2065              & 2478       \\
      \hline 
      {H4M$2.8$H}       & $0.155$         &  1722       &  2384            &   2691              & 3498        \\
      {H4M$2.8$I}       & $0.155$         &  1730       &  2284            &   2751              & 3646        \\
      {H4M$2.8$P}       & $0.155$         &  1752       &  2297            &   2768              & 3760        \\
      \hline 
      {SLyM$2.6$H}      & $0.169$         &   2288      & 2866             &  3341               & 4450       \\
      {SLyM$2.6$P}      & $0.169$         &   2301      & 2899             &  3404               & 4458       \\
      \hline
      {SLyM$2.7$H}      & $0.175$         &   2364      & 2855             &   3744              & 4558       \\          
      {SLyM$2.7$I}      & $0.175$         &   2352      & 2803             &   3714              & 4567       \\
      {SLyM$2.7$P}      & $0.175$         &   2378      & 2807             &   3930              & 4718       \\
          \hline\hline
    \end{tabular}
  \end{table}
\end{center}

Consistent with previous results~(see~e.g.~\cite{Bauswein:2015yca,Vretinaris:2019spn,Takami:2014tva}),
we note that the GW spectrum has at least  four main frequencies (oscillation modes) associated with the
fluid motion during the postmerger phase, which depend strongly on the compactness of the stars, as well
as the seed magnetic field (see~Table~\ref{table:freq_GW}). As shown in Fig.~\ref{fig:hydro_magGW}, these
oscillation modes are only present in middle- and long-lived  HMNSs. In the short-lived HMNS and prompt
collapse cases, most of the matter ($\gtrsim 99\%$ of the total rest-mass of the system; see inset in
Fig.~\ref{fig:M0_outside}) is swallowed by the BH in less than $2.5\,\rm ms$, and hence the oscillations
in the spectra are associated with the quasinormal modes of the nascent BH.

The frequency of the $f_{2-0}$ mode, which is produced by a nonlinear interaction between the quadrupole
and quasiradial modes~\cite{Bauswein:2015yca}, depends strongly on the NS compactness and is
roughly independent of the magnetic field (see~Table~\ref{table:freq_GW}). We also find that this
mode is consistent within $\lesssim 6\%$ with the ``universal'' behavior associated
with the compactness described by the fitting  formula [Eq.~(1)] in~\cite{Bauswein:2015yca}.
By contrast,  we  observe that in most of the cases, the magnetic field damps the amplitude
and the frequency of the $f_{\rm spiral}$ mode, which has been associated to the interaction between
the two stellar cores rotating about each other right after merger.  In particular, $f_{\rm spiral}$
peaks $\sim 87\, \rm Hz$ earlier in H4M2.8P than in H4M2.8H.  This suggests that magnetic viscosity
tends to accelerate the merger of the two stellar cores. This peak frequency shows a large deviation
 ($\lesssim 17\%$) from the fitting formula [Eq.~(2)] in~\cite{Bauswein:2015yca}.
On the other hand, the $f_{\rm peak}$ mode, which is related to the rotation of the nonaxisymmetric HMNS,
rises~$\sim 30-150\,\rm Hz$ in the magnetized cases beyond those in the unmagnetized ones.
This behavior is  consistent with the rotation profiles of the HMNS remnants displayed in Fig.~\ref{fig:rotatio_prof}.
We also observe that this mode is consistent to within $\lesssim 12\%$ with  the fitting formula
[Eq.~(3)] in~\cite{Bauswein:2015yca}. Finally, the $f_{2+0}$ mode in all the unmagnetized case peaks earlier
than that in the magnetized cases. Remarkably, this peak is highly affected by the pulsar-like magnetic
field~(see~Table~\ref{table:freq_GW}).  In particular, $f_{2+0}$ in SLyM2.7H peaks $\sim 160\,\rm Hz$
earlier than that in~SLyM2.7P. These results suggest that the magnetic field introduces a degeneracy
with the EOS: two different EOSs may potentially peak at the same frequency due to magnetic effects.
These results imply that magnetic effects should be taken into account to constrain the EOS.

According the the classification of the postmerger dynamics and GW spectra for transient remnants
discussed in~\cite{Bauswein:2015yca}, all our transients are Type III. In these cases, the dominant
secondary peak in the GW spectrum is $f_{\rm spiral}$, while $f_{2-0}$ is very weak~(see right panel
in~Fig.~\ref{fig:hydro_magGW}) due to low-amplitude radial oscillations of the two
NS cores rotating about each other after merger.

Finally, we probe whether the magnetic field leaves any detectable imprints on the GWs that can be measured by
the current or future GW detectors. For this purpose, we compute the match function $\mathcal{M}_{\text{\tiny{GW}}}$
\cite{Harry2018}
\begin{eqnarray}
\mathcal{M}_{\text{\tiny{GW}}} =  \underset{(\phi_c,t_c)}{{\rm max}}
\frac{\left<{h}_1|{h}_2(\phi_c,t_c)\right>}{\sqrt{\left<{h}_1|{h}_1\right>
    \left<{h}_2|{h}_2\right>}}\,,
\label{eq:match_func}
\end{eqnarray}
between two given waveforms. The maximization is taken over a large set
of phase shifts $\phi_c$ and time shifts $t_c$. Here $\left<{h}_1|{h}_2\right>$ denotes
the noise-weighted inner product
\begin{eqnarray}
  \left<h_1|h_2\right>= 4\,{\rm Re}\int_0^\infty\frac{\tilde{h}_1(f)\,
    \tilde{h}^*_2(f)}{S_h(f)}\,df\,,
\end{eqnarray}
where $h=h_+-i\,h_\times$, $\tilde{h}$ is the Fourier transform of the strain amplitude
$\sqrt{\tilde h_+(f)^2 +\tilde h_{\times}(f)^2}$ of the dominant mode $(l,m)=(2,2)$, and
$S_h(f)$ is the noise curve of a given detector~\cite{LIGOScientific:2016wof}.
As discussed on~\cite{Harry2018}, the value of the match function at which waveforms can
potentially be distinguishable depends on the SNR of the waveforms.
For a SNR of 15, signals can be indistinguishable if~$\mathcal{M}_{\text{\tiny{GW}}}\gtrsim
0.9978$, or for a SNR of 25, if the match is larger than 0.9992. 
For the SLy binaries, those whose three main spectral frequencies peak above the aLIGO
sensitive curve~(see top right panel in Fig.~\ref{fig:hydro_magGW}), we find that
at distance of $50\,\rm Mpc$ the match function is:
i) $\mathcal{M}_{\text{\tiny{GW}}} =0.9811$ between SLyM2.7H and SLyM2.7I;
ii)$\mathcal{M}_{\text{\tiny{GW}}} =0.9876$ between SLyM2.7I and SLyM2.7P;
and
iii)$\mathcal{M}_{\text{\tiny{GW}}}=0.9708$ between SLyM2.7H and SLyM2.7P
with a SNR of $\sim 1.2$ and $\sim 3.2$ for KAGRA and aLIGO, respectively, or with a SNR of
$\sim 30$ at a distance of~$\sim  2\,\rm Mpc$ for KAGRA or of~$\sim 6.0\,\rm Mpc$ for aLIGO,
given their current sensitivity \cite{LIGOScientific:2016wof}. We recall that GW170817, the
closest GW signal detected to date, had a luminosity distance of $40^{+8}_{-14}\,\rm
Mpc$~\cite{TheLIGOScientific:2017qsa}.
Current based-ground GW detectors are unlikely to  observer the imprints of the
magnetic field on the GW. However, using the expected sensitivity curve for the Einstein
Telescope~\cite{TheLIGOScientific:2017qsa}, at distance of~$50\,\rm Mpc$
we find the same $\mathcal{M}_{\text{\tiny{GW}}}$ but with a SNR of $\sim 30$, and so
it  can easily observe these imprints. Similar results are found for the  H4 and $\Gamma=2$
binaries. Next generation of GW detectors are thus required to measure the
magnetic field imprints on the gravitational radiation at the expected NSNS merger
distances.
%
\section{Conclusions}
\label{sec:conclusion}
A plethora of new GW observations from compact binary mergers is changing our understanding
of the Universe at an unprecedented rate. However, these observations have some limitations:
i) source localizations are $\gtrsim 20\,\rm deg^2$ (see e.g.~\cite{LIGOScientific:2020ibl,
  LIGOScientific:2018mvr}), preventing identification of the source environment;
ii) the merger of low-mass compact binaries cannot be detected by current ground-based detectors,
preventing our determination of their final fates; iii) GW signals contain uncertainties in
the individual masses and spins of the binary companions, etc. However, the coincident detection
of GWs with electromagnetic counterparts across the EM spectrum are useful in overcoming some of
these limitations. In particular, the detection of GW170817 along with its EM counterparts
enabled us to address several long-standing issues:
the central engines that power sGRBs, the discovery of off-axis afterglows, the unambiguous
identification of a kilonova (AT2017gfo) and the production of heavy elements
\cite{TheLIGOScientific:2017qsa, Savchenko:2017ffs,Cowperthwaite:2017dyu,Kasliwal:2017ngb,Smartt:2017fuw}.
GW170817 also demonstrated that to interpret new observations and, in particular, to apply them to
understand the physics of matter under extreme conditions, it is crucial to employ results
from theoretical modeling.

To understand the physical processes that trigger the emergence of incipient jets,
the role of magnetic fields in the dynamical ejection of matter, and 
the features of GWs from NSNS mergers, we surveyed magnetized NSNS configurations that undergo
merger followed by either delayed or prompt collapse to a BH. The binary companions are irrotational
stars modeled using a piecewise polytropic implementation of the representative nuclear EOSs SLy and H4, as
in~\cite{Read:2008iy}. The stars were endowed with an initially poloidal and dynamically weak magnetic
field that was either confined to the stellar interior or protruding from the interior into the exterior,
as in typical pulsars.

Consistent with~\cite{Ruiz:2016rai,Ruiz:2017inq}, we found that jets launched by BH + disks
originate only from NSNSs mergers that form HMNS remnants that undergo delayed collapse.
This conclusion is independent of the EOS or the magnetic field geometry. This last result  may differ
from BHNS mergers, where a jet is launched only when the NSs are suitably magnetized with a pulsar-like
interior + exterior magnetic field~\cite{prs15}.

We noticed  that the EOS  have a strong impact on the time delay $\Delta t_{\rm jet}$ between the GW
peak amplitude and the jet launching. We observed that the closer the total mass of the binary is
to the threshold value for prompt collapse, the shorter $\Delta t_{\rm jet}$. We also found that
this time strongly depends on the initial extend of the seed magnetic field. The magnetic energy
$\mathcal{M}$ in  BH + disk remnants whose progenitors are initially endowed with a pulsar-like
interior + exterior magnetic field is a factor of $\sim 20$ larger than in those endowed with a
magnetic field confined to the stellar interior.
As $\mathcal{M}$ is not enhanced after BH formation, the remnant in the latter requires more time
for magnetic pressure gradients to overcome the ram-pressure of the fall-back debris and
launch an incipient jet. 
The lifetime of the jets [$\Delta t\sim 92-150\,\rm ms$] and their outgoing Poynting luminosities
[$L_{\rm EM}\sim 10^{52\pm 1}  \rm erg/s$] are consistent with the sGRB engine lifetime
\cite{Beniamini:2020adb,Bhat:2016odd,Lien:2016zny,Svinkin:2016fho}, as well as with the BZ mechanism for
launching jets and their associated Poynting luminosities~\cite{BZeffect77}.
The luminosities  and BH accretion rates also lie within the rather narrow ``universal'' range of
values predicted in~\cite{Shapiro:2017cny} for BH + disk + jet systems formed from compact binary mergers
containing magnetized NSs. These results suggests that incipient jets are typically the final outcome of
magnetized NSNS undergoing delayed collapse to a BH.
Notice that in~\cite{Ruiz:2020via},  we discussed the evolution of $\Gamma$2M2.8 in full 3D
(no symmetries), and found that the remnant launches a magnetically-driven jet at roughly the same
time as that in the equatorial evolution. In addition, we found that the physical parameters of the
BH + disk system, such as mass of the accretion disk, BH spin, etc, are in both cases roughly the
same. These results imply that equatorial symmetry does not play any role in the final outcome of the
binary merger.  As magnetic instabilities cannot be affected by the EOS, we expect our current results to
hold in Full 3D.

We also observed that the dynamical ejection of matter amounts to~$M_{\rm esc}\sim 10^{-4}-10^{-2}M_{\odot}$
and is strongly affected by the EOS. In particular we found that: i) the softer the EOS, the larger
the amount of matter ejected following the NSNS merger. Specifically, the ejecta in SLyM2.7P is a factor of~$\sim 3$
larger than that in H42.8P ~(see Table~\ref{table:key_results_NSNS} for other cases);
ii) the ejecta can be up to a factor of $\sim 8$ larger in magnetized NSNS mergers  than that in
unmagnetized ones. It has been suggested that the magnetic field lines  of a rotating compact object
may accelerate fluid elements due to a magnetocentrifugal mechanism~\cite{1982MNRAS.199..883B}.
These combined results suggest that GRMHD studies are required to fully explain kilonova signals from
GW170817-like events. We used an analytical model~\cite{Perego:2021dpw} to compute the peak luminosity,
rise time and effective temperature of the
potential kilonova. We found that the bolometric luminosity  is $L_{\rm knova}=10^{40.6\pm0.5}\rm erg/s$
with rise times of $\tau_{\rm peak}\sim 0.4-5.1$~days~and effective temperature of~$\sim 10^{3.5}\rm\,K$.
We note that this temperature can be translated in a peak wavelength $\lambda_{\rm peak}=1.35\times
{10^3}\,{\rm nm}\,({T_{\rm peak}}/10^{3.33}\,\rm{K})^{-1}$~\cite{Perego:2021dpw}, and so
$\lambda_{\rm peak}\sim 730-1830\,\rm nm$. This EM radiation can be detected with current or planned
telescopes~\cite{2018PASP..130a5002M,Chen:2020zoq}.

Finally, we probed whether the gravitational waveforms contain measurable imprints of the 
seed magnetic field. We extended the GW spectra in the low frequency domain by appending a
{\tt TaylorT1} post-Newtonian waveform to that of our numerical simulations.  We found that the magnetic
field damps the amplitude and shifts the frequency of the main oscillations modes of the transient HMNS.
These two effects introduce a degeneracy with the EOS, since two different EOSs may have peaks at
the same frequency due to magnetic effects, and so magnetic fields should be taken into account
to constrain the EOS. In addition, we computed the match function between waveforms from systems
with the same EOS but different magnetic field content and initial geometry.
We  found that at distance of $50\rm Mpc$ only the next generation of based-ground GW detectors, such as the
Einstein Telescope, can observe imprints of the magnetic field on the GWs. 

Movies and additional 3D visualizations highlighting our simulations can be found at~\cite{Moviepage}.

%
%
\acknowledgements
We thank the Illinois Relativity REU team (H.~Jinghan, M.~Kotak, E.~Yu, and J.~Zhou) for
assistance with some of the visualizations. This work has been supported in part by National Science
Foundation (NSF) Grant PHY-1662211 and PHY-2006066, and NASA Grant 80NSSC17K0070 to the University of
Illinois at Urbana-Champaign. This work made use of the Extreme Science and Engineering Discovery
Environment (XSEDE), which is supported by National Science Foundation grant number TG-MCA99S008. This
research is part of the Blue Waters sustained-petascale computing project, which is supported by the
National Science Foundation (awards OCI-0725070 and ACI-1238993) and the State of Illinois. Blue Waters
is a joint effort of the University of Illinois at Urbana-Champaign and its National Center
for Supercomputing Applications. Resources supporting this work were also provided by the NASA High-End
Computing (HEC) Program through the NASA Advanced Supercomputing (NAS) Division at Ames Research Center.

\bibliographystyle{apsrev4-1}        
\bibliography{references}            

\begin{thebibliography}{125}%
\makeatletter
\providecommand \@ifxundefined [1]{%
 \@ifx{#1\undefined}
}%
\providecommand \@ifnum [1]{%
 \ifnum #1\expandafter \@firstoftwo
 \else \expandafter \@secondoftwo
 \fi
}%
\providecommand \@ifx [1]{%
 \ifx #1\expandafter \@firstoftwo
 \else \expandafter \@secondoftwo
 \fi
}%
\providecommand \natexlab [1]{#1}%
\providecommand \enquote  [1]{``#1''}%
\providecommand \bibnamefont  [1]{#1}%
\providecommand \bibfnamefont [1]{#1}%
\providecommand \citenamefont [1]{#1}%
\providecommand \href@noop [0]{\@secondoftwo}%
\providecommand \href [0]{\begingroup \@sanitize@url \@href}%
\providecommand \@href[1]{\@@startlink{#1}\@@href}%
\providecommand \@@href[1]{\endgroup#1\@@endlink}%
\providecommand \@sanitize@url [0]{\catcode `\\12\catcode `\$12\catcode
  `\&12\catcode `\#12\catcode `\^12\catcode `\_12\catcode `\%12\relax}%
\providecommand \@@startlink[1]{}%
\providecommand \@@endlink[0]{}%
\providecommand \url  [0]{\begingroup\@sanitize@url \@url }%
\providecommand \@url [1]{\endgroup\@href {#1}{\urlprefix }}%
\providecommand \urlprefix  [0]{URL }%
\providecommand \Eprint [0]{\href }%
\providecommand \doibase [0]{http://dx.doi.org/}%
\providecommand \selectlanguage [0]{\@gobble}%
\providecommand \bibinfo  [0]{\@secondoftwo}%
\providecommand \bibfield  [0]{\@secondoftwo}%
\providecommand \translation [1]{[#1]}%
\providecommand \BibitemOpen [0]{}%
\providecommand \bibitemStop [0]{}%
\providecommand \bibitemNoStop [0]{.\EOS\space}%
\providecommand \EOS [0]{\spacefactor3000\relax}%
\providecommand \BibitemShut  [1]{\csname bibitem#1\endcsname}%
\let\auto@bib@innerbib\@empty
\bibitem [{\citenamefont {Abbott}\ \emph
  {et~al.}(2017{\natexlab{a}})\citenamefont {Abbott} \emph
  {et~al.}}]{TheLIGOScientific:2017qsa}%
  \BibitemOpen
  \bibfield  {author} {\bibinfo {author} {\bibfnamefont {B.~P.}\ \bibnamefont
  {Abbott}} \emph {et~al.} (\bibinfo {collaboration} {Virgo, LIGO
  Scientific}),\ }\href {\doibase 10.1103/PhysRevLett.119.161101} {\bibfield
  {journal} {\bibinfo  {journal} {Phys. Rev. Lett.}\ }\textbf {\bibinfo
  {volume} {119}},\ \bibinfo {pages} {161101} (\bibinfo {year}
  {2017}{\natexlab{a}})},\ \Eprint {http://arxiv.org/abs/1710.05832}
  {arXiv:1710.05832 [gr-qc]} \BibitemShut {NoStop}%
\bibitem [{\citenamefont {{von Kienlin}}\ \emph {et~al.}(2017)\citenamefont
  {{von Kienlin}}, \citenamefont {{Meegan}},\ and\ \citenamefont
  {{Goldstein}}}]{FERMI2017GCN}%
  \BibitemOpen
  \bibfield  {author} {\bibinfo {author} {\bibfnamefont {A.}~\bibnamefont {{von
  Kienlin}}}, \bibinfo {author} {\bibfnamefont {C.}~\bibnamefont {{Meegan}}}, \
  and\ \bibinfo {author} {\bibfnamefont {A.}~\bibnamefont {{Goldstein}}},\
  }\href@noop {} {\bibfield  {journal} {\bibinfo  {journal} {GRB Coordinates
  Network, Circular Service, No.~21520, \#1 (2017)}\ }\textbf {\bibinfo
  {volume} {1520}} (\bibinfo {year} {2017})}\BibitemShut {NoStop}%
\bibitem [{\citenamefont {Savchenko}\ \emph {et~al.}(2017)\citenamefont
  {Savchenko} \emph {et~al.}}]{Savchenko:2017ffs}%
  \BibitemOpen
  \bibfield  {author} {\bibinfo {author} {\bibfnamefont {V.}~\bibnamefont
  {Savchenko}} \emph {et~al.},\ }\href {\doibase 10.3847/2041-8213/aa8f94}
  {\bibfield  {journal} {\bibinfo  {journal} {Astrophys. J.}\ }\textbf
  {\bibinfo {volume} {848}},\ \bibinfo {pages} {L15} (\bibinfo {year}
  {2017})},\ \Eprint {http://arxiv.org/abs/1710.05449} {arXiv:1710.05449
  [astro-ph.HE]} \BibitemShut {NoStop}%
\bibitem [{\citenamefont {{Narayan}}\ \emph {et~al.}(1992)\citenamefont
  {{Narayan}}, \citenamefont {{Paczynski}},\ and\ \citenamefont
  {{Piran}}}]{NaPaPi}%
  \BibitemOpen
  \bibfield  {author} {\bibinfo {author} {\bibfnamefont {R.}~\bibnamefont
  {{Narayan}}}, \bibinfo {author} {\bibfnamefont {B.}~\bibnamefont
  {{Paczynski}}}, \ and\ \bibinfo {author} {\bibfnamefont {T.}~\bibnamefont
  {{Piran}}},\ }\href {\doibase 10.1086/186493} {\bibfield  {journal} {\bibinfo
   {journal} {Astrophys. J. Letters}\ }\textbf {\bibinfo {volume} {395}},\
  \bibinfo {pages} {L83} (\bibinfo {year} {1992})}\BibitemShut {NoStop}%
\bibitem [{\citenamefont {{Eichler}}\ \emph {et~al.}(1989)\citenamefont
  {{Eichler}}, \citenamefont {{Livio}}, \citenamefont {{Piran}},\ and\
  \citenamefont {{Schramm}}}]{EiLiPiSc}%
  \BibitemOpen
  \bibfield  {author} {\bibinfo {author} {\bibfnamefont {D.}~\bibnamefont
  {{Eichler}}}, \bibinfo {author} {\bibfnamefont {M.}~\bibnamefont {{Livio}}},
  \bibinfo {author} {\bibfnamefont {T.}~\bibnamefont {{Piran}}}, \ and\
  \bibinfo {author} {\bibfnamefont {D.~N.}\ \bibnamefont {{Schramm}}},\ }\href
  {\doibase 10.1038/340126a0} {\bibfield  {journal} {\bibinfo  {journal}
  {\nat}\ }\textbf {\bibinfo {volume} {340}},\ \bibinfo {pages} {126} (\bibinfo
  {year} {1989})}\BibitemShut {NoStop}%
\bibitem [{\citenamefont {{Paczynski}}(1986)}]{Pac86ApJ}%
  \BibitemOpen
  \bibfield  {author} {\bibinfo {author} {\bibfnamefont {B.}~\bibnamefont
  {{Paczynski}}},\ }\href {\doibase 10.1086/184740} {\bibfield  {journal}
  {\bibinfo  {journal} {Astrophys. J.}\ }\textbf {\bibinfo {volume} {308}},\
  \bibinfo {pages} {L43} (\bibinfo {year} {1986})}\BibitemShut {NoStop}%
\bibitem [{\citenamefont {Ruiz}\ \emph {et~al.}(2016)\citenamefont {Ruiz},
  \citenamefont {Lang}, \citenamefont {Paschalidis},\ and\ \citenamefont
  {Shapiro}}]{Ruiz:2016rai}%
  \BibitemOpen
  \bibfield  {author} {\bibinfo {author} {\bibfnamefont {M.}~\bibnamefont
  {Ruiz}}, \bibinfo {author} {\bibfnamefont {R.~N.}\ \bibnamefont {Lang}},
  \bibinfo {author} {\bibfnamefont {V.}~\bibnamefont {Paschalidis}}, \ and\
  \bibinfo {author} {\bibfnamefont {S.~L.}\ \bibnamefont {Shapiro}},\ }\href
  {\doibase 10.3847/2041-8205/824/1/L6} {\bibfield  {journal} {\bibinfo
  {journal} {Astrophys. J.}\ }\textbf {\bibinfo {volume} {824}},\ \bibinfo
  {pages} {L6} (\bibinfo {year} {2016})}\BibitemShut {NoStop}%
\bibitem [{\citenamefont {Ruiz}\ and\ \citenamefont
  {Shapiro}(2017)}]{Ruiz:2017inq}%
  \BibitemOpen
  \bibfield  {author} {\bibinfo {author} {\bibfnamefont {M.}~\bibnamefont
  {Ruiz}}\ and\ \bibinfo {author} {\bibfnamefont {S.~L.}\ \bibnamefont
  {Shapiro}},\ }\href {\doibase 10.1103/PhysRevD.96.084063} {\bibfield
  {journal} {\bibinfo  {journal} {Phys. Rev.}\ }\textbf {\bibinfo {volume}
  {D96}},\ \bibinfo {pages} {084063} (\bibinfo {year} {2017})},\ \Eprint
  {http://arxiv.org/abs/1709.00414} {arXiv:1709.00414 [astro-ph.HE]}
  \BibitemShut {NoStop}%
\bibitem [{\citenamefont {{Paschalidis}}\ \emph {et~al.}(2015)\citenamefont
  {{Paschalidis}}, \citenamefont {{Ruiz}},\ and\ \citenamefont
  {{Shapiro}}}]{prs15}%
  \BibitemOpen
  \bibfield  {author} {\bibinfo {author} {\bibfnamefont {V.}~\bibnamefont
  {{Paschalidis}}}, \bibinfo {author} {\bibfnamefont {M.}~\bibnamefont
  {{Ruiz}}}, \ and\ \bibinfo {author} {\bibfnamefont {S.~L.}\ \bibnamefont
  {{Shapiro}}},\ }\href {\doibase 10.1088/2041-8205/806/1/L14} {\bibfield
  {journal} {\bibinfo  {journal} {Astrophys. J.}\ }\textbf {\bibinfo {volume}
  {806}},\ \bibinfo {pages} {L14} (\bibinfo {year} {2015})},\ \Eprint
  {http://arxiv.org/abs/1410.7392} {arXiv:1410.7392 [astro-ph.HE]} \BibitemShut
  {NoStop}%
\bibitem [{\citenamefont {Ruiz}\ \emph
  {et~al.}(2018{\natexlab{a}})\citenamefont {Ruiz}, \citenamefont {Shapiro},\
  and\ \citenamefont {Tsokaros}}]{Ruiz:2018wah}%
  \BibitemOpen
  \bibfield  {author} {\bibinfo {author} {\bibfnamefont {M.}~\bibnamefont
  {Ruiz}}, \bibinfo {author} {\bibfnamefont {S.~L.}\ \bibnamefont {Shapiro}}, \
  and\ \bibinfo {author} {\bibfnamefont {A.}~\bibnamefont {Tsokaros}},\ }\href
  {\doibase 10.1103/PhysRevD.98.123017} {\bibfield  {journal} {\bibinfo
  {journal} {Phys. Rev. D}\ }\textbf {\bibinfo {volume} {98}},\ \bibinfo
  {pages} {123017} (\bibinfo {year} {2018}{\natexlab{a}})},\ \Eprint
  {http://arxiv.org/abs/1810.08618} {arXiv:1810.08618 [astro-ph.HE]}
  \BibitemShut {NoStop}%
\bibitem [{\citenamefont {Abbott}\ \emph {et~al.}(2018)\citenamefont {Abbott}
  \emph {et~al.}}]{LIGOScientific:2018cki}%
  \BibitemOpen
  \bibfield  {author} {\bibinfo {author} {\bibfnamefont {B.~P.}\ \bibnamefont
  {Abbott}} \emph {et~al.} (\bibinfo {collaboration} {LIGO Scientific,
  Virgo}),\ }\href {\doibase 10.1103/PhysRevLett.121.161101} {\bibfield
  {journal} {\bibinfo  {journal} {Phys. Rev. Lett.}\ }\textbf {\bibinfo
  {volume} {121}},\ \bibinfo {pages} {161101} (\bibinfo {year} {2018})},\
  \Eprint {http://arxiv.org/abs/1805.11581} {arXiv:1805.11581 [gr-qc]}
  \BibitemShut {NoStop}%
\bibitem [{\citenamefont {Radice}\ \emph
  {et~al.}(2018{\natexlab{a}})\citenamefont {Radice}, \citenamefont {Perego},
  \citenamefont {Zappa},\ and\ \citenamefont {Bernuzzi}}]{Radice:2017lry}%
  \BibitemOpen
  \bibfield  {author} {\bibinfo {author} {\bibfnamefont {D.}~\bibnamefont
  {Radice}}, \bibinfo {author} {\bibfnamefont {A.}~\bibnamefont {Perego}},
  \bibinfo {author} {\bibfnamefont {F.}~\bibnamefont {Zappa}}, \ and\ \bibinfo
  {author} {\bibfnamefont {S.}~\bibnamefont {Bernuzzi}},\ }\href {\doibase
  10.3847/2041-8213/aaa402} {\bibfield  {journal} {\bibinfo  {journal}
  {Astrophys. J.}\ }\textbf {\bibinfo {volume} {852}},\ \bibinfo {pages} {L29}
  (\bibinfo {year} {2018}{\natexlab{a}})},\ \Eprint
  {http://arxiv.org/abs/1711.03647} {arXiv:1711.03647 [astro-ph.HE]}
  \BibitemShut {NoStop}%
\bibitem [{\citenamefont {Margalit}\ and\ \citenamefont
  {Metzger}(2017)}]{Margalit:2017dij}%
  \BibitemOpen
  \bibfield  {author} {\bibinfo {author} {\bibfnamefont {B.}~\bibnamefont
  {Margalit}}\ and\ \bibinfo {author} {\bibfnamefont {B.~D.}\ \bibnamefont
  {Metzger}},\ }\href {\doibase 10.3847/2041-8213/aa991c} {\bibfield  {journal}
  {\bibinfo  {journal} {Astrophys. J.}\ }\textbf {\bibinfo {volume} {850}},\
  \bibinfo {pages} {L19} (\bibinfo {year} {2017})},\ \Eprint
  {http://arxiv.org/abs/1710.05938} {arXiv:1710.05938 [astro-ph.HE]}
  \BibitemShut {NoStop}%
\bibitem [{\citenamefont {Ruiz}\ \emph
  {et~al.}(2018{\natexlab{b}})\citenamefont {Ruiz}, \citenamefont {Shapiro},\
  and\ \citenamefont {Tsokaros}}]{Ruiz:2017due}%
  \BibitemOpen
  \bibfield  {author} {\bibinfo {author} {\bibfnamefont {M.}~\bibnamefont
  {Ruiz}}, \bibinfo {author} {\bibfnamefont {S.~L.}\ \bibnamefont {Shapiro}}, \
  and\ \bibinfo {author} {\bibfnamefont {A.}~\bibnamefont {Tsokaros}},\ }\href
  {\doibase 10.1103/PhysRevD.97.021501} {\bibfield  {journal} {\bibinfo
  {journal} {Phys. Rev.}\ }\textbf {\bibinfo {volume} {D97}},\ \bibinfo {pages}
  {021501} (\bibinfo {year} {2018}{\natexlab{b}})},\ \Eprint
  {http://arxiv.org/abs/1711.00473} {arXiv:1711.00473 [astro-ph.HE]}
  \BibitemShut {NoStop}%
\bibitem [{\citenamefont {Rezzolla}\ \emph {et~al.}(2018)\citenamefont
  {Rezzolla}, \citenamefont {Most},\ and\ \citenamefont
  {Weih}}]{Rezzolla:2017aly}%
  \BibitemOpen
  \bibfield  {author} {\bibinfo {author} {\bibfnamefont {L.}~\bibnamefont
  {Rezzolla}}, \bibinfo {author} {\bibfnamefont {E.~R.}\ \bibnamefont {Most}},
  \ and\ \bibinfo {author} {\bibfnamefont {L.~R.}\ \bibnamefont {Weih}},\
  }\href {\doibase 10.3847/2041-8213/aaa401} {\bibfield  {journal} {\bibinfo
  {journal} {Astrophys. J.}\ }\textbf {\bibinfo {volume} {852}},\ \bibinfo
  {pages} {L25} (\bibinfo {year} {2018})},\ \bibinfo {note} {[Astrophys. J.
  Lett.852,L25(2018)]},\ \Eprint {http://arxiv.org/abs/1711.00314}
  {arXiv:1711.00314 [astro-ph.HE]} \BibitemShut {NoStop}%
\bibitem [{\citenamefont {Shibata}\ \emph {et~al.}(2019)\citenamefont
  {Shibata}, \citenamefont {Zhou}, \citenamefont {Kiuchi},\ and\ \citenamefont
  {Fujibayashi}}]{Shibata:2019ctb}%
  \BibitemOpen
  \bibfield  {author} {\bibinfo {author} {\bibfnamefont {M.}~\bibnamefont
  {Shibata}}, \bibinfo {author} {\bibfnamefont {E.}~\bibnamefont {Zhou}},
  \bibinfo {author} {\bibfnamefont {K.}~\bibnamefont {Kiuchi}}, \ and\ \bibinfo
  {author} {\bibfnamefont {S.}~\bibnamefont {Fujibayashi}},\ }\href {\doibase
  10.1103/PhysRevD.100.023015} {\bibfield  {journal} {\bibinfo  {journal}
  {Phys. Rev. D}\ }\textbf {\bibinfo {volume} {100}},\ \bibinfo {pages}
  {023015} (\bibinfo {year} {2019})},\ \Eprint
  {http://arxiv.org/abs/1905.03656} {arXiv:1905.03656 [astro-ph.HE]}
  \BibitemShut {NoStop}%
\bibitem [{\citenamefont {Cowperthwaite}\ \emph {et~al.}(2017)\citenamefont
  {Cowperthwaite} \emph {et~al.}}]{Cowperthwaite:2017dyu}%
  \BibitemOpen
  \bibfield  {author} {\bibinfo {author} {\bibfnamefont {P.~S.}\ \bibnamefont
  {Cowperthwaite}} \emph {et~al.},\ }\href {\doibase 10.3847/2041-8213/aa8fc7}
  {\bibfield  {journal} {\bibinfo  {journal} {Astrophys. J.}\ }\textbf
  {\bibinfo {volume} {848}},\ \bibinfo {pages} {L17} (\bibinfo {year}
  {2017})},\ \Eprint {http://arxiv.org/abs/1710.05840} {arXiv:1710.05840
  [astro-ph.HE]} \BibitemShut {NoStop}%
\bibitem [{\citenamefont {Kasliwal}\ \emph {et~al.}(2017)\citenamefont
  {Kasliwal} \emph {et~al.}}]{Kasliwal:2017ngb}%
  \BibitemOpen
  \bibfield  {author} {\bibinfo {author} {\bibfnamefont {M.~M.}\ \bibnamefont
  {Kasliwal}} \emph {et~al.},\ }\href {\doibase 10.1126/science.aap9455}
  {\bibfield  {journal} {\bibinfo  {journal} {Science}\ }\textbf {\bibinfo
  {volume} {358}},\ \bibinfo {pages} {1559} (\bibinfo {year} {2017})},\ \Eprint
  {http://arxiv.org/abs/1710.05436} {arXiv:1710.05436 [astro-ph.HE]}
  \BibitemShut {NoStop}%
\bibitem [{\citenamefont {Smartt}\ \emph {et~al.}(2017)\citenamefont {Smartt}
  \emph {et~al.}}]{Smartt:2017fuw}%
  \BibitemOpen
  \bibfield  {author} {\bibinfo {author} {\bibfnamefont {S.~J.}\ \bibnamefont
  {Smartt}} \emph {et~al.},\ }\href {\doibase 10.1038/nature24303} {\bibfield
  {journal} {\bibinfo  {journal} {Nature}\ }\textbf {\bibinfo {volume} {551}},\
  \bibinfo {pages} {75} (\bibinfo {year} {2017})},\ \Eprint
  {http://arxiv.org/abs/1710.05841} {arXiv:1710.05841 [astro-ph.HE]}
  \BibitemShut {NoStop}%
\bibitem [{\citenamefont {Abbott}\ \emph
  {et~al.}(2017{\natexlab{b}})\citenamefont {Abbott} \emph
  {et~al.}}]{LIGOScientific:2017adf}%
  \BibitemOpen
  \bibfield  {author} {\bibinfo {author} {\bibfnamefont {B.~P.}\ \bibnamefont
  {Abbott}} \emph {et~al.} (\bibinfo {collaboration} {LIGO Scientific, Virgo,
  1M2H, Dark Energy Camera GW-E, DES, DLT40, Las Cumbres Observatory, VINROUGE,
  MASTER}),\ }\href {\doibase 10.1038/nature24471} {\bibfield  {journal}
  {\bibinfo  {journal} {Nature}\ }\textbf {\bibinfo {volume} {551}},\ \bibinfo
  {pages} {85} (\bibinfo {year} {2017}{\natexlab{b}})},\ \Eprint
  {http://arxiv.org/abs/1710.05835} {arXiv:1710.05835 [astro-ph.CO]}
  \BibitemShut {NoStop}%
\bibitem [{\citenamefont {Dietrich}\ \emph {et~al.}(2020)\citenamefont
  {Dietrich}, \citenamefont {Coughlin}, \citenamefont {Pang}, \citenamefont
  {Bulla}, \citenamefont {Heinzel}, \citenamefont {Issa}, \citenamefont
  {Tews},\ and\ \citenamefont {Antier}}]{Dietrich:2020efo}%
  \BibitemOpen
  \bibfield  {author} {\bibinfo {author} {\bibfnamefont {T.}~\bibnamefont
  {Dietrich}}, \bibinfo {author} {\bibfnamefont {M.~W.}\ \bibnamefont
  {Coughlin}}, \bibinfo {author} {\bibfnamefont {P.~T.~H.}\ \bibnamefont
  {Pang}}, \bibinfo {author} {\bibfnamefont {M.}~\bibnamefont {Bulla}},
  \bibinfo {author} {\bibfnamefont {J.}~\bibnamefont {Heinzel}}, \bibinfo
  {author} {\bibfnamefont {L.}~\bibnamefont {Issa}}, \bibinfo {author}
  {\bibfnamefont {I.}~\bibnamefont {Tews}}, \ and\ \bibinfo {author}
  {\bibfnamefont {S.}~\bibnamefont {Antier}},\ }\href {\doibase
  10.1126/science.abb4317} {\bibfield  {journal} {\bibinfo  {journal}
  {Science}\ }\textbf {\bibinfo {volume} {370}},\ \bibinfo {pages} {1450}
  (\bibinfo {year} {2020})},\ \Eprint {http://arxiv.org/abs/2002.11355}
  {arXiv:2002.11355 [astro-ph.HE]} \BibitemShut {NoStop}%
\bibitem [{\citenamefont {{Paczy{\'n}ski}}(1993)}]{1993NYASA.688..321P}%
  \BibitemOpen
  \bibfield  {author} {\bibinfo {author} {\bibfnamefont {B.}~\bibnamefont
  {{Paczy{\'n}ski}}},\ }in\ \href {\doibase 10.1111/j.1749-6632.1993.tb43907.x}
  {\emph {\bibinfo {booktitle} {Texas/PASCOS '92: Relativistic Astrophysics and
  Particle Cosmology}}},\ Vol.\ \bibinfo {volume} {688},\ \bibinfo {editor}
  {edited by\ \bibinfo {editor} {\bibfnamefont {C.~W.}\ \bibnamefont
  {{Akerlof}}}\ and\ \bibinfo {editor} {\bibfnamefont {M.~A.}\ \bibnamefont
  {{Srednicki}}}}\ (\bibinfo {year} {1993})\ p.\ \bibinfo {pages}
  {321}\BibitemShut {NoStop}%
\bibitem [{\citenamefont {{Mao}}\ and\ \citenamefont
  {{Yi}}(1994)}]{1994ApJ...424L.131M}%
  \BibitemOpen
  \bibfield  {author} {\bibinfo {author} {\bibfnamefont {S.}~\bibnamefont
  {{Mao}}}\ and\ \bibinfo {author} {\bibfnamefont {I.}~\bibnamefont {{Yi}}},\
  }\href {\doibase 10.1086/187292} {\bibfield  {journal} {\bibinfo  {journal}
  {Astrophys. J.}\ }\textbf {\bibinfo {volume} {424}},\ \bibinfo {pages} {L131}
  (\bibinfo {year} {1994})}\BibitemShut {NoStop}%
\bibitem [{199(1998)}]{1998AIPC..428.....M}%
  \BibitemOpen
  \href@noop {} {\emph {\bibinfo {title} {Gamma-Ray Bursts, 4th Hunstville
  Symposium}}},\ \bibinfo {series} {American Institute of Physics Conference
  Series}, Vol.\ \bibinfo {volume} {428}\ (\bibinfo {year} {1998})\BibitemShut
  {NoStop}%
\bibitem [{\citenamefont {Kawamura}\ \emph {et~al.}(2016)\citenamefont
  {Kawamura}, \citenamefont {Giacomazzo}, \citenamefont {Kastaun},
  \citenamefont {Ciolfi}, \citenamefont {Endrizzi}, \citenamefont {Baiotti},\
  and\ \citenamefont {Perna}}]{Kawamura:2016nmk}%
  \BibitemOpen
  \bibfield  {author} {\bibinfo {author} {\bibfnamefont {T.}~\bibnamefont
  {Kawamura}}, \bibinfo {author} {\bibfnamefont {B.}~\bibnamefont
  {Giacomazzo}}, \bibinfo {author} {\bibfnamefont {W.}~\bibnamefont {Kastaun}},
  \bibinfo {author} {\bibfnamefont {R.}~\bibnamefont {Ciolfi}}, \bibinfo
  {author} {\bibfnamefont {A.}~\bibnamefont {Endrizzi}}, \bibinfo {author}
  {\bibfnamefont {L.}~\bibnamefont {Baiotti}}, \ and\ \bibinfo {author}
  {\bibfnamefont {R.}~\bibnamefont {Perna}},\ }\href {\doibase
  10.1103/PhysRevD.94.064012} {\bibfield  {journal} {\bibinfo  {journal} {Phys.
  Rev.}\ }\textbf {\bibinfo {volume} {D94}},\ \bibinfo {pages} {064012}
  (\bibinfo {year} {2016})},\ \Eprint {http://arxiv.org/abs/1607.01791}
  {arXiv:1607.01791 [astro-ph.HE]} \BibitemShut {NoStop}%
\bibitem [{\citenamefont {Ciolfi}\ \emph {et~al.}(2017)\citenamefont {Ciolfi},
  \citenamefont {Kastaun}, \citenamefont {Giacomazzo}, \citenamefont
  {Endrizzi}, \citenamefont {Siegel},\ and\ \citenamefont
  {Perna}}]{Ciolfi:2017uak}%
  \BibitemOpen
  \bibfield  {author} {\bibinfo {author} {\bibfnamefont {R.}~\bibnamefont
  {Ciolfi}}, \bibinfo {author} {\bibfnamefont {W.}~\bibnamefont {Kastaun}},
  \bibinfo {author} {\bibfnamefont {B.}~\bibnamefont {Giacomazzo}}, \bibinfo
  {author} {\bibfnamefont {A.}~\bibnamefont {Endrizzi}}, \bibinfo {author}
  {\bibfnamefont {D.~M.}\ \bibnamefont {Siegel}}, \ and\ \bibinfo {author}
  {\bibfnamefont {R.}~\bibnamefont {Perna}},\ }\href {\doibase
  10.1103/PhysRevD.95.063016} {\bibfield  {journal} {\bibinfo  {journal} {Phys.
  Rev.}\ }\textbf {\bibinfo {volume} {D95}},\ \bibinfo {pages} {063016}
  (\bibinfo {year} {2017})},\ \Eprint {http://arxiv.org/abs/1701.08738}
  {arXiv:1701.08738 [astro-ph.HE]} \BibitemShut {NoStop}%
\bibitem [{\citenamefont {{Rezzolla}}\ \emph {et~al.}(2011)\citenamefont
  {{Rezzolla}}, \citenamefont {{Giacomazzo}}, \citenamefont {{Baiotti}},
  \citenamefont {{Granot}}, \citenamefont {{Kouveliotou}},\ and\ \citenamefont
  {{Aloy}}}]{rgbgka11}%
  \BibitemOpen
  \bibfield  {author} {\bibinfo {author} {\bibfnamefont {L.}~\bibnamefont
  {{Rezzolla}}}, \bibinfo {author} {\bibfnamefont {B.}~\bibnamefont
  {{Giacomazzo}}}, \bibinfo {author} {\bibfnamefont {L.}~\bibnamefont
  {{Baiotti}}}, \bibinfo {author} {\bibfnamefont {J.}~\bibnamefont {{Granot}}},
  \bibinfo {author} {\bibfnamefont {C.}~\bibnamefont {{Kouveliotou}}}, \ and\
  \bibinfo {author} {\bibfnamefont {M.~A.}\ \bibnamefont {{Aloy}}},\ }\href
  {\doibase 10.1088/2041-8205/732/1/L6} {\bibfield  {journal} {\bibinfo
  {journal} {Astrophys. J. Letters}\ }\textbf {\bibinfo {volume} {732}},\
  \bibinfo {eid} {L6} (\bibinfo {year} {2011})}\BibitemShut {NoStop}%
\bibitem [{\citenamefont {Kiuchi}\ \emph {et~al.}(2014)\citenamefont {Kiuchi},
  \citenamefont {Kyutoku}, \citenamefont {Sekiguchi}, \citenamefont {Shibata},\
  and\ \citenamefont {Wada}}]{Kiuchi:2014hja}%
  \BibitemOpen
  \bibfield  {author} {\bibinfo {author} {\bibfnamefont {K.}~\bibnamefont
  {Kiuchi}}, \bibinfo {author} {\bibfnamefont {K.}~\bibnamefont {Kyutoku}},
  \bibinfo {author} {\bibfnamefont {Y.}~\bibnamefont {Sekiguchi}}, \bibinfo
  {author} {\bibfnamefont {M.}~\bibnamefont {Shibata}}, \ and\ \bibinfo
  {author} {\bibfnamefont {T.}~\bibnamefont {Wada}},\ }\href {\doibase
  10.1103/PhysRevD.90.041502} {\bibfield  {journal} {\bibinfo  {journal}
  {Phys.Rev.}\ }\textbf {\bibinfo {volume} {D90}},\ \bibinfo {pages} {041502}
  (\bibinfo {year} {2014})}\BibitemShut {NoStop}%
\bibitem [{\citenamefont {Paschalidis}(2017)}]{Paschalidis:2016agf}%
  \BibitemOpen
  \bibfield  {author} {\bibinfo {author} {\bibfnamefont {V.}~\bibnamefont
  {Paschalidis}},\ }\href {\doibase 10.1088/1361-6382/aa61ce} {\bibfield
  {journal} {\bibinfo  {journal} {Class. Quant. Grav.}\ }\textbf {\bibinfo
  {volume} {34}},\ \bibinfo {pages} {084002} (\bibinfo {year}
  {2017})}\BibitemShut {NoStop}%
\bibitem [{\citenamefont {Foucart}\ \emph {et~al.}(2021)\citenamefont
  {Foucart}, \citenamefont {Moesta}, \citenamefont {Ramirez}, \citenamefont
  {Wright}, \citenamefont {Darbha},\ and\ \citenamefont
  {Kasen}}]{Foucart:2021ikp}%
  \BibitemOpen
  \bibfield  {author} {\bibinfo {author} {\bibfnamefont {F.}~\bibnamefont
  {Foucart}}, \bibinfo {author} {\bibfnamefont {P.}~\bibnamefont {Moesta}},
  \bibinfo {author} {\bibfnamefont {T.}~\bibnamefont {Ramirez}}, \bibinfo
  {author} {\bibfnamefont {A.~J.}\ \bibnamefont {Wright}}, \bibinfo {author}
  {\bibfnamefont {S.}~\bibnamefont {Darbha}}, \ and\ \bibinfo {author}
  {\bibfnamefont {D.}~\bibnamefont {Kasen}},\ }\href@noop {} {\  (\bibinfo
  {year} {2021})},\ \Eprint {http://arxiv.org/abs/2109.00565} {arXiv:2109.00565
  [astro-ph.HE]} \BibitemShut {NoStop}%
\bibitem [{\citenamefont {M\"osta}\ \emph {et~al.}(2020)\citenamefont
  {M\"osta}, \citenamefont {Radice}, \citenamefont {Haas}, \citenamefont
  {Schnetter},\ and\ \citenamefont {Bernuzzi}}]{Mosta:2020hlh}%
  \BibitemOpen
  \bibfield  {author} {\bibinfo {author} {\bibfnamefont {P.}~\bibnamefont
  {M\"osta}}, \bibinfo {author} {\bibfnamefont {D.}~\bibnamefont {Radice}},
  \bibinfo {author} {\bibfnamefont {R.}~\bibnamefont {Haas}}, \bibinfo {author}
  {\bibfnamefont {E.}~\bibnamefont {Schnetter}}, \ and\ \bibinfo {author}
  {\bibfnamefont {S.}~\bibnamefont {Bernuzzi}},\ }\href {\doibase
  10.3847/2041-8213/abb6ef} {\bibfield  {journal} {\bibinfo  {journal}
  {Astrophys. J. Lett.}\ }\textbf {\bibinfo {volume} {901}},\ \bibinfo {pages}
  {L37} (\bibinfo {year} {2020})},\ \Eprint {http://arxiv.org/abs/2003.06043}
  {arXiv:2003.06043 [astro-ph.HE]} \BibitemShut {NoStop}%
\bibitem [{\citenamefont {Popham}\ \emph {et~al.}(1999)\citenamefont {Popham},
  \citenamefont {Woosley},\ and\ \citenamefont {Fryer}}]{Popham:1998ab}%
  \BibitemOpen
  \bibfield  {author} {\bibinfo {author} {\bibfnamefont {R.}~\bibnamefont
  {Popham}}, \bibinfo {author} {\bibfnamefont {S.~E.}\ \bibnamefont {Woosley}},
  \ and\ \bibinfo {author} {\bibfnamefont {C.}~\bibnamefont {Fryer}},\ }\href
  {\doibase 10.1086/307259} {\bibfield  {journal} {\bibinfo  {journal}
  {Astrophys. J.}\ }\textbf {\bibinfo {volume} {518}},\ \bibinfo {pages} {356}
  (\bibinfo {year} {1999})},\ \Eprint {http://arxiv.org/abs/astro-ph/9807028}
  {arXiv:astro-ph/9807028 [astro-ph]} \BibitemShut {NoStop}%
\bibitem [{\citenamefont {Di~Matteo}\ \emph {et~al.}(2002)\citenamefont
  {Di~Matteo}, \citenamefont {Perna},\ and\ \citenamefont
  {Narayan}}]{Matteo:2002ck}%
  \BibitemOpen
  \bibfield  {author} {\bibinfo {author} {\bibfnamefont {T.}~\bibnamefont
  {Di~Matteo}}, \bibinfo {author} {\bibfnamefont {R.}~\bibnamefont {Perna}}, \
  and\ \bibinfo {author} {\bibfnamefont {R.}~\bibnamefont {Narayan}},\ }\href
  {\doibase 10.1086/342832} {\bibfield  {journal} {\bibinfo  {journal}
  {Astrophys. J.}\ }\textbf {\bibinfo {volume} {579}},\ \bibinfo {pages} {706}
  (\bibinfo {year} {2002})},\ \Eprint {http://arxiv.org/abs/astro-ph/0207319}
  {arXiv:astro-ph/0207319 [astro-ph]} \BibitemShut {NoStop}%
\bibitem [{\citenamefont {Chen}\ and\ \citenamefont
  {Beloborodov}(2007)}]{Chen:2006rra}%
  \BibitemOpen
  \bibfield  {author} {\bibinfo {author} {\bibfnamefont {W.-X.}\ \bibnamefont
  {Chen}}\ and\ \bibinfo {author} {\bibfnamefont {A.~M.}\ \bibnamefont
  {Beloborodov}},\ }\href {\doibase 10.1086/508923} {\bibfield  {journal}
  {\bibinfo  {journal} {Astrophys. J.}\ }\textbf {\bibinfo {volume} {657}},\
  \bibinfo {pages} {383} (\bibinfo {year} {2007})},\ \Eprint
  {http://arxiv.org/abs/astro-ph/0607145} {arXiv:astro-ph/0607145 [astro-ph]}
  \BibitemShut {NoStop}%
\bibitem [{\citenamefont {{Lei}}\ \emph {et~al.}(2013)\citenamefont {{Lei}},
  \citenamefont {{Zhang}},\ and\ \citenamefont {{Liang}}}]{Lei2013ApJ}%
  \BibitemOpen
  \bibfield  {author} {\bibinfo {author} {\bibfnamefont {W.-H.}\ \bibnamefont
  {{Lei}}}, \bibinfo {author} {\bibfnamefont {B.}~\bibnamefont {{Zhang}}}, \
  and\ \bibinfo {author} {\bibfnamefont {E.-W.}\ \bibnamefont {{Liang}}},\
  }\href {\doibase 10.1088/0004-637X/765/2/125} {\bibfield  {journal} {\bibinfo
   {journal} {apj}\ }\textbf {\bibinfo {volume} {765}},\ \bibinfo {eid} {125}
  (\bibinfo {year} {2013})},\ \Eprint {http://arxiv.org/abs/1209.4427}
  {arXiv:1209.4427 [astro-ph.HE]} \BibitemShut {NoStop}%
\bibitem [{\citenamefont {Just}\ \emph {et~al.}(2016)\citenamefont {Just},
  \citenamefont {Obergaulinger}, \citenamefont {Janka}, \citenamefont
  {Bauswein},\ and\ \citenamefont {Schwarz}}]{Just:2015dba}%
  \BibitemOpen
  \bibfield  {author} {\bibinfo {author} {\bibfnamefont {O.}~\bibnamefont
  {Just}}, \bibinfo {author} {\bibfnamefont {M.}~\bibnamefont {Obergaulinger}},
  \bibinfo {author} {\bibfnamefont {H.~T.}\ \bibnamefont {Janka}}, \bibinfo
  {author} {\bibfnamefont {A.}~\bibnamefont {Bauswein}}, \ and\ \bibinfo
  {author} {\bibfnamefont {N.}~\bibnamefont {Schwarz}},\ }\href {\doibase
  10.3847/2041-8205/816/2/L30} {\bibfield  {journal} {\bibinfo  {journal}
  {Astrophys. J.}\ }\textbf {\bibinfo {volume} {816}},\ \bibinfo {pages} {L30}
  (\bibinfo {year} {2016})}\BibitemShut {NoStop}%
\bibitem [{\citenamefont {Zou}\ and\ \citenamefont {Piran}(2010)}]{Zou2009}%
  \BibitemOpen
  \bibfield  {author} {\bibinfo {author} {\bibfnamefont {Y.-C.}\ \bibnamefont
  {Zou}}\ and\ \bibinfo {author} {\bibfnamefont {T.}~\bibnamefont {Piran}},\
  }\href {\doibase 10.1111/j.1365-2966.2009.15863.x} {\bibfield  {journal}
  {\bibinfo  {journal} {Monthly Notices of the Royal Astronomical Society}\
  }\textbf {\bibinfo {volume} {402}},\ \bibinfo {pages} {1854} (\bibinfo {year}
  {2010})},\ \Eprint
  {http://arxiv.org/abs/https://academic.oup.com/mnras/article-pdf/402/3/1854/3126764/mnras0402-1854.pdf}
  {https://academic.oup.com/mnras/article-pdf/402/3/1854/3126764/mnras0402-1854.pdf}
  \BibitemShut {NoStop}%
\bibitem [{\citenamefont {Lei}\ \emph {et~al.}(2017)\citenamefont {Lei},
  \citenamefont {Zhang}, \citenamefont {Wu},\ and\ \citenamefont
  {Liang}}]{Lei:2017zro}%
  \BibitemOpen
  \bibfield  {author} {\bibinfo {author} {\bibfnamefont {W.-H.}\ \bibnamefont
  {Lei}}, \bibinfo {author} {\bibfnamefont {B.}~\bibnamefont {Zhang}}, \bibinfo
  {author} {\bibfnamefont {X.-F.}\ \bibnamefont {Wu}}, \ and\ \bibinfo {author}
  {\bibfnamefont {E.-W.}\ \bibnamefont {Liang}},\ }\href {\doibase
  10.3847/1538-4357/aa9074} {\bibfield  {journal} {\bibinfo  {journal}
  {Astrophys. J.}\ }\textbf {\bibinfo {volume} {849}},\ \bibinfo {pages} {47}
  (\bibinfo {year} {2017})},\ \Eprint {http://arxiv.org/abs/1708.05043}
  {arXiv:1708.05043 [astro-ph.HE]} \BibitemShut {NoStop}%
\bibitem [{\citenamefont {{Blandford}}\ and\ \citenamefont
  {{Znajek}}(1977)}]{BZeffect77}%
  \BibitemOpen
  \bibfield  {author} {\bibinfo {author} {\bibfnamefont {R.~D.}\ \bibnamefont
  {{Blandford}}}\ and\ \bibinfo {author} {\bibfnamefont {R.~L.}\ \bibnamefont
  {{Znajek}}},\ }\href {\doibase 10.1093/mnras/179.3.433} {\bibfield  {journal}
  {\bibinfo  {journal} {Mon. Not. Roy. Astron. Soc.}\ }\textbf {\bibinfo
  {volume} {179}},\ \bibinfo {pages} {433} (\bibinfo {year}
  {1977})}\BibitemShut {NoStop}%
\bibitem [{\citenamefont {Dirirsa}(2017)}]{Dirirsa:2017pgm}%
  \BibitemOpen
  \bibfield  {author} {\bibinfo {author} {\bibfnamefont {F.~F.}\ \bibnamefont
  {Dirirsa}} (\bibinfo {collaboration} {Fermi-LAT}),\ }\bibfield  {booktitle}
  {\emph {\bibinfo {booktitle} {{Proceedings, 4th Annual Conference on High
  Energy Astrophysics in Southern Africa (HEASA 2016): Cape Town, South Africa,
  August 25-26, 2016}}},\ }\href {\doibase 10.22323/1.275.0004} {\bibfield
  {journal} {\bibinfo  {journal} {PoS}\ }\textbf {\bibinfo {volume}
  {HEASA2016}},\ \bibinfo {pages} {004} (\bibinfo {year} {2017})}\BibitemShut
  {NoStop}%
\bibitem [{\citenamefont {{Douchin}}\ and\ \citenamefont
  {{Haensel}}(2001)}]{Douchin01}%
  \BibitemOpen
  \bibfield  {author} {\bibinfo {author} {\bibfnamefont {F.}~\bibnamefont
  {{Douchin}}}\ and\ \bibinfo {author} {\bibfnamefont {P.}~\bibnamefont
  {{Haensel}}},\ }\href {\doibase 10.1051/0004-6361:20011402} {\bibfield
  {journal} {\bibinfo  {journal} {Astron. Astrophys.}\ }\textbf {\bibinfo
  {volume} {380}},\ \bibinfo {pages} {151} (\bibinfo {year} {2001})},\ \Eprint
  {http://arxiv.org/abs/arXiv:astro-ph/0111092} {arXiv:astro-ph/0111092}
  \BibitemShut {NoStop}%
\bibitem [{\citenamefont {Glendenning}\ and\ \citenamefont
  {Moszkowski}(1991)}]{PhysRevLett.67.2414}%
  \BibitemOpen
  \bibfield  {author} {\bibinfo {author} {\bibfnamefont {N.~K.}\ \bibnamefont
  {Glendenning}}\ and\ \bibinfo {author} {\bibfnamefont {S.~A.}\ \bibnamefont
  {Moszkowski}},\ }\href {\doibase 10.1103/PhysRevLett.67.2414} {\bibfield
  {journal} {\bibinfo  {journal} {Phys. Rev. Lett.}\ }\textbf {\bibinfo
  {volume} {67}},\ \bibinfo {pages} {2414} (\bibinfo {year}
  {1991})}\BibitemShut {NoStop}%
\bibitem [{\citenamefont {Read}\ \emph {et~al.}(2009)\citenamefont {Read},
  \citenamefont {Lackey}, \citenamefont {Owen},\ and\ \citenamefont
  {Friedman}}]{Read:2008iy}%
  \BibitemOpen
  \bibfield  {author} {\bibinfo {author} {\bibfnamefont {J.~S.}\ \bibnamefont
  {Read}}, \bibinfo {author} {\bibfnamefont {B.~D.}\ \bibnamefont {Lackey}},
  \bibinfo {author} {\bibfnamefont {B.~J.}\ \bibnamefont {Owen}}, \ and\
  \bibinfo {author} {\bibfnamefont {J.~L.}\ \bibnamefont {Friedman}},\ }\href
  {\doibase 10.1103/PhysRevD.79.124032} {\bibfield  {journal} {\bibinfo
  {journal} {Phys. Rev.}\ }\textbf {\bibinfo {volume} {D79}},\ \bibinfo {pages}
  {124032} (\bibinfo {year} {2009})}\BibitemShut {NoStop}%
\bibitem [{\citenamefont {Fonseca}\ \emph {et~al.}(2016)\citenamefont {Fonseca}
  \emph {et~al.}}]{Fonseca:2016tux}%
  \BibitemOpen
  \bibfield  {author} {\bibinfo {author} {\bibfnamefont {E.}~\bibnamefont
  {Fonseca}} \emph {et~al.},\ }\href {\doibase 10.3847/0004-637X/832/2/167}
  {\bibfield  {journal} {\bibinfo  {journal} {Astrophys. J.}\ }\textbf
  {\bibinfo {volume} {832}},\ \bibinfo {pages} {167} (\bibinfo {year}
  {2016})},\ \Eprint {http://arxiv.org/abs/1603.00545} {arXiv:1603.00545
  [astro-ph.HE]} \BibitemShut {NoStop}%
\bibitem [{\citenamefont {Antoniadis}\ \emph {et~al.}(2013)\citenamefont
  {Antoniadis} \emph {et~al.}}]{Antoniadis:2013pzd}%
  \BibitemOpen
  \bibfield  {author} {\bibinfo {author} {\bibfnamefont {J.}~\bibnamefont
  {Antoniadis}} \emph {et~al.},\ }\href {\doibase 10.1126/science.1233232}
  {\bibfield  {journal} {\bibinfo  {journal} {Science}\ }\textbf {\bibinfo
  {volume} {340}},\ \bibinfo {pages} {6131} (\bibinfo {year} {2013})},\ \Eprint
  {http://arxiv.org/abs/1304.6875} {arXiv:1304.6875 [astro-ph.HE]} \BibitemShut
  {NoStop}%
\bibitem [{\citenamefont {Cromartie}\ \emph {et~al.}(2019)\citenamefont
  {Cromartie} \emph {et~al.}}]{NANOGrav:2019jur}%
  \BibitemOpen
  \bibfield  {author} {\bibinfo {author} {\bibfnamefont {H.~T.}\ \bibnamefont
  {Cromartie}} \emph {et~al.} (\bibinfo {collaboration} {NANOGrav}),\ }\href
  {\doibase 10.1038/s41550-019-0880-2} {\bibfield  {journal} {\bibinfo
  {journal} {Nature Astron.}\ }\textbf {\bibinfo {volume} {4}},\ \bibinfo
  {pages} {72} (\bibinfo {year} {2019})},\ \Eprint
  {http://arxiv.org/abs/1904.06759} {arXiv:1904.06759 [astro-ph.HE]}
  \BibitemShut {NoStop}%
\bibitem [{\citenamefont {Ruiz}\ \emph {et~al.}(2019)\citenamefont {Ruiz},
  \citenamefont {Tsokaros}, \citenamefont {Paschalidis},\ and\ \citenamefont
  {Shapiro}}]{Ruiz:2019ezy}%
  \BibitemOpen
  \bibfield  {author} {\bibinfo {author} {\bibfnamefont {M.}~\bibnamefont
  {Ruiz}}, \bibinfo {author} {\bibfnamefont {A.}~\bibnamefont {Tsokaros}},
  \bibinfo {author} {\bibfnamefont {V.}~\bibnamefont {Paschalidis}}, \ and\
  \bibinfo {author} {\bibfnamefont {S.~L.}\ \bibnamefont {Shapiro}},\ }\href
  {\doibase 10.1103/PhysRevD.99.084032} {\bibfield  {journal} {\bibinfo
  {journal} {Phys. Rev. D}\ }\textbf {\bibinfo {volume} {99}},\ \bibinfo
  {pages} {084032} (\bibinfo {year} {2019})},\ \Eprint
  {http://arxiv.org/abs/1902.08636} {arXiv:1902.08636 [astro-ph.HE]}
  \BibitemShut {NoStop}%
\bibitem [{\citenamefont {Ciolfi}\ \emph {et~al.}(2019)\citenamefont {Ciolfi},
  \citenamefont {Kastaun}, \citenamefont {Kalinani},\ and\ \citenamefont
  {Giacomazzo}}]{Ciolfi:2019fie}%
  \BibitemOpen
  \bibfield  {author} {\bibinfo {author} {\bibfnamefont {R.}~\bibnamefont
  {Ciolfi}}, \bibinfo {author} {\bibfnamefont {W.}~\bibnamefont {Kastaun}},
  \bibinfo {author} {\bibfnamefont {J.~V.}\ \bibnamefont {Kalinani}}, \ and\
  \bibinfo {author} {\bibfnamefont {B.}~\bibnamefont {Giacomazzo}},\ }\href
  {\doibase 10.1103/PhysRevD.100.023005} {\bibfield  {journal} {\bibinfo
  {journal} {Phys. Rev. D}\ }\textbf {\bibinfo {volume} {100}},\ \bibinfo
  {pages} {023005} (\bibinfo {year} {2019})},\ \Eprint
  {http://arxiv.org/abs/1904.10222} {arXiv:1904.10222 [astro-ph.HE]}
  \BibitemShut {NoStop}%
\bibitem [{\citenamefont {Beniamini}\ \emph {et~al.}(2020)\citenamefont
  {Beniamini}, \citenamefont {Duran}, \citenamefont {Petropoulou},\ and\
  \citenamefont {Giannios}}]{Beniamini:2020adb}%
  \BibitemOpen
  \bibfield  {author} {\bibinfo {author} {\bibfnamefont {P.}~\bibnamefont
  {Beniamini}}, \bibinfo {author} {\bibfnamefont {R.~B.}\ \bibnamefont
  {Duran}}, \bibinfo {author} {\bibfnamefont {M.}~\bibnamefont {Petropoulou}},
  \ and\ \bibinfo {author} {\bibfnamefont {D.}~\bibnamefont {Giannios}},\
  }\href {\doibase 10.3847/2041-8213/ab9223} {\bibfield  {journal} {\bibinfo
  {journal} {Astrophys. J. Lett.}\ }\textbf {\bibinfo {volume} {895}},\
  \bibinfo {pages} {L33} (\bibinfo {year} {2020})},\ \Eprint
  {http://arxiv.org/abs/2001.00950} {arXiv:2001.00950 [astro-ph.HE]}
  \BibitemShut {NoStop}%
\bibitem [{\citenamefont {Bhat}\ \emph {et~al.}(2016)\citenamefont {Bhat} \emph
  {et~al.}}]{Bhat:2016odd}%
  \BibitemOpen
  \bibfield  {author} {\bibinfo {author} {\bibfnamefont {P.~N.}\ \bibnamefont
  {Bhat}} \emph {et~al.},\ }\href {\doibase 10.3847/0067-0049/223/2/28}
  {\bibfield  {journal} {\bibinfo  {journal} {Astrophys. J. Suppl.}\ }\textbf
  {\bibinfo {volume} {223}},\ \bibinfo {pages} {28} (\bibinfo {year} {2016})},\
  \Eprint {http://arxiv.org/abs/1603.07612} {arXiv:1603.07612 [astro-ph.HE]}
  \BibitemShut {NoStop}%
\bibitem [{\citenamefont {Lien}\ \emph {et~al.}(2016)\citenamefont {Lien} \emph
  {et~al.}}]{Lien:2016zny}%
  \BibitemOpen
  \bibfield  {author} {\bibinfo {author} {\bibfnamefont {A.}~\bibnamefont
  {Lien}} \emph {et~al.},\ }\href {\doibase 10.3847/0004-637X/829/1/7}
  {\bibfield  {journal} {\bibinfo  {journal} {Astrophys. J.}\ }\textbf
  {\bibinfo {volume} {829}},\ \bibinfo {pages} {7} (\bibinfo {year} {2016})},\
  \Eprint {http://arxiv.org/abs/1606.01956} {arXiv:1606.01956 [astro-ph.HE]}
  \BibitemShut {NoStop}%
\bibitem [{\citenamefont {Svinkin}\ \emph {et~al.}(2016)\citenamefont
  {Svinkin}, \citenamefont {Frederiks}, \citenamefont {Aptekar}, \citenamefont
  {Golenetskii}, \citenamefont {Pal'shin}, \citenamefont {Oleynik},
  \citenamefont {Tsvetkova}, \citenamefont {Ulanov}, \citenamefont {Cline},\
  and\ \citenamefont {Hurley}}]{Svinkin:2016fho}%
  \BibitemOpen
  \bibfield  {author} {\bibinfo {author} {\bibfnamefont {D.~S.}\ \bibnamefont
  {Svinkin}}, \bibinfo {author} {\bibfnamefont {D.~D.}\ \bibnamefont
  {Frederiks}}, \bibinfo {author} {\bibfnamefont {R.~L.}\ \bibnamefont
  {Aptekar}}, \bibinfo {author} {\bibfnamefont {S.~V.}\ \bibnamefont
  {Golenetskii}}, \bibinfo {author} {\bibfnamefont {V.~D.}\ \bibnamefont
  {Pal'shin}}, \bibinfo {author} {\bibfnamefont {P.~P.}\ \bibnamefont
  {Oleynik}}, \bibinfo {author} {\bibfnamefont {A.~E.}\ \bibnamefont
  {Tsvetkova}}, \bibinfo {author} {\bibfnamefont {M.~V.}\ \bibnamefont
  {Ulanov}}, \bibinfo {author} {\bibfnamefont {T.~L.}\ \bibnamefont {Cline}}, \
  and\ \bibinfo {author} {\bibfnamefont {K.}~\bibnamefont {Hurley}},\ }\href
  {\doibase 10.3847/0067-0049/224/1/10} {\bibfield  {journal} {\bibinfo
  {journal} {Astrophys. J. Suppl.}\ }\textbf {\bibinfo {volume} {224}},\
  \bibinfo {pages} {10} (\bibinfo {year} {2016})},\ \Eprint
  {http://arxiv.org/abs/1603.06832} {arXiv:1603.06832 [astro-ph.HE]}
  \BibitemShut {NoStop}%
\bibitem [{\citenamefont {{Blandford}}\ and\ \citenamefont
  {{Payne}}(1982)}]{1982MNRAS.199..883B}%
  \BibitemOpen
  \bibfield  {author} {\bibinfo {author} {\bibfnamefont {R.~D.}\ \bibnamefont
  {{Blandford}}}\ and\ \bibinfo {author} {\bibfnamefont {D.~G.}\ \bibnamefont
  {{Payne}}},\ }\href {\doibase 10.1093/mnras/199.4.883} {\bibfield  {journal}
  {\bibinfo  {journal} {Monthly Notices of the Royal Astronomical Society}\
  }\textbf {\bibinfo {volume} {199}},\ \bibinfo {pages} {883} (\bibinfo {year}
  {1982})}\BibitemShut {NoStop}%
\bibitem [{\citenamefont {{Perego}}\ \emph {et~al.}(2021)\citenamefont
  {{Perego}}, \citenamefont {{Thielemann}},\ and\ \citenamefont
  {{Cescutti}}}]{Perego:2021dpw}%
  \BibitemOpen
  \bibfield  {author} {\bibinfo {author} {\bibfnamefont {A.}~\bibnamefont
  {{Perego}}}, \bibinfo {author} {\bibfnamefont {F.~K.}\ \bibnamefont
  {{Thielemann}}}, \ and\ \bibinfo {author} {\bibfnamefont {G.}~\bibnamefont
  {{Cescutti}}},\ }\enquote {\bibinfo {title} {{r-Process Nucleosynthesis from
  Compact Binary Mergers}},}\ in\ \href {\doibase
  10.1007/978-981-15-4702-7\_13-1} {\emph {\bibinfo {booktitle} {Handbook of
  Gravitational Wave Astronomy. Edited by C. Bambi}}}\ (\bibinfo {year}
  {2021})\ p.~\bibinfo {pages} {1}\BibitemShut {NoStop}%
\bibitem [{\citenamefont {{Matthews}}\ \emph {et~al.}(2018)\citenamefont
  {{Matthews}} \emph {et~al.}}]{2018PASP..130a5002M}%
  \BibitemOpen
  \bibfield  {author} {\bibinfo {author} {\bibfnamefont {L.~D.}\ \bibnamefont
  {{Matthews}}} \emph {et~al.},\ }\href {\doibase 10.1088/1538-3873/aa9c3d}
  {\bibfield  {journal} {\bibinfo  {journal} {PASP}\ }\textbf {\bibinfo
  {volume} {130}},\ \bibinfo {pages} {015002} (\bibinfo {year} {2018})},\
  \Eprint {http://arxiv.org/abs/1711.06770} {arXiv:1711.06770 [astro-ph.IM]}
  \BibitemShut {NoStop}%
\bibitem [{\citenamefont {Chen}\ \emph {et~al.}(2021)\citenamefont {Chen},
  \citenamefont {Cowperthwaite}, \citenamefont {Metzger},\ and\ \citenamefont
  {Berger}}]{Chen:2020zoq}%
  \BibitemOpen
  \bibfield  {author} {\bibinfo {author} {\bibfnamefont {H.-Y.}\ \bibnamefont
  {Chen}}, \bibinfo {author} {\bibfnamefont {P.~S.}\ \bibnamefont
  {Cowperthwaite}}, \bibinfo {author} {\bibfnamefont {B.~D.}\ \bibnamefont
  {Metzger}}, \ and\ \bibinfo {author} {\bibfnamefont {E.}~\bibnamefont
  {Berger}},\ }\href {\doibase 10.3847/2041-8213/abdab0} {\bibfield  {journal}
  {\bibinfo  {journal} {Astrophys. J. Lett.}\ }\textbf {\bibinfo {volume}
  {908}},\ \bibinfo {pages} {L4} (\bibinfo {year} {2021})},\ \Eprint
  {http://arxiv.org/abs/2011.01211} {arXiv:2011.01211 [astro-ph.CO]}
  \BibitemShut {NoStop}%
\bibitem [{\citenamefont {Harry}\ and\ \citenamefont
  {Hinderer}(2018)}]{Harry2018}%
  \BibitemOpen
  \bibfield  {author} {\bibinfo {author} {\bibfnamefont {I.}~\bibnamefont
  {Harry}}\ and\ \bibinfo {author} {\bibfnamefont {T.}~\bibnamefont
  {Hinderer}},\ }\href {\doibase 10.1088/1361-6382/aac7e3} {\bibfield
  {journal} {\bibinfo  {journal} {Classical and Quantum Gravity}\ }\textbf
  {\bibinfo {volume} {35}},\ \bibinfo {pages} {145010} (\bibinfo {year}
  {2018})}\BibitemShut {NoStop}%
\bibitem [{\citenamefont {Abbott}\ \emph
  {et~al.}(2017{\natexlab{c}})\citenamefont {Abbott} \emph
  {et~al.}}]{LIGOScientific:2016wof}%
  \BibitemOpen
  \bibfield  {author} {\bibinfo {author} {\bibfnamefont {B.~P.}\ \bibnamefont
  {Abbott}} \emph {et~al.} (\bibinfo {collaboration} {LIGO Scientific}),\
  }\href {\doibase 10.1088/1361-6382/aa51f4} {\bibfield  {journal} {\bibinfo
  {journal} {Class. Quant. Grav.}\ }\textbf {\bibinfo {volume} {34}},\ \bibinfo
  {pages} {044001} (\bibinfo {year} {2017}{\natexlab{c}})},\ \Eprint
  {http://arxiv.org/abs/1607.08697} {arXiv:1607.08697 [astro-ph.IM]}
  \BibitemShut {NoStop}%
\bibitem [{\citenamefont {Ruiz}\ \emph
  {et~al.}(2020{\natexlab{a}})\citenamefont {Ruiz}, \citenamefont {Tsokaros},\
  and\ \citenamefont {Shapiro}}]{Ruiz:2020via}%
  \BibitemOpen
  \bibfield  {author} {\bibinfo {author} {\bibfnamefont {M.}~\bibnamefont
  {Ruiz}}, \bibinfo {author} {\bibfnamefont {A.}~\bibnamefont {Tsokaros}}, \
  and\ \bibinfo {author} {\bibfnamefont {S.~L.}\ \bibnamefont {Shapiro}},\
  }\href {\doibase 10.1103/PhysRevD.101.064042} {\bibfield  {journal} {\bibinfo
   {journal} {Phys. Rev. D}\ }\textbf {\bibinfo {volume} {101}},\ \bibinfo
  {pages} {064042} (\bibinfo {year} {2020}{\natexlab{a}})},\ \Eprint
  {http://arxiv.org/abs/2001.09153} {arXiv:2001.09153 [astro-ph.HE]}
  \BibitemShut {NoStop}%
\bibitem [{\citenamefont {Tsokaros}\ \emph
  {et~al.}(2020{\natexlab{a}})\citenamefont {Tsokaros}, \citenamefont {Ruiz},
  \citenamefont {Shapiro}, \citenamefont {Sun},\ and\ \citenamefont
  {Ury\={u}}}]{Tsokaros:2019lnx}%
  \BibitemOpen
  \bibfield  {author} {\bibinfo {author} {\bibfnamefont {A.}~\bibnamefont
  {Tsokaros}}, \bibinfo {author} {\bibfnamefont {M.}~\bibnamefont {Ruiz}},
  \bibinfo {author} {\bibfnamefont {S.~L.}\ \bibnamefont {Shapiro}}, \bibinfo
  {author} {\bibfnamefont {L.}~\bibnamefont {Sun}}, \ and\ \bibinfo {author}
  {\bibfnamefont {K.}~\bibnamefont {Ury\={u}}},\ }\href {\doibase
  10.1103/PhysRevLett.124.071101} {\bibfield  {journal} {\bibinfo  {journal}
  {Phys. Rev. Lett.}\ }\textbf {\bibinfo {volume} {124}},\ \bibinfo {pages}
  {071101} (\bibinfo {year} {2020}{\natexlab{a}})},\ \Eprint
  {http://arxiv.org/abs/1911.06865} {arXiv:1911.06865 [astro-ph.HE]}
  \BibitemShut {NoStop}%
\bibitem [{\citenamefont {Etienne}\ \emph {et~al.}(2010)\citenamefont
  {Etienne}, \citenamefont {Liu},\ and\ \citenamefont
  {Shapiro}}]{Etienne:2010ui}%
  \BibitemOpen
  \bibfield  {author} {\bibinfo {author} {\bibfnamefont {Z.~B.}\ \bibnamefont
  {Etienne}}, \bibinfo {author} {\bibfnamefont {Y.~T.}\ \bibnamefont {Liu}}, \
  and\ \bibinfo {author} {\bibfnamefont {S.~L.}\ \bibnamefont {Shapiro}},\
  }\href {\doibase 10.1103/PhysRevD.82.084031} {\bibfield  {journal} {\bibinfo
  {journal} {Phys.Rev.}\ }\textbf {\bibinfo {volume} {D82}},\ \bibinfo {pages}
  {084031} (\bibinfo {year} {2010})}\BibitemShut {NoStop}%
\bibitem [{CactusConfigs()}]{CactusConfigs}%
  \BibitemOpen
  CactusConfigs,\ \href@noop {} {}\bibinfo {note} {Cactus Machine
  Configurations: {\tt
  http://www.cactuscode.org/Documentation/Configurations.html}}\BibitemShut
  {NoStop}%
\bibitem [{\citenamefont {Schnetter}\ \emph {et~al.}(2004)\citenamefont
  {Schnetter}, \citenamefont {Hawley},\ and\ \citenamefont {Hawke}}]{Carpet}%
  \BibitemOpen
  \bibfield  {author} {\bibinfo {author} {\bibfnamefont {E.}~\bibnamefont
  {Schnetter}}, \bibinfo {author} {\bibfnamefont {S.~H.}\ \bibnamefont
  {Hawley}}, \ and\ \bibinfo {author} {\bibfnamefont {I.}~\bibnamefont
  {Hawke}},\ }\href {\doibase 10.1088/0264-9381/21/6/014} {\bibfield  {journal}
  {\bibinfo  {journal} {Class. Quantum Grav.}\ }\textbf {\bibinfo {volume}
  {21}},\ \bibinfo {pages} {1465} (\bibinfo {year} {2004})},\ \Eprint
  {http://arxiv.org/abs/arXiv:gr-qc/0310042} {arXiv:gr-qc/0310042} \BibitemShut
  {NoStop}%
\bibitem [{Carpet()}]{carpetweb}%
  \BibitemOpen
  Carpet,\ \href {{http://www.carpetcode.org/}} {}\bibinfo {note} {{Carpet Code
  homepage}}\BibitemShut {NoStop}%
\bibitem [{\citenamefont {Shibata}\ and\ \citenamefont
  {Nakamura}(1995)}]{shibnak95}%
  \BibitemOpen
  \bibfield  {author} {\bibinfo {author} {\bibfnamefont {M.}~\bibnamefont
  {Shibata}}\ and\ \bibinfo {author} {\bibfnamefont {T.}~\bibnamefont
  {Nakamura}},\ }\href {\doibase 10.1103/PhysRevD.52.5428} {\bibfield
  {journal} {\bibinfo  {journal} {Phys. Rev. D}\ }\textbf {\bibinfo {volume}
  {52}},\ \bibinfo {pages} {5428} (\bibinfo {year} {1995})}\BibitemShut
  {NoStop}%
\bibitem [{\citenamefont {Baumgarte}\ and\ \citenamefont {Shapiro}(1999)}]{BS}%
  \BibitemOpen
  \bibfield  {author} {\bibinfo {author} {\bibfnamefont {T.~W.}\ \bibnamefont
  {Baumgarte}}\ and\ \bibinfo {author} {\bibfnamefont {S.~L.}\ \bibnamefont
  {Shapiro}},\ }\href {\doibase 10.1103/PhysRevD.59.024007} {\bibfield
  {journal} {\bibinfo  {journal} {Phys. Rev.}\ }\textbf {\bibinfo {volume}
  {D59}},\ \bibinfo {pages} {024007} (\bibinfo {year} {1999})},\ \Eprint
  {http://arxiv.org/abs/gr-qc/9810065} {arXiv:gr-qc/9810065 [gr-qc]}
  \BibitemShut {NoStop}%
\bibitem [{\citenamefont {Etienne}\ \emph {et~al.}(2008)\citenamefont
  {Etienne}, \citenamefont {Faber}, \citenamefont {Liu}, \citenamefont
  {Shapiro}, \citenamefont {Taniguchi},\ and\ \citenamefont
  {Baumgarte}}]{Etienne:2007jg}%
  \BibitemOpen
  \bibfield  {author} {\bibinfo {author} {\bibfnamefont {Z.~B.}\ \bibnamefont
  {Etienne}}, \bibinfo {author} {\bibfnamefont {J.~A.}\ \bibnamefont {Faber}},
  \bibinfo {author} {\bibfnamefont {Y.~T.}\ \bibnamefont {Liu}}, \bibinfo
  {author} {\bibfnamefont {S.~L.}\ \bibnamefont {Shapiro}}, \bibinfo {author}
  {\bibfnamefont {K.}~\bibnamefont {Taniguchi}}, \ and\ \bibinfo {author}
  {\bibfnamefont {T.~W.}\ \bibnamefont {Baumgarte}},\ }\href {\doibase
  10.1103/PhysRevD.77.084002} {\bibfield  {journal} {\bibinfo  {journal} {Phys.
  Rev.}\ }\textbf {\bibinfo {volume} {D77}},\ \bibinfo {pages} {084002}
  (\bibinfo {year} {2008})},\ \Eprint {http://arxiv.org/abs/0712.2460}
  {arXiv:0712.2460 [astro-ph]} \BibitemShut {NoStop}%
\bibitem [{\citenamefont {{Baker}}\ \emph {et~al.}(2006)\citenamefont
  {{Baker}}, \citenamefont {{Centrella}}, \citenamefont {{Choi}}, \citenamefont
  {{Koppitz}},\ and\ \citenamefont {{van Meter}}}]{goddard06}%
  \BibitemOpen
  \bibfield  {author} {\bibinfo {author} {\bibfnamefont {J.~G.}\ \bibnamefont
  {{Baker}}}, \bibinfo {author} {\bibfnamefont {J.}~\bibnamefont
  {{Centrella}}}, \bibinfo {author} {\bibfnamefont {D.-I.}\ \bibnamefont
  {{Choi}}}, \bibinfo {author} {\bibfnamefont {M.}~\bibnamefont {{Koppitz}}}, \
  and\ \bibinfo {author} {\bibfnamefont {J.}~\bibnamefont {{van Meter}}},\
  }\href {\doibase 10.1103/PhysRevD.73.104002} {\bibfield  {journal} {\bibinfo
  {journal} {Phys. Rev. D}\ }\textbf {\bibinfo {volume} {73}},\ \bibinfo
  {pages} {104002} (\bibinfo {year} {2006})}\BibitemShut {NoStop}%
\bibitem [{\citenamefont {Farris}\ \emph {et~al.}(2012)\citenamefont {Farris},
  \citenamefont {Gold}, \citenamefont {Paschalidis}, \citenamefont {Etienne},\
  and\ \citenamefont {Shapiro}}]{Farris:2012ux}%
  \BibitemOpen
  \bibfield  {author} {\bibinfo {author} {\bibfnamefont {B.~D.}\ \bibnamefont
  {Farris}}, \bibinfo {author} {\bibfnamefont {R.}~\bibnamefont {Gold}},
  \bibinfo {author} {\bibfnamefont {V.}~\bibnamefont {Paschalidis}}, \bibinfo
  {author} {\bibfnamefont {Z.~B.}\ \bibnamefont {Etienne}}, \ and\ \bibinfo
  {author} {\bibfnamefont {S.~L.}\ \bibnamefont {Shapiro}},\ }\href {\doibase
  10.1103/PhysRevLett.109.221102} {\bibfield  {journal} {\bibinfo  {journal}
  {Phys.Rev.Lett.}\ }\textbf {\bibinfo {volume} {109}},\ \bibinfo {pages}
  {221102} (\bibinfo {year} {2012})}\BibitemShut {NoStop}%
\bibitem [{\citenamefont {Etienne}\ \emph
  {et~al.}(2012{\natexlab{a}})\citenamefont {Etienne}, \citenamefont
  {Paschalidis}, \citenamefont {Liu},\ and\ \citenamefont
  {Shapiro}}]{Etienne:2011re}%
  \BibitemOpen
  \bibfield  {author} {\bibinfo {author} {\bibfnamefont {Z.~B.}\ \bibnamefont
  {Etienne}}, \bibinfo {author} {\bibfnamefont {V.}~\bibnamefont
  {Paschalidis}}, \bibinfo {author} {\bibfnamefont {Y.~T.}\ \bibnamefont
  {Liu}}, \ and\ \bibinfo {author} {\bibfnamefont {S.~L.}\ \bibnamefont
  {Shapiro}},\ }\href {\doibase 10.1103/PhysRevD.85.024013} {\bibfield
  {journal} {\bibinfo  {journal} {Phys.Rev.}\ }\textbf {\bibinfo {volume}
  {D85}},\ \bibinfo {pages} {024013} (\bibinfo {year}
  {2012}{\natexlab{a}})}\BibitemShut {NoStop}%
\bibitem [{\citenamefont {Tsokaros}\ \emph {et~al.}(2015)\citenamefont
  {Tsokaros}, \citenamefont {Uryū},\ and\ \citenamefont
  {Rezzolla}}]{Tsokaros:2015fea}%
  \BibitemOpen
  \bibfield  {author} {\bibinfo {author} {\bibfnamefont {A.}~\bibnamefont
  {Tsokaros}}, \bibinfo {author} {\bibfnamefont {K.}~\bibnamefont {Uryū}}, \
  and\ \bibinfo {author} {\bibfnamefont {L.}~\bibnamefont {Rezzolla}},\ }\href
  {\doibase 10.1103/PhysRevD.91.104030} {\bibfield  {journal} {\bibinfo
  {journal} {Phys. Rev.}\ }\textbf {\bibinfo {volume} {D91}},\ \bibinfo {pages}
  {104030} (\bibinfo {year} {2015})},\ \Eprint
  {http://arxiv.org/abs/1502.05674} {arXiv:1502.05674 [gr-qc]} \BibitemShut
  {NoStop}%
\bibitem [{\citenamefont {Tsokaros}\ \emph {et~al.}(2018)\citenamefont
  {Tsokaros}, \citenamefont {Uryu}, \citenamefont {Ruiz},\ and\ \citenamefont
  {Shapiro}}]{Tsokaros:2018dqs}%
  \BibitemOpen
  \bibfield  {author} {\bibinfo {author} {\bibfnamefont {A.}~\bibnamefont
  {Tsokaros}}, \bibinfo {author} {\bibfnamefont {K.}~\bibnamefont {Uryu}},
  \bibinfo {author} {\bibfnamefont {M.}~\bibnamefont {Ruiz}}, \ and\ \bibinfo
  {author} {\bibfnamefont {S.~L.}\ \bibnamefont {Shapiro}},\ }\href {\doibase
  10.1103/PhysRevD.98.124019} {\bibfield  {journal} {\bibinfo  {journal} {Phys.
  Rev.}\ }\textbf {\bibinfo {volume} {D98}},\ \bibinfo {pages} {124019}
  (\bibinfo {year} {2018})},\ \Eprint {http://arxiv.org/abs/1809.08237}
  {arXiv:1809.08237 [gr-qc]} \BibitemShut {NoStop}%
\bibitem [{\citenamefont {Riley}\ \emph {et~al.}(2021)\citenamefont {Riley}
  \emph {et~al.}}]{Riley:2021pdl}%
  \BibitemOpen
  \bibfield  {author} {\bibinfo {author} {\bibfnamefont {T.~E.}\ \bibnamefont
  {Riley}} \emph {et~al.},\ }\href@noop {} {\  (\bibinfo {year} {2021})},\
  \Eprint {http://arxiv.org/abs/2105.06980} {arXiv:2105.06980 [astro-ph.HE]}
  \BibitemShut {NoStop}%
\bibitem [{\citenamefont {Pang}\ \emph {et~al.}(2021)\citenamefont {Pang},
  \citenamefont {Tews}, \citenamefont {Coughlin}, \citenamefont {Bulla},
  \citenamefont {Van Den~Broeck},\ and\ \citenamefont
  {Dietrich}}]{Pang:2021jta}%
  \BibitemOpen
  \bibfield  {author} {\bibinfo {author} {\bibfnamefont {P.~T.~H.}\
  \bibnamefont {Pang}}, \bibinfo {author} {\bibfnamefont {I.}~\bibnamefont
  {Tews}}, \bibinfo {author} {\bibfnamefont {M.~W.}\ \bibnamefont {Coughlin}},
  \bibinfo {author} {\bibfnamefont {M.}~\bibnamefont {Bulla}}, \bibinfo
  {author} {\bibfnamefont {C.}~\bibnamefont {Van Den~Broeck}}, \ and\ \bibinfo
  {author} {\bibfnamefont {T.}~\bibnamefont {Dietrich}},\ }\href@noop {} {\
  (\bibinfo {year} {2021})},\ \Eprint {http://arxiv.org/abs/2105.08688}
  {arXiv:2105.08688 [astro-ph.HE]} \BibitemShut {NoStop}%
\bibitem [{\citenamefont {Miller}\ \emph {et~al.}(2019)\citenamefont {Miller}
  \emph {et~al.}}]{Miller:2019cac}%
  \BibitemOpen
  \bibfield  {author} {\bibinfo {author} {\bibfnamefont {M.~C.}\ \bibnamefont
  {Miller}} \emph {et~al.},\ }\href {\doibase 10.3847/2041-8213/ab50c5}
  {\bibfield  {journal} {\bibinfo  {journal} {Astrophys. J. Lett.}\ }\textbf
  {\bibinfo {volume} {887}},\ \bibinfo {pages} {L24} (\bibinfo {year}
  {2019})},\ \Eprint {http://arxiv.org/abs/1912.05705} {arXiv:1912.05705
  [astro-ph.HE]} \BibitemShut {NoStop}%
\bibitem [{\citenamefont {Tsokaros}\ \emph
  {et~al.}(2020{\natexlab{b}})\citenamefont {Tsokaros}, \citenamefont {Ruiz},\
  and\ \citenamefont {Shapiro}}]{Tsokaros:2020hli}%
  \BibitemOpen
  \bibfield  {author} {\bibinfo {author} {\bibfnamefont {A.}~\bibnamefont
  {Tsokaros}}, \bibinfo {author} {\bibfnamefont {M.}~\bibnamefont {Ruiz}}, \
  and\ \bibinfo {author} {\bibfnamefont {S.~L.}\ \bibnamefont {Shapiro}},\
  }\href {\doibase 10.3847/1538-4357/abc421} {\bibfield  {journal} {\bibinfo
  {journal} {Astrophys. J.}\ }\textbf {\bibinfo {volume} {905}},\ \bibinfo
  {pages} {48} (\bibinfo {year} {2020}{\natexlab{b}})},\ \Eprint
  {http://arxiv.org/abs/2007.05526} {arXiv:2007.05526 [astro-ph.HE]}
  \BibitemShut {NoStop}%
\bibitem [{\citenamefont {Shibata}\ and\ \citenamefont
  {Taniguchi}(2006)}]{Shibata:2006nm}%
  \BibitemOpen
  \bibfield  {author} {\bibinfo {author} {\bibfnamefont {M.}~\bibnamefont
  {Shibata}}\ and\ \bibinfo {author} {\bibfnamefont {K.}~\bibnamefont
  {Taniguchi}},\ }\href {\doibase 10.1103/PhysRevD.73.064027} {\bibfield
  {journal} {\bibinfo  {journal} {Phys. Rev.}\ }\textbf {\bibinfo {volume}
  {D73}},\ \bibinfo {pages} {064027} (\bibinfo {year} {2006})}\BibitemShut
  {NoStop}%
\bibitem [{\citenamefont {Bauswein}\ \emph {et~al.}(2020)\citenamefont
  {Bauswein}, \citenamefont {Blacker}, \citenamefont {Vijayan}, \citenamefont
  {Stergioulas}, \citenamefont {Chatziioannou}, \citenamefont {Clark},
  \citenamefont {Bastian}, \citenamefont {Blaschke}, \citenamefont {Cierniak},\
  and\ \citenamefont {Fischer}}]{Bauswein:2020aag}%
  \BibitemOpen
  \bibfield  {author} {\bibinfo {author} {\bibfnamefont {A.}~\bibnamefont
  {Bauswein}}, \bibinfo {author} {\bibfnamefont {S.}~\bibnamefont {Blacker}},
  \bibinfo {author} {\bibfnamefont {V.}~\bibnamefont {Vijayan}}, \bibinfo
  {author} {\bibfnamefont {N.}~\bibnamefont {Stergioulas}}, \bibinfo {author}
  {\bibfnamefont {K.}~\bibnamefont {Chatziioannou}}, \bibinfo {author}
  {\bibfnamefont {J.~A.}\ \bibnamefont {Clark}}, \bibinfo {author}
  {\bibfnamefont {N.-U.~F.}\ \bibnamefont {Bastian}}, \bibinfo {author}
  {\bibfnamefont {D.~B.}\ \bibnamefont {Blaschke}}, \bibinfo {author}
  {\bibfnamefont {M.}~\bibnamefont {Cierniak}}, \ and\ \bibinfo {author}
  {\bibfnamefont {T.}~\bibnamefont {Fischer}},\ }\href {\doibase
  10.1103/PhysRevLett.125.141103} {\bibfield  {journal} {\bibinfo  {journal}
  {Phys. Rev. Lett.}\ }\textbf {\bibinfo {volume} {125}},\ \bibinfo {pages}
  {141103} (\bibinfo {year} {2020})},\ \Eprint
  {http://arxiv.org/abs/2004.00846} {arXiv:2004.00846 [astro-ph.HE]}
  \BibitemShut {NoStop}%
\bibitem [{\citenamefont {{Baumgarte}}\ \emph {et~al.}(2000)\citenamefont
  {{Baumgarte}}, \citenamefont {{Shapiro}},\ and\ \citenamefont
  {{Shibata}}}]{BaShSh}%
  \BibitemOpen
  \bibfield  {author} {\bibinfo {author} {\bibfnamefont {T.~W.}\ \bibnamefont
  {{Baumgarte}}}, \bibinfo {author} {\bibfnamefont {S.~L.}\ \bibnamefont
  {{Shapiro}}}, \ and\ \bibinfo {author} {\bibfnamefont {M.}~\bibnamefont
  {{Shibata}}},\ }\href {\doibase 10.1086/312425} {\bibfield  {journal}
  {\bibinfo  {journal} {Astrophys. J. Letters}\ }\textbf {\bibinfo {volume}
  {528}},\ \bibinfo {pages} {L29} (\bibinfo {year} {2000})},\ \Eprint
  {http://arxiv.org/abs/astro-ph/9910565} {astro-ph/9910565} \BibitemShut
  {NoStop}%
\bibitem [{\citenamefont {Bauswein}\ \emph {et~al.}(2010)\citenamefont
  {Bauswein}, \citenamefont {Janka},\ and\ \citenamefont
  {Oechslin}}]{Bauswein:2010dn}%
  \BibitemOpen
  \bibfield  {author} {\bibinfo {author} {\bibfnamefont {A.}~\bibnamefont
  {Bauswein}}, \bibinfo {author} {\bibfnamefont {H.~T.}\ \bibnamefont {Janka}},
  \ and\ \bibinfo {author} {\bibfnamefont {R.}~\bibnamefont {Oechslin}},\
  }\href {\doibase 10.1103/PhysRevD.82.084043} {\bibfield  {journal} {\bibinfo
  {journal} {Phys. Rev. D}\ }\textbf {\bibinfo {volume} {82}},\ \bibinfo
  {pages} {084043} (\bibinfo {year} {2010})},\ \Eprint
  {http://arxiv.org/abs/1006.3315} {arXiv:1006.3315 [astro-ph.SR]} \BibitemShut
  {NoStop}%
\bibitem [{\citenamefont {Paschalidis}\ \emph {et~al.}(2011)\citenamefont
  {Paschalidis}, \citenamefont {Liu}, \citenamefont {Etienne},\ and\
  \citenamefont {Shapiro}}]{Paschalidis:2011ez}%
  \BibitemOpen
  \bibfield  {author} {\bibinfo {author} {\bibfnamefont {V.}~\bibnamefont
  {Paschalidis}}, \bibinfo {author} {\bibfnamefont {Y.~T.}\ \bibnamefont
  {Liu}}, \bibinfo {author} {\bibfnamefont {Z.}~\bibnamefont {Etienne}}, \ and\
  \bibinfo {author} {\bibfnamefont {S.~L.}\ \bibnamefont {Shapiro}},\ }\href
  {\doibase 10.1103/PhysRevD.84.104032} {\bibfield  {journal} {\bibinfo
  {journal} {Phys. Rev. D}\ }\textbf {\bibinfo {volume} {84}},\ \bibinfo
  {pages} {104032} (\bibinfo {year} {2011})},\ \Eprint
  {http://arxiv.org/abs/1109.5177} {arXiv:1109.5177 [astro-ph.HE]} \BibitemShut
  {NoStop}%
\bibitem [{\citenamefont {Aguilera-Miret}\ \emph {et~al.}(2020)\citenamefont
  {Aguilera-Miret}, \citenamefont {Vigan\`o}, \citenamefont {Carrasco},
  \citenamefont {Mi\~nano},\ and\ \citenamefont
  {Palenzuela}}]{Aguilera-Miret:2020dhz}%
  \BibitemOpen
  \bibfield  {author} {\bibinfo {author} {\bibfnamefont {R.}~\bibnamefont
  {Aguilera-Miret}}, \bibinfo {author} {\bibfnamefont {D.}~\bibnamefont
  {Vigan\`o}}, \bibinfo {author} {\bibfnamefont {F.}~\bibnamefont {Carrasco}},
  \bibinfo {author} {\bibfnamefont {B.}~\bibnamefont {Mi\~nano}}, \ and\
  \bibinfo {author} {\bibfnamefont {C.}~\bibnamefont {Palenzuela}},\ }\href
  {\doibase 10.1103/PhysRevD.102.103006} {\bibfield  {journal} {\bibinfo
  {journal} {Phys. Rev. D}\ }\textbf {\bibinfo {volume} {102}},\ \bibinfo
  {pages} {103006} (\bibinfo {year} {2020})},\ \Eprint
  {http://arxiv.org/abs/2009.06669} {arXiv:2009.06669 [gr-qc]} \BibitemShut
  {NoStop}%
\bibitem [{\citenamefont {Kiuchi}\ \emph {et~al.}(2015)\citenamefont {Kiuchi},
  \citenamefont {Cerdá-Durán}, \citenamefont {Kyutoku}, \citenamefont
  {Sekiguchi},\ and\ \citenamefont {Shibata}}]{Kiuchi:2015sga}%
  \BibitemOpen
  \bibfield  {author} {\bibinfo {author} {\bibfnamefont {K.}~\bibnamefont
  {Kiuchi}}, \bibinfo {author} {\bibfnamefont {P.}~\bibnamefont
  {Cerdá-Durán}}, \bibinfo {author} {\bibfnamefont {K.}~\bibnamefont
  {Kyutoku}}, \bibinfo {author} {\bibfnamefont {Y.}~\bibnamefont {Sekiguchi}},
  \ and\ \bibinfo {author} {\bibfnamefont {M.}~\bibnamefont {Shibata}},\ }\href
  {\doibase 10.1103/PhysRevD.92.124034} {\bibfield  {journal} {\bibinfo
  {journal} {Phys. Rev.}\ }\textbf {\bibinfo {volume} {D92}},\ \bibinfo {pages}
  {124034} (\bibinfo {year} {2015})}\BibitemShut {NoStop}%
\bibitem [{\citenamefont {Etienne}\ \emph
  {et~al.}(2012{\natexlab{b}})\citenamefont {Etienne}, \citenamefont {Liu},
  \citenamefont {Paschalidis},\ and\ \citenamefont {Shapiro}}]{Etienne:2011ea}%
  \BibitemOpen
  \bibfield  {author} {\bibinfo {author} {\bibfnamefont {Z.~B.}\ \bibnamefont
  {Etienne}}, \bibinfo {author} {\bibfnamefont {Y.~T.}\ \bibnamefont {Liu}},
  \bibinfo {author} {\bibfnamefont {V.}~\bibnamefont {Paschalidis}}, \ and\
  \bibinfo {author} {\bibfnamefont {S.~L.}\ \bibnamefont {Shapiro}},\ }\href
  {\doibase 10.1103/PhysRevD.85.064029} {\bibfield  {journal} {\bibinfo
  {journal} {Phys.Rev.}\ }\textbf {\bibinfo {volume} {D85}},\ \bibinfo {pages}
  {064029} (\bibinfo {year} {2012}{\natexlab{b}})}\BibitemShut {NoStop}%
\bibitem [{\citenamefont {Tsokaros}\ \emph {et~al.}(2019)\citenamefont
  {Tsokaros}, \citenamefont {Ruiz}, \citenamefont {Paschalidis}, \citenamefont
  {Shapiro},\ and\ \citenamefont {Ury\={u}}}]{Tsokaros:2019anx}%
  \BibitemOpen
  \bibfield  {author} {\bibinfo {author} {\bibfnamefont {A.}~\bibnamefont
  {Tsokaros}}, \bibinfo {author} {\bibfnamefont {M.}~\bibnamefont {Ruiz}},
  \bibinfo {author} {\bibfnamefont {V.}~\bibnamefont {Paschalidis}}, \bibinfo
  {author} {\bibfnamefont {S.~L.}\ \bibnamefont {Shapiro}}, \ and\ \bibinfo
  {author} {\bibfnamefont {K.}~\bibnamefont {Ury\={u}}},\ }\href {\doibase
  10.1103/PhysRevD.100.024061} {\bibfield  {journal} {\bibinfo  {journal}
  {Phys. Rev. D}\ }\textbf {\bibinfo {volume} {100}},\ \bibinfo {pages}
  {024061} (\bibinfo {year} {2019})},\ \Eprint
  {http://arxiv.org/abs/1906.00011} {arXiv:1906.00011 [gr-qc]} \BibitemShut
  {NoStop}%
\bibitem [{\citenamefont {{Thornburg}}(2004)}]{ahfinderdirect}%
  \BibitemOpen
  \bibfield  {author} {\bibinfo {author} {\bibfnamefont {J.}~\bibnamefont
  {{Thornburg}}},\ }\href {\doibase 10.1088/0264-9381/21/2/026} {\bibfield
  {journal} {\bibinfo  {journal} {Class.~Quant.~Grav.}\ }\textbf {\bibinfo
  {volume} {21}},\ \bibinfo {pages} {743} (\bibinfo {year} {2004})}\BibitemShut
  {NoStop}%
\bibitem [{\citenamefont {{Dreyer}}\ \emph {et~al.}(2003)\citenamefont
  {{Dreyer}}, \citenamefont {{Krishnan}}, \citenamefont {{Shoemaker}},\ and\
  \citenamefont {{Schnetter}}}]{dkss03}%
  \BibitemOpen
  \bibfield  {author} {\bibinfo {author} {\bibfnamefont {O.}~\bibnamefont
  {{Dreyer}}}, \bibinfo {author} {\bibfnamefont {B.}~\bibnamefont
  {{Krishnan}}}, \bibinfo {author} {\bibfnamefont {D.}~\bibnamefont
  {{Shoemaker}}}, \ and\ \bibinfo {author} {\bibfnamefont {E.}~\bibnamefont
  {{Schnetter}}},\ }\href {\doibase 10.1103/PhysRevD.67.024018} {\bibfield
  {journal} {\bibinfo  {journal} {Phys. Rev. D}\ }\textbf {\bibinfo {volume}
  {67}},\ \bibinfo {pages} {024018} (\bibinfo {year} {2003})}\BibitemShut
  {NoStop}%
\bibitem [{\citenamefont {Ruiz}\ \emph {et~al.}(2008)\citenamefont {Ruiz},
  \citenamefont {Takahashi}, \citenamefont {Alcubierre},\ and\ \citenamefont
  {Nunez}}]{Ruiz:2007yx}%
  \BibitemOpen
  \bibfield  {author} {\bibinfo {author} {\bibfnamefont {M.}~\bibnamefont
  {Ruiz}}, \bibinfo {author} {\bibfnamefont {R.}~\bibnamefont {Takahashi}},
  \bibinfo {author} {\bibfnamefont {M.}~\bibnamefont {Alcubierre}}, \ and\
  \bibinfo {author} {\bibfnamefont {D.}~\bibnamefont {Nunez}},\ }\href
  {\doibase 10.1007/s10714-007-0570-8, 10.1007/s10714-008-0684-7} {\bibfield
  {journal} {\bibinfo  {journal} {Gen. Rel. Grav.}\ }\textbf {\bibinfo {volume}
  {40}},\ \bibinfo {pages} {2467} (\bibinfo {year} {2008})}\BibitemShut
  {NoStop}%
\bibitem [{\citenamefont {Shibata}\ and\ \citenamefont
  {Hotokezaka}(2019)}]{Shibata:2019wef}%
  \BibitemOpen
  \bibfield  {author} {\bibinfo {author} {\bibfnamefont {M.}~\bibnamefont
  {Shibata}}\ and\ \bibinfo {author} {\bibfnamefont {K.}~\bibnamefont
  {Hotokezaka}},\ }\href {\doibase 10.1146/annurev-nucl-101918-023625}
  {\bibfield  {journal} {\bibinfo  {journal} {Ann. Rev. Nucl. Part. Sci.}\
  }\textbf {\bibinfo {volume} {69}},\ \bibinfo {pages} {41} (\bibinfo {year}
  {2019})},\ \Eprint {http://arxiv.org/abs/1908.02350} {arXiv:1908.02350
  [astro-ph.HE]} \BibitemShut {NoStop}%
\bibitem [{\citenamefont {Radice}\ \emph
  {et~al.}(2018{\natexlab{b}})\citenamefont {Radice}, \citenamefont {Perego},
  \citenamefont {Hotokezaka}, \citenamefont {Fromm}, \citenamefont {Bernuzzi},\
  and\ \citenamefont {Roberts}}]{Radice:2018pdn}%
  \BibitemOpen
  \bibfield  {author} {\bibinfo {author} {\bibfnamefont {D.}~\bibnamefont
  {Radice}}, \bibinfo {author} {\bibfnamefont {A.}~\bibnamefont {Perego}},
  \bibinfo {author} {\bibfnamefont {K.}~\bibnamefont {Hotokezaka}}, \bibinfo
  {author} {\bibfnamefont {S.~A.}\ \bibnamefont {Fromm}}, \bibinfo {author}
  {\bibfnamefont {S.}~\bibnamefont {Bernuzzi}}, \ and\ \bibinfo {author}
  {\bibfnamefont {L.~F.}\ \bibnamefont {Roberts}},\ }\href {\doibase
  10.3847/1538-4357/aaf054} {\bibfield  {journal} {\bibinfo  {journal}
  {Astrophys. J.}\ }\textbf {\bibinfo {volume} {869}},\ \bibinfo {pages} {130}
  (\bibinfo {year} {2018}{\natexlab{b}})},\ \Eprint
  {http://arxiv.org/abs/1809.11161} {arXiv:1809.11161 [astro-ph.HE]}
  \BibitemShut {NoStop}%
\bibitem [{\citenamefont {Abbott}\ \emph
  {et~al.}(2017{\natexlab{d}})\citenamefont {Abbott} \emph
  {et~al.}}]{LIGOScientific:2017pwl}%
  \BibitemOpen
  \bibfield  {author} {\bibinfo {author} {\bibfnamefont {B.~P.}\ \bibnamefont
  {Abbott}} \emph {et~al.} (\bibinfo {collaboration} {LIGO Scientific,
  Virgo}),\ }\href {\doibase 10.3847/2041-8213/aa9478} {\bibfield  {journal}
  {\bibinfo  {journal} {Astrophys. J. Lett.}\ }\textbf {\bibinfo {volume}
  {850}},\ \bibinfo {pages} {L39} (\bibinfo {year} {2017}{\natexlab{d}})},\
  \Eprint {http://arxiv.org/abs/1710.05836} {arXiv:1710.05836 [astro-ph.HE]}
  \BibitemShut {NoStop}%
\bibitem [{\citenamefont {C\^ot\'e}\ \emph {et~al.}(2018)\citenamefont
  {C\^ot\'e} \emph {et~al.}}]{Cote:2017evr}%
  \BibitemOpen
  \bibfield  {author} {\bibinfo {author} {\bibfnamefont {B.}~\bibnamefont
  {C\^ot\'e}} \emph {et~al.},\ }\href {\doibase 10.3847/1538-4357/aaad67}
  {\bibfield  {journal} {\bibinfo  {journal} {Astrophys. J.}\ }\textbf
  {\bibinfo {volume} {855}},\ \bibinfo {pages} {99} (\bibinfo {year} {2018})},\
  \Eprint {http://arxiv.org/abs/1710.05875} {arXiv:1710.05875 [astro-ph.GA]}
  \BibitemShut {NoStop}%
\bibitem [{\citenamefont {Shakura}\ and\ \citenamefont
  {Sunyaev}(1973)}]{Shakura73}%
  \BibitemOpen
  \bibfield  {author} {\bibinfo {author} {\bibfnamefont {N.~I.}\ \bibnamefont
  {Shakura}}\ and\ \bibinfo {author} {\bibfnamefont {R.~A.}\ \bibnamefont
  {Sunyaev}},\ }\href@noop {} {\bibfield  {journal} {\bibinfo  {journal}
  {Astronomy and Astrophysics}\ }\textbf {\bibinfo {volume} {24}},\ \bibinfo
  {pages} {337} (\bibinfo {year} {1973})}\BibitemShut {NoStop}%
\bibitem [{\citenamefont {{Penna}}\ \emph {et~al.}(2010)\citenamefont
  {{Penna}}, \citenamefont {{McKinney}}, \citenamefont {{Narayan}},
  \citenamefont {{Tchekhovskoy}}, \citenamefont {{Shafee}},\ and\ \citenamefont
  {{McClintock}}}]{FASTEST_GROWING_MRI_WAVELENGTH}%
  \BibitemOpen
  \bibfield  {author} {\bibinfo {author} {\bibfnamefont {R.~F.}\ \bibnamefont
  {{Penna}}}, \bibinfo {author} {\bibfnamefont {J.~C.}\ \bibnamefont
  {{McKinney}}}, \bibinfo {author} {\bibfnamefont {R.}~\bibnamefont
  {{Narayan}}}, \bibinfo {author} {\bibfnamefont {A.}~\bibnamefont
  {{Tchekhovskoy}}}, \bibinfo {author} {\bibfnamefont {R.}~\bibnamefont
  {{Shafee}}}, \ and\ \bibinfo {author} {\bibfnamefont {J.~E.}\ \bibnamefont
  {{McClintock}}},\ }\href {\doibase 10.1111/j.1365-2966.2010.17170.x}
  {\bibfield  {journal} {\bibinfo  {journal} {mnras}\ }\textbf {\bibinfo
  {volume} {408}},\ \bibinfo {pages} {752} (\bibinfo {year}
  {2010})}\BibitemShut {NoStop}%
\bibitem [{\citenamefont {{Etienne}}\ \emph {et~al.}(2012)\citenamefont
  {{Etienne}}, \citenamefont {{Paschalidis}},\ and\ \citenamefont
  {{Shapiro}}}]{UIUC_PAPER2}%
  \BibitemOpen
  \bibfield  {author} {\bibinfo {author} {\bibfnamefont {Z.~B.}\ \bibnamefont
  {{Etienne}}}, \bibinfo {author} {\bibfnamefont {V.}~\bibnamefont
  {{Paschalidis}}}, \ and\ \bibinfo {author} {\bibfnamefont {S.~L.}\
  \bibnamefont {{Shapiro}}},\ }\href {\doibase 10.1103/PhysRevD.86.084026}
  {\bibfield  {journal} {\bibinfo  {journal} {Phys. Rev. D}\ }\textbf {\bibinfo
  {volume} {86}},\ \bibinfo {eid} {084026} (\bibinfo {year}
  {2012})}\BibitemShut {NoStop}%
\bibitem [{\citenamefont {Sano}\ \emph {et~al.}(2004)\citenamefont {Sano},
  \citenamefont {Inutsuka}, \citenamefont {Turner},\ and\ \citenamefont
  {Stone}}]{Sano:2003bf}%
  \BibitemOpen
  \bibfield  {author} {\bibinfo {author} {\bibfnamefont {T.}~\bibnamefont
  {Sano}}, \bibinfo {author} {\bibfnamefont {S.-i.}\ \bibnamefont {Inutsuka}},
  \bibinfo {author} {\bibfnamefont {N.~J.}\ \bibnamefont {Turner}}, \ and\
  \bibinfo {author} {\bibfnamefont {J.~M.}\ \bibnamefont {Stone}},\ }\href
  {\doibase 10.1086/382184} {\bibfield  {journal} {\bibinfo  {journal}
  {Astrophys. J.}\ }\textbf {\bibinfo {volume} {605}},\ \bibinfo {pages} {321}
  (\bibinfo {year} {2004})},\ \Eprint {http://arxiv.org/abs/astro-ph/0312480}
  {arXiv:astro-ph/0312480 [astro-ph]} \BibitemShut {NoStop}%
\bibitem [{\citenamefont {Shiokawa}\ \emph {et~al.}(2012)\citenamefont
  {Shiokawa}, \citenamefont {Dolence}, \citenamefont {Gammie},\ and\
  \citenamefont {Noble}}]{Shiokawa:2011ih}%
  \BibitemOpen
  \bibfield  {author} {\bibinfo {author} {\bibfnamefont {H.}~\bibnamefont
  {Shiokawa}}, \bibinfo {author} {\bibfnamefont {J.~C.}\ \bibnamefont
  {Dolence}}, \bibinfo {author} {\bibfnamefont {C.~F.}\ \bibnamefont {Gammie}},
  \ and\ \bibinfo {author} {\bibfnamefont {S.~C.}\ \bibnamefont {Noble}},\
  }\href {\doibase 10.1088/0004-637X/744/2/187} {\bibfield  {journal} {\bibinfo
   {journal} {Astrophys. J.}\ }\textbf {\bibinfo {volume} {744}},\ \bibinfo
  {pages} {187} (\bibinfo {year} {2012})},\ \Eprint
  {http://arxiv.org/abs/1111.0396} {arXiv:1111.0396 [astro-ph.HE]} \BibitemShut
  {NoStop}%
\bibitem [{\citenamefont {Farris}\ \emph {et~al.}(2010)\citenamefont {Farris},
  \citenamefont {Liu},\ and\ \citenamefont {Shapiro}}]{Farris:2009mt}%
  \BibitemOpen
  \bibfield  {author} {\bibinfo {author} {\bibfnamefont {B.~D.}\ \bibnamefont
  {Farris}}, \bibinfo {author} {\bibfnamefont {Y.~T.}\ \bibnamefont {Liu}}, \
  and\ \bibinfo {author} {\bibfnamefont {S.~L.}\ \bibnamefont {Shapiro}},\
  }\href {\doibase 10.1103/PhysRevD.81.084008} {\bibfield  {journal} {\bibinfo
  {journal} {Phys.Rev.}\ }\textbf {\bibinfo {volume} {D81}},\ \bibinfo {pages}
  {084008} (\bibinfo {year} {2010})}\BibitemShut {NoStop}%
\bibitem [{\citenamefont {{Giacomazzo}}\ \emph {et~al.}(2011)\citenamefont
  {{Giacomazzo}}, \citenamefont {{Rezzolla}},\ and\ \citenamefont
  {{Baiotti}}}]{BrunoMagNSNS}%
  \BibitemOpen
  \bibfield  {author} {\bibinfo {author} {\bibfnamefont {B.}~\bibnamefont
  {{Giacomazzo}}}, \bibinfo {author} {\bibfnamefont {L.}~\bibnamefont
  {{Rezzolla}}}, \ and\ \bibinfo {author} {\bibfnamefont {L.}~\bibnamefont
  {{Baiotti}}},\ }\href {\doibase 10.1103/PhysRevD.83.044014} {\bibfield
  {journal} {\bibinfo  {journal} {Phys. Rev. D}\ }\textbf {\bibinfo {volume}
  {83}},\ \bibinfo {eid} {044014} (\bibinfo {year} {2011})}\BibitemShut
  {NoStop}%
\bibitem [{\citenamefont {Shibata}\ and\ \citenamefont
  {Uryu}(2002)}]{Shibata:2002jb}%
  \BibitemOpen
  \bibfield  {author} {\bibinfo {author} {\bibfnamefont {M.}~\bibnamefont
  {Shibata}}\ and\ \bibinfo {author} {\bibfnamefont {K.}~\bibnamefont {Uryu}},\
  }\href {\doibase 10.1143/PTP.107.265} {\bibfield  {journal} {\bibinfo
  {journal} {Prog. Theor. Phys.}\ }\textbf {\bibinfo {volume} {107}},\ \bibinfo
  {pages} {265} (\bibinfo {year} {2002})}\BibitemShut {NoStop}%
\bibitem [{\citenamefont {Shibata}\ \emph {et~al.}(2003)\citenamefont
  {Shibata}, \citenamefont {Taniguchi},\ and\ \citenamefont {Uryu}}]{STU1}%
  \BibitemOpen
  \bibfield  {author} {\bibinfo {author} {\bibfnamefont {M.}~\bibnamefont
  {Shibata}}, \bibinfo {author} {\bibfnamefont {K.}~\bibnamefont {Taniguchi}},
  \ and\ \bibinfo {author} {\bibfnamefont {K.}~\bibnamefont {Uryu}},\ }\href
  {\doibase 10.1103/PhysRevD.68.084020} {\bibfield  {journal} {\bibinfo
  {journal} {Phys. Rev.}\ }\textbf {\bibinfo {volume} {D68}},\ \bibinfo {pages}
  {084020} (\bibinfo {year} {2003})},\ \Eprint
  {http://arxiv.org/abs/gr-qc/0310030} {arXiv:gr-qc/0310030 [gr-qc]}
  \BibitemShut {NoStop}%
\bibitem [{\citenamefont {{Shibata}}\ and\ \citenamefont
  {{Taniguchi}}(2006)}]{ST}%
  \BibitemOpen
  \bibfield  {author} {\bibinfo {author} {\bibfnamefont {M.}~\bibnamefont
  {{Shibata}}}\ and\ \bibinfo {author} {\bibfnamefont {K.}~\bibnamefont
  {{Taniguchi}}},\ }\href {\doibase 10.1103/PhysRevD.73.064027} {\bibfield
  {journal} {\bibinfo  {journal} {Phys. Rev. D}\ }\textbf {\bibinfo {volume}
  {73}},\ \bibinfo {pages} {064027} (\bibinfo {year} {2006})}\BibitemShut
  {NoStop}%
\bibitem [{\citenamefont {{Ciolfi}}\ \emph {et~al.}(2011)\citenamefont
  {{Ciolfi}}, \citenamefont {{Lander}}, \citenamefont {{Manca}},\ and\
  \citenamefont {{Rezzolla}}}]{Stable_NS_cannot_have_poloidal_fields2}%
  \BibitemOpen
  \bibfield  {author} {\bibinfo {author} {\bibfnamefont {R.}~\bibnamefont
  {{Ciolfi}}}, \bibinfo {author} {\bibfnamefont {S.~K.}\ \bibnamefont
  {{Lander}}}, \bibinfo {author} {\bibfnamefont {G.~M.}\ \bibnamefont
  {{Manca}}}, \ and\ \bibinfo {author} {\bibfnamefont {L.}~\bibnamefont
  {{Rezzolla}}},\ }\href {\doibase 10.1088/2041-8205/736/1/L6} {\bibfield
  {journal} {\bibinfo  {journal} {Astrophys. J. Letters}\ }\textbf {\bibinfo
  {volume} {736}},\ \bibinfo {eid} {L6} (\bibinfo {year} {2011})}\BibitemShut
  {NoStop}%
\bibitem [{\citenamefont {{Lasky}}\ \emph {et~al.}(2011)\citenamefont
  {{Lasky}}, \citenamefont {{Zink}}, \citenamefont {{Kokkotas}},\ and\
  \citenamefont {{Glampedakis}}}]{Stable_NS_cannot_have_poloidal_fields1}%
  \BibitemOpen
  \bibfield  {author} {\bibinfo {author} {\bibfnamefont {P.~D.}\ \bibnamefont
  {{Lasky}}}, \bibinfo {author} {\bibfnamefont {B.}~\bibnamefont {{Zink}}},
  \bibinfo {author} {\bibfnamefont {K.~D.}\ \bibnamefont {{Kokkotas}}}, \ and\
  \bibinfo {author} {\bibfnamefont {K.}~\bibnamefont {{Glampedakis}}},\ }\href
  {\doibase 10.1088/2041-8205/735/1/L20} {\bibfield  {journal} {\bibinfo
  {journal} {Astrophys. J. Letters}\ }\textbf {\bibinfo {volume} {735}},\
  \bibinfo {eid} {L20} (\bibinfo {year} {2011})}\BibitemShut {NoStop}%
\bibitem [{\citenamefont {Sun}\ \emph {et~al.}(2018)\citenamefont {Sun},
  \citenamefont {Ruiz},\ and\ \citenamefont {Shapiro}}]{Sun:2018gcl}%
  \BibitemOpen
  \bibfield  {author} {\bibinfo {author} {\bibfnamefont {L.}~\bibnamefont
  {Sun}}, \bibinfo {author} {\bibfnamefont {M.}~\bibnamefont {Ruiz}}, \ and\
  \bibinfo {author} {\bibfnamefont {S.~L.}\ \bibnamefont {Shapiro}},\
  }\href@noop {} {\  (\bibinfo {year} {2018})},\ \Eprint
  {http://arxiv.org/abs/1812.03176} {arXiv:1812.03176 [astro-ph.HE]}
  \BibitemShut {NoStop}%
\bibitem [{\citenamefont {Kiuchi}\ \emph {et~al.}(2018)\citenamefont {Kiuchi},
  \citenamefont {Kyutoku}, \citenamefont {Sekiguchi},\ and\ \citenamefont
  {Shibata}}]{Kiuchi:2017zzg}%
  \BibitemOpen
  \bibfield  {author} {\bibinfo {author} {\bibfnamefont {K.}~\bibnamefont
  {Kiuchi}}, \bibinfo {author} {\bibfnamefont {K.}~\bibnamefont {Kyutoku}},
  \bibinfo {author} {\bibfnamefont {Y.}~\bibnamefont {Sekiguchi}}, \ and\
  \bibinfo {author} {\bibfnamefont {M.}~\bibnamefont {Shibata}},\ }\href
  {\doibase 10.1103/PhysRevD.97.124039} {\bibfield  {journal} {\bibinfo
  {journal} {Phys. Rev.}\ }\textbf {\bibinfo {volume} {D97}},\ \bibinfo {pages}
  {124039} (\bibinfo {year} {2018})},\ \Eprint
  {http://arxiv.org/abs/1710.01311} {arXiv:1710.01311 [astro-ph.HE]}
  \BibitemShut {NoStop}%
\bibitem [{\citenamefont {{Kiuchi}}\ \emph {et~al.}(2012)\citenamefont
  {{Kiuchi}}, \citenamefont {{Kyutoku}},\ and\ \citenamefont
  {{Shibata}}}]{kks12}%
  \BibitemOpen
  \bibfield  {author} {\bibinfo {author} {\bibfnamefont {K.}~\bibnamefont
  {{Kiuchi}}}, \bibinfo {author} {\bibfnamefont {K.}~\bibnamefont {{Kyutoku}}},
  \ and\ \bibinfo {author} {\bibfnamefont {M.}~\bibnamefont {{Shibata}}},\
  }\href {\doibase 10.1103/PhysRevD.86.064008} {\bibfield  {journal} {\bibinfo
  {journal} {Phys. Rev. D}\ }\textbf {\bibinfo {volume} {86}},\ \bibinfo {eid}
  {064008} (\bibinfo {year} {2012})},\ \Eprint {http://arxiv.org/abs/1207.6444}
  {arXiv:1207.6444 [astro-ph.HE]} \BibitemShut {NoStop}%
\bibitem [{\citenamefont {Thorne}\ \emph {et~al.}(1986)\citenamefont {Thorne},
  \citenamefont {Price},\ and\ \citenamefont {Macdonald}}]{Thorne86}%
  \BibitemOpen
  \bibfield  {author} {\bibinfo {author} {\bibfnamefont {K.~S.}\ \bibnamefont
  {Thorne}}, \bibinfo {author} {\bibfnamefont {R.~H.}\ \bibnamefont {Price}}, \
  and\ \bibinfo {author} {\bibfnamefont {D.~A.}\ \bibnamefont {Macdonald}},\
  }\href@noop {} {\emph {\bibinfo {title} {The Membrane Paradigm}}}\ (\bibinfo
  {publisher} {Yale University Press},\ \bibinfo {address} {New Haven},\
  \bibinfo {year} {1986})\BibitemShut {NoStop}%
\bibitem [{\citenamefont {{Komissarov}}(2001)}]{Komissarov2001}%
  \BibitemOpen
  \bibfield  {author} {\bibinfo {author} {\bibfnamefont {S.~S.}\ \bibnamefont
  {{Komissarov}}},\ }\href {\doibase 10.1046/j.1365-8711.2001.04863.x}
  {\bibfield  {journal} {\bibinfo  {journal} {Mon. Not. Roy. Astron. Soc.}\
  }\textbf {\bibinfo {volume} {326}},\ \bibinfo {pages} {L41} (\bibinfo {year}
  {2001})}\BibitemShut {NoStop}%
\bibitem [{\citenamefont {{McKinney}}\ and\ \citenamefont
  {{Gammie}}(2004)}]{2004ApJ...611..977M}%
  \BibitemOpen
  \bibfield  {author} {\bibinfo {author} {\bibfnamefont {J.~C.}\ \bibnamefont
  {{McKinney}}}\ and\ \bibinfo {author} {\bibfnamefont {C.~F.}\ \bibnamefont
  {{Gammie}}},\ }\href {\doibase 10.1086/422244} {\bibfield  {journal}
  {\bibinfo  {journal} {\apj}\ }\textbf {\bibinfo {volume} {611}},\ \bibinfo
  {pages} {977} (\bibinfo {year} {2004})}\BibitemShut {NoStop}%
\bibitem [{\citenamefont {{Vlahakis}}\ and\ \citenamefont
  {{K{\"o}nigl}}(2003)}]{Vlahakis2003}%
  \BibitemOpen
  \bibfield  {author} {\bibinfo {author} {\bibfnamefont {N.}~\bibnamefont
  {{Vlahakis}}}\ and\ \bibinfo {author} {\bibfnamefont {A.}~\bibnamefont
  {{K{\"o}nigl}}},\ }\href {\doibase 10.1086/378226} {\enquote {\bibinfo
  {title} {{Relativistic Magnetohydrodynamics with Application to Gamma-Ray
  Burst Outflows. I. Theory and Semianalytic Trans-Alfv{\'e}nic Solutions}},}\
  } (\bibinfo {year} {2003})\BibitemShut {NoStop}%
\bibitem [{\citenamefont {Shapiro}(2017)}]{Shapiro:2017cny}%
  \BibitemOpen
  \bibfield  {author} {\bibinfo {author} {\bibfnamefont {S.~L.}\ \bibnamefont
  {Shapiro}},\ }\href {\doibase 10.1103/PhysRevD.95.101303} {\bibfield
  {journal} {\bibinfo  {journal} {Phys. Rev.}\ }\textbf {\bibinfo {volume}
  {D95}},\ \bibinfo {pages} {101303} (\bibinfo {year} {2017})},\ \Eprint
  {http://arxiv.org/abs/1705.04695} {arXiv:1705.04695 [astro-ph.HE]}
  \BibitemShut {NoStop}%
\bibitem [{\citenamefont {Li}\ \emph {et~al.}(2016)\citenamefont {Li},
  \citenamefont {Zhang},\ and\ \citenamefont {Lü}}]{Li:2016pes}%
  \BibitemOpen
  \bibfield  {author} {\bibinfo {author} {\bibfnamefont {Y.}~\bibnamefont
  {Li}}, \bibinfo {author} {\bibfnamefont {B.}~\bibnamefont {Zhang}}, \ and\
  \bibinfo {author} {\bibfnamefont {H.-J.}\ \bibnamefont {Lü}},\ }\href
  {\doibase 10.3847/0067-0049/227/1/7} {\bibfield  {journal} {\bibinfo
  {journal} {Astrophys. J. Suppl.}\ }\textbf {\bibinfo {volume} {227}},\
  \bibinfo {pages} {7} (\bibinfo {year} {2016})},\ \Eprint
  {http://arxiv.org/abs/1608.03383} {arXiv:1608.03383 [astro-ph.HE]}
  \BibitemShut {NoStop}%
\bibitem [{\citenamefont {Metzger}(2017)}]{Metzger:2016pju}%
  \BibitemOpen
  \bibfield  {author} {\bibinfo {author} {\bibfnamefont {B.~D.}\ \bibnamefont
  {Metzger}},\ }\href {\doibase 10.1007/s41114-017-0006-z} {\bibfield
  {journal} {\bibinfo  {journal} {Living Rev. Rel.}\ }\textbf {\bibinfo
  {volume} {20}},\ \bibinfo {pages} {3} (\bibinfo {year} {2017})},\ \Eprint
  {http://arxiv.org/abs/1610.09381} {arXiv:1610.09381 [astro-ph.HE]}
  \BibitemShut {NoStop}%
\bibitem [{\citenamefont {Li}\ and\ \citenamefont
  {Paczynski}(1998)}]{Li:1998bw}%
  \BibitemOpen
  \bibfield  {author} {\bibinfo {author} {\bibfnamefont {L.-X.}\ \bibnamefont
  {Li}}\ and\ \bibinfo {author} {\bibfnamefont {B.}~\bibnamefont {Paczynski}},\
  }\href {\doibase 10.1086/311680} {\bibfield  {journal} {\bibinfo  {journal}
  {Astrophys. J. Lett.}\ }\textbf {\bibinfo {volume} {507}},\ \bibinfo {pages}
  {L59} (\bibinfo {year} {1998})},\ \Eprint
  {http://arxiv.org/abs/astro-ph/9807272} {arXiv:astro-ph/9807272} \BibitemShut
  {NoStop}%
\bibitem [{\citenamefont {{Tanaka}}\ and\ \citenamefont
  {{Hotokezaka}}(2013)}]{2013ApJ...775..113T}%
  \BibitemOpen
  \bibfield  {author} {\bibinfo {author} {\bibfnamefont {M.}~\bibnamefont
  {{Tanaka}}}\ and\ \bibinfo {author} {\bibfnamefont {K.}~\bibnamefont
  {{Hotokezaka}}},\ }\href {\doibase 10.1088/0004-637X/775/2/113} {\bibfield
  {journal} {\bibinfo  {journal} {\apj}\ }\textbf {\bibinfo {volume} {775}},\
  \bibinfo {eid} {113} (\bibinfo {year} {2013})},\ \Eprint
  {http://arxiv.org/abs/1306.3742} {arXiv:1306.3742 [astro-ph.HE]} \BibitemShut
  {NoStop}%
\bibitem [{\citenamefont {{Barnes}}\ and\ \citenamefont
  {{Kasen}}(2013)}]{Barnes:2013wka}%
  \BibitemOpen
  \bibfield  {author} {\bibinfo {author} {\bibfnamefont {J.}~\bibnamefont
  {{Barnes}}}\ and\ \bibinfo {author} {\bibfnamefont {D.}~\bibnamefont
  {{Kasen}}},\ }\href {\doibase 10.1088/0004-637X/775/1/18} {\bibfield
  {journal} {\bibinfo  {journal} {Astrophys. J.}\ }\textbf {\bibinfo {volume}
  {775}},\ \bibinfo {eid} {18} (\bibinfo {year} {2013})},\ \Eprint
  {http://arxiv.org/abs/1303.5787} {arXiv:1303.5787 [astro-ph.HE]} \BibitemShut
  {NoStop}%
\bibitem [{\citenamefont {{Ajith}}\ \emph {et~al.}(2007)\citenamefont
  {{Ajith}}, \citenamefont {{Boyle}}, \citenamefont {{Brown}}, \citenamefont
  {{Fairhurst}}, \citenamefont {{Hannam}}, \citenamefont {{Hinder}},
  \citenamefont {{Husa}}, \citenamefont {{Krishnan}}, \citenamefont {{Mercer}},
  \citenamefont {{Ohme}}, \citenamefont {{Ott}}, \citenamefont {{Read}},
  \citenamefont {{Santamaria}},\ and\ \citenamefont
  {{Whelan}}}]{nijidataformat}%
  \BibitemOpen
  \bibfield  {author} {\bibinfo {author} {\bibfnamefont {P.}~\bibnamefont
  {{Ajith}}}, \bibinfo {author} {\bibfnamefont {M.}~\bibnamefont {{Boyle}}},
  \bibinfo {author} {\bibfnamefont {D.~A.}\ \bibnamefont {{Brown}}}, \bibinfo
  {author} {\bibfnamefont {S.}~\bibnamefont {{Fairhurst}}}, \bibinfo {author}
  {\bibfnamefont {M.}~\bibnamefont {{Hannam}}}, \bibinfo {author}
  {\bibfnamefont {I.}~\bibnamefont {{Hinder}}}, \bibinfo {author}
  {\bibfnamefont {S.}~\bibnamefont {{Husa}}}, \bibinfo {author} {\bibfnamefont
  {B.}~\bibnamefont {{Krishnan}}}, \bibinfo {author} {\bibfnamefont {R.~A.}\
  \bibnamefont {{Mercer}}}, \bibinfo {author} {\bibfnamefont {F.}~\bibnamefont
  {{Ohme}}}, \bibinfo {author} {\bibfnamefont {C.~D.}\ \bibnamefont {{Ott}}},
  \bibinfo {author} {\bibfnamefont {J.~S.}\ \bibnamefont {{Read}}}, \bibinfo
  {author} {\bibfnamefont {L.}~\bibnamefont {{Santamaria}}}, \ and\ \bibinfo
  {author} {\bibfnamefont {J.~T.}\ \bibnamefont {{Whelan}}},\ }\href@noop {}
  {\bibfield  {journal} {\bibinfo  {journal} {arXiv e-prints}\ } (\bibinfo
  {year} {2007})},\ \Eprint {http://arxiv.org/abs/0709.0093} {arXiv:0709.0093
  [gr-qc]} \BibitemShut {NoStop}%
\bibitem [{\citenamefont {Ruiz}\ \emph
  {et~al.}(2020{\natexlab{b}})\citenamefont {Ruiz}, \citenamefont
  {Paschalidis}, \citenamefont {Tsokaros},\ and\ \citenamefont
  {Shapiro}}]{Ruiz:2020elr}%
  \BibitemOpen
  \bibfield  {author} {\bibinfo {author} {\bibfnamefont {M.}~\bibnamefont
  {Ruiz}}, \bibinfo {author} {\bibfnamefont {V.}~\bibnamefont {Paschalidis}},
  \bibinfo {author} {\bibfnamefont {A.}~\bibnamefont {Tsokaros}}, \ and\
  \bibinfo {author} {\bibfnamefont {S.~L.}\ \bibnamefont {Shapiro}},\ }\href
  {\doibase 10.1103/PhysRevD.102.124077} {\bibfield  {journal} {\bibinfo
  {journal} {Phys. Rev. D}\ }\textbf {\bibinfo {volume} {102}},\ \bibinfo
  {pages} {124077} (\bibinfo {year} {2020}{\natexlab{b}})},\ \Eprint
  {http://arxiv.org/abs/2011.08863} {arXiv:2011.08863 [astro-ph.HE]}
  \BibitemShut {NoStop}%
\bibitem [{\citenamefont {Bauswein}\ and\ \citenamefont
  {Stergioulas}(2015)}]{Bauswein:2015yca}%
  \BibitemOpen
  \bibfield  {author} {\bibinfo {author} {\bibfnamefont {A.}~\bibnamefont
  {Bauswein}}\ and\ \bibinfo {author} {\bibfnamefont {N.}~\bibnamefont
  {Stergioulas}},\ }\href {\doibase 10.1103/PhysRevD.91.124056} {\bibfield
  {journal} {\bibinfo  {journal} {Phys. Rev. D}\ }\textbf {\bibinfo {volume}
  {91}},\ \bibinfo {pages} {124056} (\bibinfo {year} {2015})},\ \Eprint
  {http://arxiv.org/abs/1502.03176} {arXiv:1502.03176 [astro-ph.SR]}
  \BibitemShut {NoStop}%
\bibitem [{\citenamefont {Vretinaris}\ \emph {et~al.}(2020)\citenamefont
  {Vretinaris}, \citenamefont {Stergioulas},\ and\ \citenamefont
  {Bauswein}}]{Vretinaris:2019spn}%
  \BibitemOpen
  \bibfield  {author} {\bibinfo {author} {\bibfnamefont {S.}~\bibnamefont
  {Vretinaris}}, \bibinfo {author} {\bibfnamefont {N.}~\bibnamefont
  {Stergioulas}}, \ and\ \bibinfo {author} {\bibfnamefont {A.}~\bibnamefont
  {Bauswein}},\ }\href {\doibase 10.1103/PhysRevD.101.084039} {\bibfield
  {journal} {\bibinfo  {journal} {Phys. Rev. D}\ }\textbf {\bibinfo {volume}
  {101}},\ \bibinfo {pages} {084039} (\bibinfo {year} {2020})},\ \Eprint
  {http://arxiv.org/abs/1910.10856} {arXiv:1910.10856 [gr-qc]} \BibitemShut
  {NoStop}%
\bibitem [{\citenamefont {Takami}\ \emph {et~al.}(2015)\citenamefont {Takami},
  \citenamefont {Rezzolla},\ and\ \citenamefont {Baiotti}}]{Takami:2014tva}%
  \BibitemOpen
  \bibfield  {author} {\bibinfo {author} {\bibfnamefont {K.}~\bibnamefont
  {Takami}}, \bibinfo {author} {\bibfnamefont {L.}~\bibnamefont {Rezzolla}}, \
  and\ \bibinfo {author} {\bibfnamefont {L.}~\bibnamefont {Baiotti}},\ }\href
  {\doibase 10.1103/PhysRevD.91.064001} {\bibfield  {journal} {\bibinfo
  {journal} {Phys. Rev. D}\ }\textbf {\bibinfo {volume} {91}},\ \bibinfo
  {pages} {064001} (\bibinfo {year} {2015})},\ \Eprint
  {http://arxiv.org/abs/1412.3240} {arXiv:1412.3240 [gr-qc]} \BibitemShut
  {NoStop}%
\bibitem [{\citenamefont {Abbott}\ \emph {et~al.}(2021)\citenamefont {Abbott}
  \emph {et~al.}}]{LIGOScientific:2020ibl}%
  \BibitemOpen
  \bibfield  {author} {\bibinfo {author} {\bibfnamefont {R.}~\bibnamefont
  {Abbott}} \emph {et~al.} (\bibinfo {collaboration} {LIGO Scientific,
  Virgo}),\ }\href {\doibase 10.1103/PhysRevX.11.021053} {\bibfield  {journal}
  {\bibinfo  {journal} {Phys. Rev. X}\ }\textbf {\bibinfo {volume} {11}},\
  \bibinfo {pages} {021053} (\bibinfo {year} {2021})},\ \Eprint
  {http://arxiv.org/abs/2010.14527} {arXiv:2010.14527 [gr-qc]} \BibitemShut
  {NoStop}%
\bibitem [{\citenamefont {Abbott}\ \emph {et~al.}(2019)\citenamefont {Abbott}
  \emph {et~al.}}]{LIGOScientific:2018mvr}%
  \BibitemOpen
  \bibfield  {author} {\bibinfo {author} {\bibfnamefont {B.~P.}\ \bibnamefont
  {Abbott}} \emph {et~al.} (\bibinfo {collaboration} {LIGO Scientific,
  Virgo}),\ }\href {\doibase 10.1103/PhysRevX.9.031040} {\bibfield  {journal}
  {\bibinfo  {journal} {Phys. Rev. X}\ }\textbf {\bibinfo {volume} {9}},\
  \bibinfo {pages} {031040} (\bibinfo {year} {2019})},\ \Eprint
  {http://arxiv.org/abs/1811.12907} {arXiv:1811.12907 [astro-ph.HE]}
  \BibitemShut {NoStop}%
\bibitem [{Mov()}]{Moviepage}%
  \BibitemOpen
  \href {http://research.physics.illinois.edu/cta/movies/NSNS_2021} {\enquote
  {\bibinfo {title} {3d visualizations and movies},}\ }\BibitemShut {NoStop}%
\end{thebibliography}%
\end{document}